\documentclass[prb,twocolumn,notitlepage,superscriptaddress]{revtex4-2}
\usepackage{amsmath,amsfonts,amssymb,amsthm,epsfig,array}
\usepackage{dsfont}
\usepackage{slashed}
\usepackage{graphics}
\usepackage{float}
\usepackage{verbatim}
\usepackage{color}
\usepackage[dvipsnames]{xcolor}
\usepackage[hidelinks,colorlinks,linkcolor=blue,
citecolor=blue,urlcolor=blue]{hyperref}
\usepackage[titletoc,title]{appendix}
\usepackage{multirow}
\usepackage{bibentry}
\usepackage{tabularx}
\usepackage[mathscr]{euscript}
\usepackage{mathtools}

\newcommand{\bsl}[1]{\boldsymbol{#1}}

\newcommand{\shpa}{{\mkern3mu\vphantom{\perp}\vrule depth 0pt\mkern2mu\vrule depth 0pt\mkern3mu}}
\newcommand{\shpacap}{\raisebox{0.5pt}{$\scriptscriptstyle\parallel$}}

\renewcommand{\mod}{\,\mathrm{mod}\,}

\newcommand{\bra}[1]{\langle #1|}
\newcommand{\ket}[1]{|#1 \rangle}
\newcommand{\braket}[2]{\left\langle #1 | #2  \right\rangle}

\newcommand{\ii}{\mathrm{i}}

\newcommand{\dsZ}{\mathbb{Z}}
\newcommand{\dsN}{\mathbb{N}}
\newcommand{\dsR}{\mathbb{R}}

\newcommand{\Tr}{\mathop{\mathrm{Tr}}}
\renewcommand{\Re}{\mathop{\mathrm{Re}}}
\renewcommand{\Im}{\mathop{\mathrm{Im}}}

\newcommand{\SO}{\mathrm{SO}}
\newcommand{\SU}{\mathrm{SU}}
\newcommand{\U}{\mathrm{U}}

\newcommand{\eqnref}[1]{Eq.\,\eqref{#1}}
\newcommand{\figref}[1]{Fig.\,\ref{#1}}
\newcommand{\tabref}[1]{Tab.\,\ref{#1}}
\newcommand{\secref}[1]{Sec.\,\ref{#1}}
\newcommand{\appref}[1]{Appendix.\,\ref{#1}}
\newcommand{\refcite}[1]{Ref.\,\cite{#1}}

\newcommand{\mat}[1]{\left(\begin{matrix}#1\end{matrix}\right)}

\newcommand{\eq}[1]{\begin{equation} #1 \end{equation}}
\newcommand{\eql}[1]{\begin{widetext}\begin{align}\begin{split} #1 \end{split}\end{align}\end{widetext}}
\newcommand{\eqa}[1]{\begin{align}\begin{split} #1 \end{split}\end{align}}

\usepackage{environ}
\NewEnviron{eqs}{%
\begin{equation}\begin{split}
    \BODY
\end{split}\end{equation}
}

\let\oldAA\AA
\renewcommand{\AA}{\text{\normalfont\oldAA}}
\newcommand{\sgn}[1]{\text{sgn}(#1)}
\newcommand{\ie}{{\emph{i.e.}}}
\newcommand{\eg}{{\emph{e.g.}}}
\newcommand{\TR}{\mathcal{T}}
\newcommand{\cc}{\mathcal{K}}
\newcommand{\s}{\mathrm{s}}

\newcommand{\W}{\mathcal{W}}

\newcommand{\E}{\mathcal{E}}

\renewcommand{\P}{\mathcal{P}}
\newcommand{\PT}{\mathcal{PT}}
\newcommand{\pf}{\text{pf}}
\newcommand{\M}{\mathcal{M}}
\newcommand{\N}{\mathcal{N}}
\renewcommand{\H}{\mathcal{H}}

\newcommand{\NLs}{NL$^*$}

\begin{document}
\title{Euler Obstructed Cooper Pairing: Nodal Superconductivity and Hinge Majorana Zero Modes}
\author{Jiabin Yu}
\email{jiabinyu@umd.edu}
\affiliation{Condensed Matter Theory Center and Joint Quantum Institute, Department of Physics,
University of Maryland, College Park, MD 20742, USA}
\author{Yu-An Chen}
\affiliation{Condensed Matter Theory Center and Joint Quantum Institute, Department of Physics,
University of Maryland, College Park, MD 20742, USA}
\author{Sankar Das Sarma}
\affiliation{Condensed Matter Theory Center and Joint Quantum Institute, Department of Physics,
University of Maryland, College Park, MD 20742, USA}

\begin{abstract}
    Since the proposal of monopole Cooper pairing in \refcite{Li2018WSMObstructedPairing}, considerable research efforts have been dedicated to the study of Cooper pairing order parameters constrained (or obstructed) by the nontrivial normal-state band topology at Fermi surfaces in 3D systems. 
    {In the current work, we generalize the topologically obstructed pairings between Chern states (like the monopole Cooper pairing) by proposing Euler obstructed Cooper pairing in 3D systems.}
    The Euler obstructed Cooper pairing widely exists between two Fermi surfaces with nontrivial band topology characterized by nonzero Euler numbers; such Fermi surfaces can exist in 3D $\P\TR$-protected spinless-Dirac/nodal-line semimetals with negligible spin-orbit coupling, where $\P\TR$ is the space-time inversion symmetry.
    An Euler obstructed pairing channel must have pairing nodes on the pairing-relevant Fermi surfaces, and the total winding number of the pairing nodes is determined by the sum or difference of the Euler numbers on the Fermi surfaces. 
    In particular, we find that when the normal state is time-reversal invariant and the pairing is weak, a sufficiently-dominant Euler obstructed pairing channel with zero total momentum leads to nodal superconductivity.
    If the Fermi surface splitting is small, the resultant nodal superconductor hosts hinge Majorana zero modes.
    The possible dominance of the Euler obstructed pairing channel near the superconducting transition and the robustness of the hinge Majorana zero modes against disorder are explicitly demonstrated using effective or tight-binding models.
    {Our work presents the first class of higher-order nodal superconductivity originating from the topologically obstructed Cooper pairing.}
\end{abstract}

\maketitle


\section{Introduction}

In the Bardeen–Cooper–Schrieffer (BCS) theory of superconductivity, the topology of normal-state bands~\cite{Hasan2010TI,Qi2010TITSC} is not considered.
Yet, the discovery of 3D topological semimetals~\cite{Wan2011WSM,Burkov2016TSM,Yan2017WSM,Bernevig2018TSM,Vishwanath2018RMPTSM} revealed that the low-energy electronic bands on normal-state Fermi surfaces (FSs) could have nontrivial topology.
Then, a fundamental question naturally arises: what is the interplay between the normal-state band topology and Cooper pairing order parameters? 
More specifically, does the nontrivial band topology on FSs impose any constraints on the possible Cooper pairing order parameters, regardless of the detailed form of the interactions?
Given the increasing experimental attention on the superconductivity in topological semimetals~\cite{Schoop2015ScinTSM,Bian2016SCinTSM,Qi2016SCinWSM,Aggarwal2016SCinDSM,Guguchia2017SCinWSM,Guguchia2019SCinDSM,Yuan2019SCinTSM,Wu2020SCinTSM,Guan2021SCinTSM,Yamada2021SCinNLSM}, the question is also experimentally relevant, in addition to being of conceptual significance.

The question was first addressed for the zero-total-momentum Cooper pairing order parameter in magnetic centrosymmetric Weyl semimetals~ \cite{Li2018WSMObstructedPairing,Murakami2003BerryPhaseMSC}.
It was shown that the nonzero Chern numbers of the low-energy bands on FSs~\cite{Wan2011WSM,TKNN} (i) require the pairing gap function to have zeros (\ie, pairing nodes) on the FSs, and (ii) provide an obstruction to a smooth pairing gap function on the FSs for certain gauge choices of the normal-state basis~\cite{Li2018WSMObstructedPairing}.
The topologically obstructed pairing was called monopole Cooper pairing~\cite{Li2018WSMObstructedPairing}, because monopole Harmonics~\cite{Wu1976MonopoleHarmonics} are needed to fully characterize the pairing matrix.
In the weak-pairing limit, the nodes of the monopole Cooper pairing lead to Weyl points in the spectrum of the Bogoliubov-de Gennes (BdG) Hamiltonian, resulting in a Weyl superconductor (SC) with chiral surface modes~\cite{Li2018WSMObstructedPairing}.
The vortex structure~\cite{Sun2020WSMMSCVortex}, fractional monopole charge~\cite{Li2020MonopoleSCHalfq}, and various other aspects~\cite{Munoz2020MonopoleSC,Park2020SCHofBut} of the monopole Cooper pairing on FSs (as well as the generalization to other order parameters~\cite{Borow2020MonopoleCDWWSM}) were later discussed.
Recently, $\dsZ_2$ obstructed Cooper pairing was proposed in Dirac semimetals~\cite{Sun2020Z2PairingObstruction}, where the normal-state bands on the FSs have nontrivial $\dsZ_2$ topology~\cite{Kane2005Z2} protected by time-reversal (TR) symmetry or a TR-like symmetry; the nodal SC resulting from the $\dsZ_2$ obstructed pairing has helical surface modes.

All previous studies on the topologically obstructed Cooper pairing in 3D systems share the following two features.
First, the normal-state FS band topology is limited to the Altland–Zirnbauer (AZ) ten-fold classification~\cite{Kitaev2009TenFoldWayTITSC,Ryu2010TenFoldWayTITSC,Chiu2016RMPTopoClas} (class A for \refcite{Li2018WSMObstructedPairing} and class AII for \refcite{Sun2020Z2PairingObstruction}).
Second, the boundary signatures of the nodal SC resulting from the topologically obstructed pairing are first-order (\ie, having boundary modes with codimension 1).
An important open question is, therefore, whether Cooper pairing order parameters in 3D systems (i) can be topologically obstructed by the normal-state band topology beyond the AZ classification, and (ii) can lead to nodal SCs with higher-order (\ie, beyond first order) boundary modes.

In this work, we answer this question in the affirmative (for both parts) by proposing the `\emph{Euler obstructed Cooper pairing}.'
The normal-state platforms for the pairing are the $\P\TR$-invariant ``monopole-charged" nodal-line~\cite{Fang2015NLSM,Bouhon2017MonopoleNL,Li2017NLMagnon,Bzdusek2017AZInversionNodal,Nomura2018GraphdiyneNLSM,Ahn2018MonopoleNLSM,Tiwari2020PTNodalLine} and spinless-Dirac~\cite{Lenggenhager2021TPBBC} semimetals with negligible spin-orbit coupling (SOC), where the nontrivial band topology on FSs (more precisely, the nontrivial topology of the top two occupied or partially occupied bands on FSs) is characterized by the Euler class~\cite{Zhao2017PTRealCN,Ahn2019TBGFragile,Slager2019Wilson,Bouhon2020FragileGeometric} or the Wilson loop winding number~\cite{Fang2015NLSM,Ahn2018MonopoleNLSM}.
The band topology is beyond the AZ classification. 

To understand the Euler obstructed Cooper pairing, we introduce a gauge-invariant expression that captures the topology characterized by the Euler class, which we refer to as the Euler number in order to distinguish it from the original expression of the Euler class. (See \secref{sec:normal_state}.)
Between any two FSs with nonzero Euler numbers (the two FSs can be the same), we find that any $\P\TR$-invariant Cooper pairing order parameter can always be split into two channels, as long as the pairing is spin-singlet or is spin-triplet with a momentum-independent spin direction.
Each of the two channels carries its own Euler index, which is determined by the sum or difference of the Euler numbers on the two FSs, and at least one channel has a nonzero Euler index.
We show that a nonzero Euler index requires the corresponding pairing channel to have pairing nodes on the FSs and further determines the total winding number of the pairing nodes.
We also show that a nonzero Euler index provides an obstruction to a smooth pairing gap function of the corresponding pairing channel on the FSs for certain gauges of the normal-state basis.
Thus, a pairing channel with a nonzero Euler index is defined to be Euler obstructed. (See \secref{sec:EOCP}.)

{
Interestingly, as discussed in \refcite{Slager2020EulerOptical,Xie2020TopologyBoundSCTBG,Bouhon2020WeylNonabelian}, there is a special gauge for the normal-state basis on the FSs with nonzero Euler number, which is called Chern gauge.
In the Chern gauge, the basis states have Chern numbers determined by the Euler number, and the Euler obstructed Cooper pairing becomes the pairing between Chern states and thus can be viewed as a $\PT$-invariant generalization of the monopole Cooper pairing proposed in \refcite{Li2018WSMObstructedPairing}.
In other words, the effect of normal-state Euler number on the Cooper pairing order parameter can be straightforwardly seen in the Chern gauge using the theory of the monopole Cooper pairing proposed in \refcite{Li2018WSMObstructedPairing}.
Nevertheless, the connection between the Euler obstructed Cooper pairing and the monopole Cooper pairing only holds for the Chern gauge.
For the Euler obstructed Cooper pairing, the obstruction to smooth
pairing matrix does not rely on the Chern gauge, and
the formalism that describes the topologically enforced
pairing nodes is completely gauge-invariant, leading us to the usage of the terminology ``Euler obstructed Cooper pairing".

Furthermore, besides the 3D systems, nonzero normal-state Euler numbers also exist in 2D systems like magic-angle twisted bilayer graphene (MATBG)~\cite{Cao2018TBGMott,Cao2018TBGSC,Po2018TBGMottSC,Zou2018TBGFragile,Ahn2019TBGFragile}.
Before our work, it has been found that in the Chern gauge of the normal-state basis, the mean-field Cooper pairing order parameters in MATBG can be split into trivial and nontrivial channels based on the Chern numbers of the paired states, and the pairing gap functions of the nontrivial channels must have zeros~\cite{Zaletel2020Sep30SoftModesTBG,Zaletel2020SkyrmionTBG,Zaletel2020AHTBG}.
However, it was unclear whether the constraints imposed by the normal-state Euler numbers on the Cooper pairing order parameters hold beyond the Chern gauge.
Our formalism answered this question in the affirmative for 3D systems by showing that the splitting into a trivial channel and the Euler obstructed channel, as well as the pairing nodes in the Euler obstructed channel, do not rely on the Chern gauge.
This makes the discussion in the Chern gauge just a special case of our general formalism based on Euler numbers, also justifying the usage of the terminology ``Euler obstructed Cooper pairing".
}

We study the physical properties of the Euler obstructed pairing with two additional physical conditions---(i) the $\PT$-invariant normal states are TR-invariant since superconductivity is typically suppressed by the TR-breaking effects like magnetism, and (ii) the pairing is weak and has zero total momentum, which is often the case in BCS-type SCs.
With these conditions, we find that the Euler obstructed pairing channel is generally accompanied by a trivial pairing channel on the FSs.
However, as long as the Euler obstructed pairing channel sufficiently dominates (or equivalently the trivial channel is perturbatively weak compared to the obstructed channel) on the FSs, the SC must be nodal~\footnote{Throughout the work, we always neglect the fine-tuned cases unless specified otherwise.}. 
In particular, if the FS splitting is small (typically compared to the chemical potential), the pairing nodes of the dominant Euler obstructed pairing channel directly lead to zero-energy monopole-charged BdG nodal rings/points, and thus give rise to hinge Majorana zero modes (MZMs). 
Based on an explicit tight-binding-model calculation, we show that the hinge MZMs are robust against chemical-potential disorder and magnetic disorder that preserve the symmetries on average, even if the disorder strength is not perturbatively small. (See \secref{sec:nodalSC}-\ref{sec:hingeMZM}.)
We also demonstrate that the Euler obstructed pairing channel can dominate in a large range of parameter values by explicitly solving the linearized gap equation for an effective model. (See \secref{sec:LGE}.)

Although the higher-order~\cite{Benalcazar2017HOTI,Schindler2018HOTI} nodal superconductivity has been studied in \refcite{Ahn2020HOTSCPT,Zhang2020HODSC,Tiwari2020HODSC,Rui2021HOWSC}, the topologically obstructed Cooper pairing has not been revealed in those cases.
Our work presents a class of higher-order nodal superconductivity that originates from the topologically obstructed Cooper pairing.

\section{Normal-State Platforms and Gauge-invariant Euler number}
\label{sec:normal_state}

In this section, we first discuss the normal-state platforms for the Euler obstructed Cooper pairing, and then introduce the gauge-invariant Euler number.
Details can be found in \appref{app:review_NLSM_Euler}-\ref{app:reformulate_Euler}.
Throughout the entire work, the normal states that we consider are always 3D systems that have the lattice translation symmetry, the $\PT$ symmetry, and the spin $\SU(2)$ symmetry (which often occurs in systems with negligible SOC), unless specified otherwise. 
We suppress the spin index for the study of the normal states in this section.

\subsection{Normal-State Platforms}

In this subsection, we will review the monopole-charged spinless-Dirac semimetals proposed in  \refcite{Lenggenhager2021TPBBC} and the monopole-charged nodal-line semimetals studied in \refcite{Fang2015NLSM,Bouhon2017MonopoleNL,Li2017NLMagnon,Bzdusek2017AZInversionNodal,Nomura2018GraphdiyneNLSM,Ahn2018MonopoleNLSM,Tiwari2020PTNodalLine}, which serve as normal-state platforms for the Euler obstructed Cooper pairing.

\begin{figure}[t]
    \centering
    \includegraphics[width=\columnwidth]{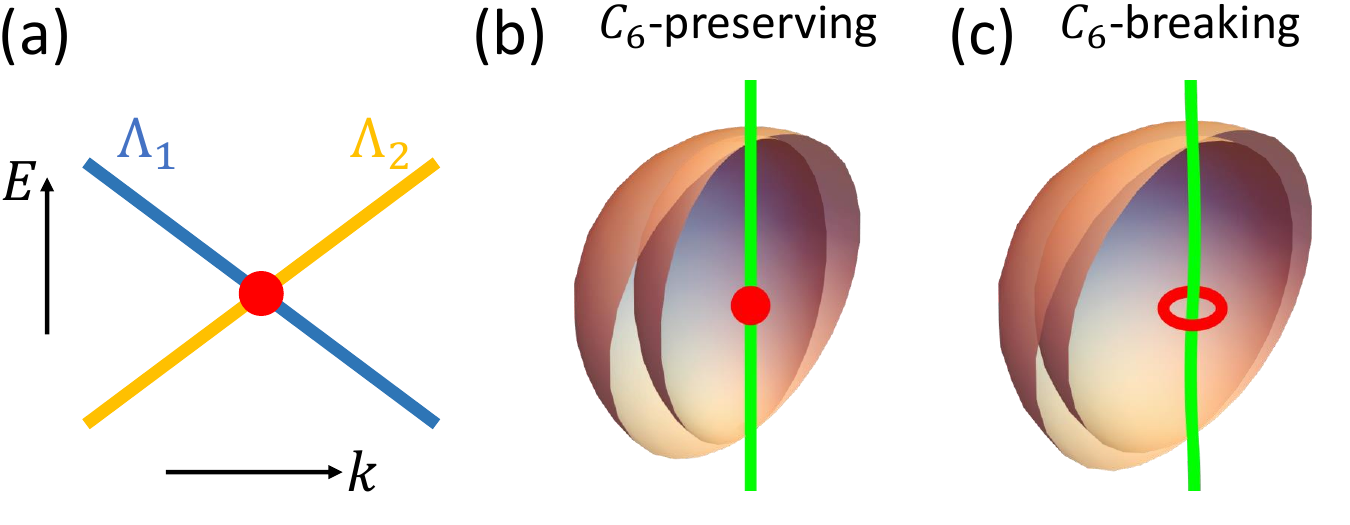}
    \caption{Schematic plots for normal-state platforms, for which the spin index is suppressed.
    In (a), we show a four-fold degenerate MSDP (red) given by the crossing between a doubly degenerate $\Lambda_1$ band (blue) and a doubly degenerate $\Lambda_2$ band (orange) on a $C_6$-invariant axis, where $E$ labels the energy and $k$ is the momentum along the axis.
    In (b), we show a MSDP (red) in the momentum space in the presence of $C_6$ symmetry.
    The green line labels the {\NLs} (at which the top two occupied bands touch) along the $C_6$-invariant axis, and the two orange surfaces are two sphere-like FSs for the chemical potential below the MSDP.
    In (c), we show the nodal structure after perturbatively breaking the $C_6$ symmetry of (b).
    The perturbation expands the MSDP into a MNL (red), but only causes a minor deformation of the {\NLs} (green) and the two sphere-like FSs (orange).
    In (b-c), we only plot half the FSs in order to show their touching at the {\NLs}.
    }
    \label{fig:FS}
\end{figure}

Let us start with a normal-state system that has an extra six-fold rotation $C_6$ symmetry.
The symmetry group spanned by $\PT$ and $C_6$ has (among others) two 2D irreducible co-representations (ICRs)~\cite{Bradley2009MathSSP,Bernevig2020MagneticTQC,Lenggenhager2021TPMNL,Lenggenhager2021TPBBC}, labeled by $\Lambda_1$ and $\Lambda_2$, and the spinless Dirac point corresponds to the four-fold degenerate band crossing between the two 2D ICRs along a $C_6$-invariant axis~\cite{Lenggenhager2021TPBBC}, as schematically shown in \figref{fig:FS}(a).
The spinless Dirac point has a nonzero $\P\TR$-protected $\dsZ_2$ monopole charge ~\cite{Lenggenhager2021TPBBC}, and thus can be called the ``monopole-charged" spinless Dirac point (MSDP).  (See \appref{app:review_NLSM_Euler}-\ref{app:spinless_DSM} for details.)
The MSDP would have eight-fold degeneracy if including spin.

When the MSDP(s) occurs between the empty and occupied bands, the system is a monopole-charged spinless-Dirac semimetal~\cite{Lenggenhager2021TPBBC}.
In a monopole-charged spinless-Dirac semimetal, the top two occupied bands coincide along a line, labeled as {\NLs}, that penetrates the MSDP, and {\NLs} lies on the $C_6$-invariant axis as schematically shown in \figref{fig:FS}(b).
With the chemical potential slightly below the energy of the MSDP, there should be two FSs given by the top two occupied bands, touching at the {\NLs}. 
Roughly speaking, the FSs can be either sphere-like (\figref{fig:FS}(b)) or hyperbolic, similar to the distinction between the type-I~\cite{Wan2011WSM} and type-II~\cite{Soluyanov2015TypeIIWSM} Weyl semimetals.
In this work, we will focus on the sphere-like FSs.

For sufficiently small FSs, the top two occupied bands must be isolated from other bands on each sphere-like FS in \figref{fig:FS}(b).
Here we treat each sphere-like FS as a 2D manifold in the momentum space; on each FS, the topmost occupied band exactly lies at the chemical potential, while the second topmost occupied band can have energy lower than the chemical potential at some momenta.
Then, according to \refcite{Zhao2017PTRealCN,Ahn2018MonopoleNLSM}, the Euler class is well defined for the top two occupied bands on each FS.
{
To show this, let us use $\ket{u_{\bsl{k},n}^{eig}}$ ($n=1,2$) to label the eigenstates of $e^{-\ii \bsl{k}\cdot\bsl{r}} H e^{\ii \bsl{k}\cdot\bsl{r}}$ for the top two occupied bands on one FS $\M$, where $\bsl{k}$ is the Bloch momentum.
We use 
\eq{
\label{eq:basis}
\ket{u_{\bsl{k},a}} = \sum_{n=1,2 } \ket{u_{\bsl{k},n}^{eig}} \widetilde{R}_{na}(\bsl{k})
}
to label the basis of the top two occupied bands, where $a=1,2$ and $\widetilde{R}$ is unitary.
$\ket{u_{\bsl{k},a}}$ is not a Bloch state; rather it is the periodic part of a Bloch state.
For the convenience of later discussion, we will define 
\eq{
\label{eq:vec_basis}
\ket{u_{\bsl{k}}}=(\ket{u_{\bsl{k},1}},\ket{u_{\bsl{k},2}})\ .
}
The unitary $\widetilde{R}$ in \eqnref{eq:basis} is not unique; different choices of $\widetilde{R}$ correspond to different gauges for the normal-state basis $\ket{u_{\bsl{k}}}$.}
Therefore, the basis has a $\bsl{k}$-dependent $\U(2)$ gauge freedom 
\eq{
\label{eq:U2_gauge}
\ket{u_{\bsl{k}}}\rightarrow \ket{u_{\bsl{k}}} R_{\bsl{k}}\ ,
}
where $R_{\bsl{k}}\in \U(2)$.

Imposing the reality condition $\PT \ket{u_{\bsl{k}}} = \ket{u_{\bsl{k}}} $ and choosing an orientation would restrict $R_{\bsl{k}}\in \SO(2)$.
{In this way, we obtained a real oriented gauge for $\ket{u_{\bsl{k}}}$, noted as  $\ket{u_{\bsl{k}}^{RO}}$.
Then, according to \refcite{Zhao2017PTRealCN,Ahn2019TBGFragile}, the Euler class for the real oriented gauge $\ket{u_{\bsl{k}}^{RO}}$ on $\M$ is captured by
\eq{
\label{eq:e_2_original}
e_2=\frac{1}{2\pi}\int_{\M} d\bsl{S}\cdot\bsl{f}_{\bsl{k}}
\ ,
}
where $\bsl{f}_{\bsl{k}}= \nabla_{\bsl{k}} \times \bsl{a}_{\bsl{k}}$ is the real curvature, and $\bsl{a}_{\bsl{k}}=  \pf[\bra{{u_{\bsl{k}}^{RO}}} \nabla_{\bsl{k}} \ket{{u_{\bsl{k}}^{RO}}}]
=
\frac{1}{2}\Tr[-\ii\tau_y\bra{{u_{\bsl{k}}^{RO}}} \nabla_{\bsl{k}} \ket{{u_{\bsl{k}}^{RO}}}]$ with $\tau$'s the Pauli matrices for the spinless normal-state basis in \eqnref{eq:basis}.
$e_2$ is defined only for the real oriented gauges of the normal-state basis.}
For different real oriented gauges, the sign of the $e_2$ varies, while $|e_2|$ stays invariant.
Thus, we mainly use $|e_2|$ instead of $e_2$ in the following.
 (See more details in \appref{app:review_NLSM_Euler}.)

$|e_2|$ does not rely on the $C_6$ symmetry, but the presence of $C_6$ allows us to determine $|e_2|$ based on the symmetry representations, according to \refcite{Ahn2019TBGFragile,Lenggenhager2021TPBBC}.
Specifically, because the top two occupied bands have different 2D ICRs at the two intersection points between the FS and the $C_6$ axis, $|e_2|$ must be odd for any real oriented gauge~\cite{Ahn2019TBGFragile,Lenggenhager2021TPBBC}.
Therefore, on each small enough FS in \figref{fig:FS}(b), the top two occupied bands are isolated, and have odd Euler class $|e_2|$.
The odd $|e_2|$ on small FSs in \figref{fig:FS}(b) agrees with the nonzero $\dsZ_2$ monopole charge of the MSDP~\cite{Ahn2019TBGFragile,Lenggenhager2021TPBBC}.
Furthermore, the nonzero $|e_2|$ requires the top two occupied bands to touch each other on each FS~\cite{Ahn2019TBGFragile,Po2018TBGMottSC,Zou2018TBGFragile}, agreeing with the touching between two FSs. (See \appref{app:review_NLSM_Euler}-\ref{app:spinless_DSM} for details.)

So far, we have reviewed the monopole-charged spinless-Dirac semimetals with the $C_6$ symmetry, and have demonstrated the existence of FSs with odd $|e_2|$ in them for the chemical potential slightly below the MSDP(s).
We emphasize that even if we decrease the chemical potential, the FSs in \figref{fig:FS}(b) should still have odd $|e_2|$, as long as the FSs stay sphere-like and the top two occupied bands stay isolated from other bands on the FSs.
In particular, the existence of the sphere-like FSs with odd $|e_2|$ does not rely on the $C_6$ symmetry.
To show this, let us include an infinitesimal $C_6$-breaking perturbation.
Owing to the nonzero $\dsZ_2$ monopole charge, a MSDP cannot be gapped but turns into a two-fold degenerate monopole-charged nodal line (MNL), as shown in \figref{fig:FS}(c), and the system become a monopole-charged nodal-line semimetal~\cite{Fang2015NLSM,Bouhon2017MonopoleNL,Li2017NLMagnon,Bzdusek2017AZInversionNodal,Nomura2018GraphdiyneNLSM,Ahn2018MonopoleNLSM,Tiwari2020PTNodalLine}.
Furthermore, the {\NLs} still exists and penetrates the MNL owing to the nonzero $\dsZ_2$ monopole charge~\cite{Ahn2018MonopoleNLSM}, and we still have two sphere-like FSs that touch each other at the {\NLs} as the reminiscent of FSs in \figref{fig:FS}(b).
The top two occupied bands on each FS stay isolated and thus still have odd $|e_2|$.

We conclude that the sphere-like FSs with nonzero $|e_2|$ can appear in both the $C_6$-protected monopole-charged spinless-Dirac semimetals and the $C_6$-breaking monopole-charged nodal-line semimetals.
Such FSs are essential for the later discussion of the Euler obstructed Cooper pairing.
Typically, we may expect that there are more than one set of two touching FSs with nonzero $|e_2|$ in one semimetal, since the nonzero $\dsZ_2$ monopole charge always requires both MSDPs and MNLs to appear in pairs~\cite{Ahn2018MonopoleNLSM}.
In the following, we will not impose the $C_6$ symmetry unless specified otherwise, because the $C_6$-invariant case can often be viewed as a special case of the general discussion without imposing $C_6$.

\subsection{Gauge-Invariant Euler Number}
\label{sec:Euler number}

In the above discussion, we use $|e_2|$ (\eqnref{eq:e_2_original}) to capture the Euler class.
The evaluation of $|e_2|$ requires picking a real oriented gauge.
However, given a generic Hamiltonian, picking a real oriented gauge for an isolated set of two bands is not trivial, and requires sophisticated efforts~\cite{Bouhon2020WeylNonabelian}.
Therefore, it is more convenient to have a generalization of $|e_2|$ that is invariant under the gauge transformation \eqnref{eq:U2_gauge}.
Although the Wilson loop winding number~\cite{Ahn2018MonopoleNLSM,Ahn2019TBGFragile,Xie2020TopologyBoundSCTBG} can be viewed as one gauge-invariant generalization of $|e_2|$, it does not provide the $\bsl{k}$-resolved information that is required for the later discussion.
To resolve this issue, we will construct a new gauge-invariant generalization of $|e_2|$ to capture the topology characterized by the Euler class.

To do so, {let us re-write $\bsl{f}_{\bsl{k}}$ in \eqnref{eq:e_2_original} as
\eq{
\bsl{f}_{\bsl{k}}= \frac{1}{\sqrt{2}}\Tr[Q_{\bsl{k}}^{RO} \nabla_{\bsl{k}}P_{\bsl{k}}\times\nabla_{\bsl{k}}P_{\bsl{k}}]\ ,
}
where 
\eqa{
Q_{\bsl{k}}^{RO} & =\frac{1}{\sqrt{2}}\ket{u_{\bsl{k}}^{RO}}(-\ii\tau_y)\bra{u_{\bsl{k}}^{RO}} \\
& =\frac{-\ii}{\sqrt{2}} \sum_{a_1,a_2} \ket{u_{\bsl{k},a_1}^{RO}}\bra{u_{\bsl{k},a_2}^{RO}} (\tau_y)_{a_1 a_2}\ ,
}
$\ket{u_{\bsl{k}}^{RO}}$ labels the real-oriented basis of the top two occupied bands on any sphere-like FS $\M$ in \figref{fig:FS}(b-c), ``$RO$" stands for the real oriented gauge, and $P_{\bsl{k}}=\ket{u_{\bsl{k}}}\bra{u_{\bsl{k}}}$ is the gauge-invariant projection operator with $\ket{u_{\bsl{k}}}$ the basis in arbitrary gauges.
Clearly, $|e_2|$ relies on the real oriented gauge because $Q_{\bsl{k}}^{RO}$ in $\bsl{f}_{\bsl{k}}$ does.
Thus, to generalize the $|e_2|$ to all gauges, we need to generalize $Q_{\bsl{k}}^{RO}$ to gauges other than the real oriented gauge.
First, we replace $\ket{u_{\bsl{k}}^{RO}}$ and $\bra{u_{\bsl{k}}^{RO}}$ in $Q_{\bsl{k}}^{RO}$ by $\ket{u_{\bsl{k}}}$ and $\bra{u_{\bsl{k}}^{\PT}}$, respectively, resulting in 
\eq{
\label{eq:intermediate_Q}
\frac{1}{\sqrt{2}}\ket{u_{\bsl{k}}}(-\ii\tau_y)\bra{u_{\bsl{k}}^{\PT}}\ ,
}
where $\ket{u_{\bsl{k}}^{\PT}}=\PT\ket{u_{\bsl{k}}}$.
The usage of $\bra{u_{\bsl{k}}^{\PT}}$ in \eqnref{eq:intermediate_Q} brings one simplification: it makes \eqnref{eq:intermediate_Q} only gain a $\U(1)$ factor
\eq{
\frac{1}{\sqrt{2}}\ket{u_{\bsl{k}}}(-\ii\tau_y)\bra{u_{\bsl{k}}^{\PT}} \rightarrow \det(R_{\bsl{k}})\frac{1}{\sqrt{2}}\ket{u_{\bsl{k}}}(-\ii\tau_y)\bra{u_{\bsl{k}}^{\PT}}
}
under the gauge transformation \eqnref{eq:U2_gauge}.
But \eqnref{eq:intermediate_Q} is not enough to construct a gauge-invariant expression for $|e_2|$, since it has $\bsl{k}$-dependent gauge freedom.
The remaining task is to convert $\det(R_{\bsl{k}})$ to a $\bsl{k}$-independent factor.

To do so, we define an $\eta_{\bsl{k}_0}(\bsl{k})$ factor as}
\eqa{
\label{eq:eta_path_indep}
& \eta_{\bsl{k}_0}(\bsl{k})=\sqrt{\det[\braket{u^{\PT}_{\bsl{k}_0}}{u_{\bsl{k}_0}}]}\det[W(\bsl{k}_0\xrightarrow{\gamma} \bsl{k})]\ ,
}
where {$\braket{u^{\PT}_{\bsl{k}_0}}{u_{\bsl{k}_0}}$ is a $2\times 2$ matrix since $\ket{u_{\bsl{k}_0}}$ has two components as defined in \eqnref{eq:vec_basis},} $\gamma$ is any path on $\M$ from $\bsl{k}_0$ to $\bsl{k}$,
\eq{
\label{eq:parallel_transport}
W(\bsl{k}_0\xrightarrow[]{\gamma} \bsl{k})=\lim_{L\rightarrow\infty}\bra{u_{\bsl{k}_0}} P_{\bsl{k}_1} P_{\bsl{k}_2} ...P_{\bsl{k}_{L-1}} \ket{u_{\bsl{k}}}
}
is the parallel transport or the Wilson line~\cite{Soluyanov2011WannierZ2,Soluyanov2012SmoothGaugeZ2,Dai2011Z2WilsonLoop,Li2020TIHOTIGaugeInvLines}, and $\bsl{k}_1,...,\bsl{k}_{L-1}$ are arranged sequentially along $\gamma$ from $\bsl{k}_0$ to $\bsl{k}$.
Both $\bsl{k}$ and $\bsl{k}_0$ live on $\M$, and we treat $\bsl{k}_0$ as a base point.
{Under the gauge transformation \eqnref{eq:U2_gauge}, we have 
\eq{
\eta_{\bsl{k}_0}(\bsl{k})\rightarrow \eta_{\bsl{k}_0}(\bsl{k}) \chi(R_{\bsl{k}_0})  \det[R_{\bsl{k}}]\ ,
}
where 
\eq{
\chi(R_{\bsl{k}_0})=\frac{\sqrt{\det[\braket{u^{\PT}_{\bsl{k}_0}}{u_{\bsl{k}_0}}]\det[R_{\bsl{k}_0}]^2} }{\sqrt{\det[\braket{u^{\PT}_{\bsl{k}_0}}{u_{\bsl{k}_0}}]} \det[R_{\bsl{k}_0}] } =\pm 1\ .
}
The above gauge transformation rule comes from 
\eqa{
& \det[W(\bsl{k}_0\xrightarrow{\gamma} \bsl{k})] \rightarrow \det[W(\bsl{k}_0\xrightarrow{\gamma} \bsl{k})] \det[R_{\bsl{k}_0}]^* \det[R_{\bsl{k}}]\\
& \text{ under \eqnref{eq:U2_gauge}}\ , 
} 
and
\eqa{
& \sqrt{\det[\braket{u^{\PT}_{\bsl{k}_0}}{u_{\bsl{k}_0}}]} \rightarrow \sqrt{\det[\braket{u^{\PT}_{\bsl{k}_0}}{u_{\bsl{k}_0}}]} \chi(R_{\bsl{k}_0})\det[R_{\bsl{k}_0}] \\
&\text{ under \eqnref{eq:U2_gauge}}\ .
}
From the gauge transformations rule, we can see $\det[W(\bsl{k}_0\xrightarrow{\gamma} \bsl{k})]$ in $\eta_{\bsl{k}_0}(\bsl{k})$ can convert $\det[R_{\bsl{k}}]$ to $\det[R_{\bsl{k}_0}]$, and $\sqrt{\det[\braket{u^{\PT}_{\bsl{k}_0}}{u_{\bsl{k}_0}}]}$ in $\eta_{\bsl{k}_0}(\bsl{k})$ can convert $\det[R_{\bsl{k}_0}]$ to $\chi(R_{\bsl{k}_0})$, resulting that $\eta_{\bsl{k}_0}(\bsl{k})$ can be used to convert the $\bsl{k}$-dependent $\det[R_{\bsl{k}}]$ to a sign factor $\chi(R_{\bsl{k}_0})$ that only depends on the base point.
Then, we can multiply \eqnref{eq:intermediate_Q} by $\eta_{\bsl{k}_0}^*(\bsl{k})$ to get a quantity that only gets a $\bsl{k}$-independent sign factor under \eqnref{eq:U2_gauge}, which is}
\eq{
\label{eq:Q_k0}
Q_{\bsl{k}_0}(\bsl{k})=-\frac{\eta_{\bsl{k}_0}^*(\bsl{k})}{\sqrt{2}}\ket{u_{\bsl{k}}}\ii \tau_y\bra{u_{\bsl{k}}^{\PT}}\ ,
}
where $Q_{\bsl{k}_0}(\bsl{k})$ is individually defined for each sphere-like FS.
{
As expected, under gauge transformation \eqnref{eq:U2_gauge}, $Q_{\bsl{k}_0}(\bsl{k})$ only gets a $\bsl{k}$-independent sign factor as
\eq{
Q_{\bsl{k}_0}(\bsl{k})\rightarrow \chi(R_{\bsl{k}_0})Q_{\bsl{k}_0}(\bsl{k}) \text{ under \eqnref{eq:U2_gauge}}\ .
}
}

{
$\eta_{\bsl{k}_0}(\bsl{k})$ introduces a new gauge freedom, which is the shift of the bases point: if we shift the base point $\bsl{k}_0$ to $\bsl{k}_0'\in\M$, we have 
\eq{
\label{eq:eta_base_point_change}
\eta_{\bsl{k}_0'}(\bsl{k})=\chi_{\bsl{k}_0',\bsl{k}_0}\eta_{\bsl{k}_0}(\bsl{k})\ ,
}
where 
\eq{
\label{eq:eta_base_point_change_factor}
\chi_{\bsl{k}_0',\bsl{k}_0}=\eta_{\bsl{k}_0'}(\bsl{k}_0)\frac{1}{\sqrt{\det[\braket{u^{\PT}_{\bsl{k}_0}}{u_{\bsl{k}_0}}]}}=\pm 1\ .
} 
So the shift of the base point also gives an extra sign factor that is independent of $\bsl{k}$ as
\eq{
Q_{\bsl{k}_0'}(\bsl{k}) = \chi_{\bsl{k}_0',\bsl{k}_0}Q_{\bsl{k}_0}(\bsl{k})\ .
}
Furthermore, $Q_{\bsl{k}_0}(\bsl{k})$ would reduce to $Q_{\bsl{k}}^{RO}$ for any real oriented gauge, meaning that $Q_{\bsl{k}_0}(\bsl{k})$ (i) is a generalization of $Q_{\bsl{k}}^{RO}$ to all gauges and (ii) only has $\bsl{k}$-independent gauge freedom.
Therefore, we can use $Q_{\bsl{k}_0}(\bsl{k})$ to construct a gauge-invariant expression for $|e_2|$.}

Before constructing such an expression, we mention that since $\det[W(\bsl{k}_0\xrightarrow{\gamma} \bsl{k})]$ is path-independent (\ie, independent of $\gamma$) after fixing $\bsl{k}_0$ and $\bsl{k}$, \eqnref{eq:Q_k0} is path independent. 
The path independent nature makes efficient the numerical evaluation of \eqnref{eq:Q_k0} in practice. (See \appref{app:reformulate_Euler} for more details.)

Now we construct a gauge-invariant expression for $|e_2|$.
With $Q_{\bsl{k}_0}(\bsl{k})$, we can define the generalization of the real curvature in \eqnref{eq:e_2_original} to all gauges as
\eq{
\label{eq:Phi}
\bsl{\Phi}_{\bsl{k}_0}(\bsl{k})=\frac{1}{\sqrt{2}}\Tr[Q_{\bsl{k}_0}(\bsl{k})\nabla_{\bsl{k}}P(\bsl{k})\times\nabla_{\bsl{k}}P(\bsl{k})]\ ,
}
which is real.
We further define a non-negative integer quantity as 
\eq{
\label{eq:N}
\N=|\N_{\bsl{k}_0}|\ ,
}
where 
\eq{
\label{eq:N_k0}
\N_{\bsl{k}_0}=\frac{1}{2\pi}\int_{\M} d\bsl{S}\cdot \bsl{\Phi}_{\bsl{k}_0}(\bsl{k})
}
is integer-valued.
$\bsl{\Phi}_{\bsl{k}_0}(\bsl{k})$ and $\N$ generalize the real curvature and $|e_2|$, respectively, because (i) $\bsl{\Phi}_{\bsl{k}_0}(\bsl{k})$ and $\N$ respectively equal to the real curvature and $|e_2|$ for any real oriented gauge, and (ii) they are well-defined on $\M$ for all gauges.
As mentioned above, under gauge transformations (\eqnref{eq:U2_gauge}) or base-point changes (changing $\bsl{k}_0$), $Q_{\bsl{k}_0}(\bsl{k})$ only gets a $\bsl{k}$-independent sign factor, and thus so do $\bsl{\Phi}_{\bsl{k}_0}(\bsl{k})$ and $\N_{\bsl{k}_0}$.
Then, $\N$ is gauge-invariant and base-point-independent, meaning that $\N$ is a gauge-invariant generalization of $|e_2|$.
We refer to $\N$ as the Euler number to distinguish it from the Euler class $|e_2|$. 
(See \appref{app:reformulate_Euler} for more details.)

On all the sphere-like FSs with nonzero $|e_2|$ discussed above, the top two occupied bands must have nonzero Euler number $\N$.
So we know the sphere-like FSs with nonzero Euler numbers can exist in both the monopole-charged spinless-Dirac semimetal and the monopole-charged nodal-line semimetal.
On any sphere-like FS $\M$ with nonzero Euler number $\N$, we know the gauge-dependent and base-point-dependent $\N_{\bsl{k}_0}$ is also nonzero, allowing us to define a gauge-invariant and base-point-independent $Q$ operator  and $\bsl{\Phi}$ as
\eqa{
\label{eq:Q_Phi}
& Q(\bsl{k})=\frac{\N_{\bsl{k}_0}}{\N} Q_{\bsl{k}_0}(\bsl{k}) \\
& \bsl{\Phi}(\bsl{k})=\frac{\N_{\bsl{k}_0}}{\N} \bsl{\Phi}_{\bsl{k}_0}(\bsl{k})\ .
}
The gauge-invariant Euler number $\N$ (\eqnref{eq:N}), the $Q$ operator, and $\bsl{\Phi}$  (\eqnref{eq:Q_Phi}), which are absent in the formalism of $|e_2|$ and the Wilson loop, are crucial to the discussion of the Euler obstructed Cooper pairing in the next section.
Based on the formalism presented above, we also generalize the modified Nielsen-Ninomiya theorem proposed in \refcite{Ahn2019TBGFragile} to all gauges. (See details in \appref{app:reformulate_Euler}.)

\begin{figure}[t]
    \centering
    \includegraphics[width=0.9\columnwidth]{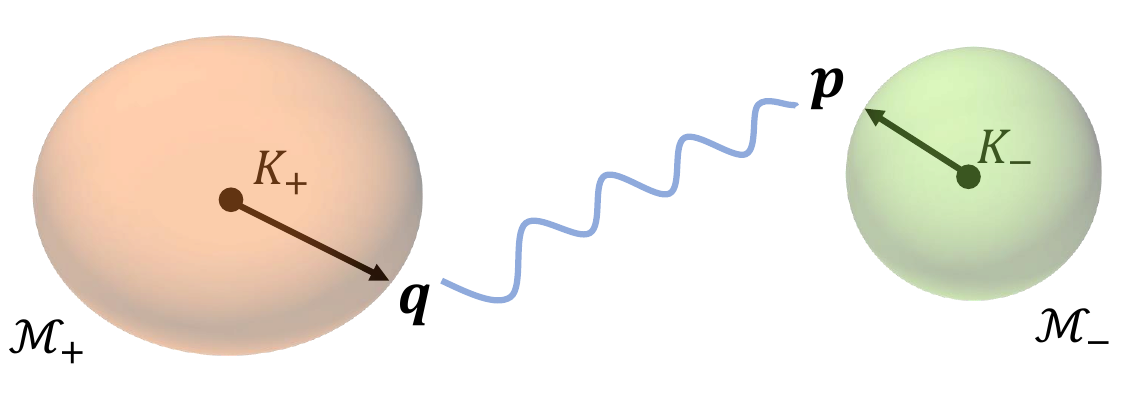}
    \caption{Schematic plots for the Cooper pairs between two sphere-like FSs $\M_\pm$ with nonzero Euler numbers.
    $\bsl{K}_\pm$ are the ``centers" of $\M_\pm$, $\bsl{q}=\bsl{k}-\bsl{K}_+\in\M_{+}$, and $\bsl{p}=\bsl{p}(\bsl{q})\in\M_{-}$.
    In particular, $\bsl{p}(\bsl{q})$ labels a one-to-one correspondence from $\M_{+}$ to $\M_{-}$ that is smooth and has smooth inverse.
    }
    \label{fig:pairing_FS}
\end{figure}

\section{Euler Obstructed Cooper Pairing}
\label{sec:EOCP}

In this section, we discuss the Euler obstructed Cooper pairing in general.
For studying the Cooper pairing, we will keep the spin index, making each normal-state band doubly degenerate.

As schematically shown in \figref{fig:pairing_FS}, we consider two generic sphere-like FSs $\M_{\pm}$ with nonzero Euler numbers $\N_\pm\neq 0$, which may exist in the monopole-charged spinless-Dirac semimetal and the monopole-charged nodal-line semimetal as discussed in the last section.
In particular, we focus on the Cooper pairs between $\M_{\pm}$.
In general, Cooper pairing can only occur to the normal-state electrons with energies close to the chemical potential $\mu$, or more explicitly, within certain superconductivity cutoff $\epsilon_c$ measured from the chemical potential.
Then, on $\M_\pm$, both the top two occupied normal-state bands, each of which is spin-doubly-degenerate now, should be considered, since both of them intersect with the chemical potential owing to band touching required by the nonzero Euler numbers. (See \appref{app:reformulate_Euler} for details.)
On the other hand, we consider the case where the superconductivity cutoff $\epsilon_c$ is much smaller than the gap above and below the top two occupied normal-state bands on $\M_\pm$, meaning that the other normal-state bands can be omitted for the study of Cooper pairing.

Because of the spin double degeneracy, the basis for the top two occupied normal-state bands now has the tensor-product form $\ket{u_{\pm,\bsl{q},a}}\otimes\ket{s}$, where $\ket{u_{\pm,\bsl{q},a}}$ ($a=1,2$) is the above-discussed spinless basis, and $\ket{s}$ ($s=\uparrow,\downarrow$) is the spin basis.
Here $\bsl{q}=\bsl{k}-\bsl{K}_\pm\in\M_{\pm}$, and $\bsl{K}_\pm$ are roughly the ``centers" of the sphere-like $\M_\pm$.
Throughout the work, we explicitly write out the tensor product operation $\otimes$ between the spin part and the rest, just to highlight the spin degree of freedom.
We use $c^{\dagger}_{\pm,\bsl{q},a,s}$ to label the creation operators of the corresponding spinful Bloch basis---the Bloch basis whose periodic part is $\ket{u_{\pm,\bsl{q},a}}\otimes\ket{s}$. 
With $c^{\dagger}_{\pm,\bsl{q}}=(...c^{\dagger}_{\pm,\bsl{q},a,s}...)$, we can write the mean-field Cooper pairing operator that pairs one electron at $\bsl{q}\in\M_{+}$ to another electron at $\bsl{p}=\bsl{p}(\bsl{q})\in\M_{-}$ as
\eq{
\label{eq:H_pairing}
H_{pairing}=\sum_{\bsl{q}\in \M_{+}} c^{\dagger}_{+,\bsl{q}} \Delta(\bsl{q})\otimes\Pi (c^{\dagger}_{-,\bsl{p}})^T + h.c.\ ,
}
where $\Delta(\bsl{q})$ is the $2\times 2$ pairing matrix for the pseudo-spin index $a$, $\Pi$ stands for the $2\times 2$ pairing matrix for the spin index $s$, and $\bsl{p}(\bsl{q})$ is a diffeomorphism from $\M_{+}$ to $\M_{-}$.
In this section, we do not need to specify the exact form of $\bsl{p}(\bsl{q})$ in \eqnref{eq:H_pairing}, and the pairing for $\bsl{q}\in\M_-$ has been included owing to the fermionic anticommutation relation. 
Furthermore, we allow the two FSs to be the same ($\M_+=\M_-$), and thus  \eqnref{eq:H_pairing} is quite general, applicable to both intra-FS and inter-FS Cooper pairing.

Owing to the spin $\SU(2)$ symmetry of the normal state, we can study the spin-singlet and spin-triplet pairings separately.
For spin-singlet pairing, we have
\eq{
\label{eq:Pi_form_spin_singlet}
\Pi=\ii s_y
}
with $s_{x,y,z}$ the Pauli matrices for the spin index, and \eqnref{eq:H_pairing} is in the most general form.
For spin-triplet pairing, we will always choose 
\eq{
\label{eq:Pi_form_spin_triplet}
\Pi=-\bsl{s}\cdot\hat{\bsl{n}} s_y
}
with $\hat{\bsl{n}}$ a $\bsl{q}$-independent unit vector for simplicity.
Therefore, we will always choose $\Pi$ in \eqnref{eq:H_pairing} to be $\bsl{q}$-independent.
Furthermore, we consider the case where the pairing does not spontaneously break the $\PT$ symmetry, indicating that 
\eq{
\label{eq:Delta_PT}
U_{+}(\bsl{q})\Delta^*(\bsl{q}) U_{-}^T(\bsl{p}) = \Delta(\bsl{q})
}
with $\PT c^{\dagger}_{\pm,\bsl{q}} (\PT)^{-1} = c^{\dagger}_{\pm,\bsl{q}}U_{\pm}(\bsl{q})\otimes \ii s_y $.
Next, we will discuss how the $\PT$-invariant pairing \eqnref{eq:H_pairing} is obstructed by the nonzero Euler numbers on $\M_\pm$.

The gauge transformation of the pairing matrix is crucial for the study of topological obstruction.
Similar to \eqnref{eq:U2_gauge}, the spinless basis $\ket{u_{\pm,\bsl{q}}}$ and the pseudo-spin part of the creation operator have $\U(2)$ gauge freedom as
\eqa{
\label{eq:U2_gauge_u_pm}
& \ket{u_{\pm,\bsl{q}}}\rightarrow \ket{u_{\pm,\bsl{q}}} R_{\pm}(\bsl{q}) \\
& c^\dagger_{\pm,\bsl{q},s}\rightarrow c^\dagger_{\pm,\bsl{q},s} R_{\pm}(\bsl{q})\ ,
}
where $R_{\pm}(\bsl{q})$ are $\U(2)$ gauge transformations.
As a result, the pseudo-spin pairing matrix generally has a $\U(2)\times \U(2)$ gauge freedom \eq{
\label{eq:Delta_U2U2}
\Delta(\bsl{q})\rightarrow R_+^\dagger(\bsl{q})\Delta(\bsl{q}) R_-^*(\bsl{p})\ .
}

{We aim to find gauge-invariant $\PT$-invariant channels of the pairing operator (\eqnref{eq:H_pairing}) that have nontrivial properties imposed by the nonzero normal-state Euler number.
Before deriving the general formalism, let us first get some intuition by looking at a special case where the normal states has both $\P$ and $\TR$ symmetries, $\bsl{p}(\bsl{q})=-\bsl{q}$, and we choose the real oriented gauge $\ket{u_{\pm,\bsl{q}}^{RO}}$ such that $\P\ket{u_{+,\bsl{q}}^{RO}}=\ket{u_{-,-\bsl{q}}^{RO}}$.
In this case, we have
\eq{
\Delta(\bsl{q})= \bsl{d}_{\perp}(\bsl{q}) \cdot (\tau_0, \ii \tau_y ) + \bsl{d}_{\shpa}(\bsl{q})\cdot\bsl{\tau}_{\shpa}\ ,
}
where $\bsl{\tau}_{\shpa}=(\tau_z,\tau_x)$.
If we change the real oriented gauge but keep $\P\ket{u_{+,\bsl{q}}^{RO}}=\ket{u_{-,-\bsl{q}}^{RO}}$ and $e_{2,\pm}$ invariant, the gauge transformation is effectively in $\SO(2)$, and thus can be viewed as a rotation within a 2D plane.
Under this transformation, $\bsl{d}_{\perp}(\bsl{q})$ does not change and thus can be viewed as being perpendicular to the 2D plane, while $\bsl{d}_{\shpa}(\bsl{q})$ transforms as a vector and thus can be viewed as being parallel to the 2D plane.
Then, it is natural to split $\Delta(\bsl{q})$ into the following two channels
\eqa{
\label{eq:channel_splitting_RO}
& \Delta_\perp(\bsl{q}) = \bsl{d}_{\perp}(\bsl{q}) \cdot (\tau_0, \ii \tau_y ) \\
& \Delta_\shpa(\bsl{q}) = \bsl{d}_{\shpa}(\bsl{q})\cdot\bsl{\tau}_{\shpa}\ ,
}
and to use $\shpa$ and $\perp$ to label them.
Both channels are $\PT$-invariant since $\bsl{d}_{\perp}(\bsl{q})$ and $\bsl{d}_{\shpa}(\bsl{q})$ are real.
The different gauge transformation rules of $\Delta_\perp$ and $\Delta_\shpa$ discussed above give the intuition that $\Delta_\perp$ and $\Delta_\shpa$ might have different topological properties, implying that at least one of the two channels is nontrivial.
}

Following the intuition, we now derive the general formalism without extra constraints on the normal states, $\bsl{p}(\bsl{q})$ and the gauges.
To do so, we first combine $\ket{u_{\pm,\bsl{q}}}$ with the pairing matrix $\Delta(\bsl{q})$ to construct a gauge-invariant operator  
\eq{ 
\label{eq:P_Delta}
P_{\Delta}(\bsl{q})=\ket{u_{+,\bsl{q}}} \Delta(\bsl{q}) \bra{u_{-,\bsl{p}}^{PT}}\ .
}
We define the $Q_{\pm}(\bsl{q})$ operator for $\ket{u_{\pm,\bsl{q}}}$ on $\M_\pm$ according to \eqnref{eq:Q_Phi}.
Then, we can {generalize the channel splitting in \eqnref{eq:channel_splitting_RO} by} splitting $P_{\Delta}(\bsl{q})$ into two channels $ P_{\Delta}=P_{\perp} + P_{\shpa}$ with
\eqa{
\label{eq:P_Delta_b_two_channels}
P_b(\bsl{q})& =\frac{1}{2}P_{\Delta}(\bsl{q})-\zeta (-1)^b Q_+(\bsl{q}) P_{\Delta}(\bsl{q}) Q_-(\bsl{p}) \\
& = \ket{u_{+,\bsl{q}}} \Delta_b(\bsl{q}) \bra{u_{-,\bsl{p}}^{PT}}\ ,
}
where $b=\perp,\shpa$, $(-1)^\perp=1$, and $(-1)^\shpa=-1$, and {$\Delta_b(\bsl{q})$ will return to \eqnref{eq:channel_splitting_RO} if imposing the extra constraints for \eqnref{eq:channel_splitting_RO}.} (See \appref{app:Euler_Obstructed_Pairing_General} for details.)
{$\zeta$ is a sign factor that is an intrinsic property of $\bsl{p}(\bsl{q})$.
Recall that $\bsl{p}(\bsl{q})$ is a diffeomorfism from $\M_{+}$ to $\M_-$, and we have assigned the normal directions of $\M_{\pm}$ to point outward.
$\bsl{p}(\bsl{q})$ would map the assigned normal direction of $\M_{+}$ to a normal direction of $\M_{-}$ (up to re-scaling).
Then, $\zeta=1$ if the mapped normal direction of $\M_{-}$ is the same as the assigned normal direction of $\M_{-}$; $\zeta=-1$ otherwise.}
For examples, if Cooper pairing happens between two TR-related electrons, we have $\bsl{p}(\bsl{q})=-\bsl{q}$, resulting in $\zeta=-1$.
Here we have chosen (and will always choose) the normal direction of any sphere-like manifold to point outward.
As a result, we have
\eq{
\label{eq:spliting_Delta_b_two_channels}
\Delta(\bsl{q}) = \Delta_\perp(\bsl{q}) +\Delta_\shpa(\bsl{q})\ .
}
As $P_{b}$ is gauge-invariant, $\Delta_b(\bsl{q})$ has the same $\U(2)\times\U(2)$ gauge transformation rule as \eqnref{eq:Delta_U2U2}.
{The channel splitting is orthogonal since 
\eq{
\frac{1}{2}\Tr[P_{b}(\bsl{k}) P_{b'}^\dagger(\bsl{k})] =\frac{1}{2}\Tr[\Delta_{b}(\bsl{k}) \Delta_{b'}^\dagger(\bsl{k})]  = \delta_{bb'} |\Delta_b(\bsl{q})|^2\ ,
}
where 
\eq{
|\Delta_b(\bsl{q})|=\sqrt{\Tr[\Delta_b(\bsl{q})\Delta_b^\dagger(\bsl{q})]/2}
} 
is the gauge-invariant amplitude of $\Delta_b(\bsl{q})$.}
Acting $\PT$ on $P_{b}$ also suggests $\Delta_b(\bsl{q})$ satisfies \eqnref{eq:Delta_PT}. (See detials in \appref{app:Euler_Obstructed_Pairing_General}.)
Therefore, we have {orthogonally} split the pseudo-spin pairing matrix $\Delta(\bsl{q})$ into two channels $\Delta_b(\bsl{q})$ that have the same gauge transformation rule and $\PT$-invariant rule as $\Delta(\bsl{q})$, meaning that we have split the pairing operator (\eqnref{eq:H_pairing}) into two gauge-invariant $\PT$-invariant channels as $H_{pairing}=H_{\perp,pairing}+H_{\shpa,pairing}$ with
\eq{
\label{eq:H_b_pairing}
H_{b,pairing}=\sum_{\bsl{q}\in \M_{+}} c^{\dagger}_{+,\bsl{q}} \Delta_b(\bsl{q})\otimes\Pi (c^{\dagger}_{-,\bsl{p}})^T + h.c.\ .
}

Next, we clarify the topological properties of the two channels.
The channel splitting of the pairing operator based on the gauge group allows us to assign winding numbers to the pairing nodes of each channel.
To do so, we first construct the gauge-invariant 
\eq{
\hat{P}_b(\bsl{q}) =  P_b(\bsl{q})/|\Delta_b(\bsl{q})|\ .
}
We then define a vector field $\bsl{v}_b$ for each channel $\Delta_b$ as
\eq{
\label{eq:v_b}
\bsl{v}_b(\bsl{q})=\frac{1}{\sqrt{2}}\Tr[Q_{+,\bsl{q}} \hat{P}_b(\bsl{q}) \nabla_{\bsl{q}} \hat{P}_b^\dagger(\bsl{q}) ]\ , 
}
which is also gauge invariant.
In particular, the singular behavior of $\bsl{v}_b$ only occurs at the pairing nodes of the $\Delta_b$ channel on $\M_+$ (or equivalently at $\bsl{q}\in \M_+$ that satisfies $|\Delta_b(\bsl{q})|= 0$), and we use $\bsl{q}_{b,i}$ to label the pairing nodes of $\Delta_b$, where $i$ ranges over all pairing nodes.
We further define $D_{b,i}$ as a disk-like open region on $\M_+$ that (i) contains $\bsl{q}_{b,i}$ and (ii) does not contain any other pairing nodes of $\Delta_b$, and then define the winding number of the pairing node $\bsl{q}_{b,i}$ as
\eqa{
\label{eq:W_b_i}
\W_{b,i}& =\frac{1}{2\pi} \int_{\partial D_{b,i}} d\bsl{q}\cdot \bsl{v}_b +\frac{1}{2\pi}\int_{D_{b,i}} d\bsl{S}_+\cdot \Phi_+ \\
& \quad - \frac{(-1)^b}{2\pi} \int_{\bsl{p}(D_{b,i})}d\bsl{S}_-\cdot \Phi_- \ ,
}
where $\partial D_{b,i}$ the boundary of $D_{b,i}$, $\Phi_{\pm}$ are derived for $\ket{u_{\pm,\bsl{q}}}$ on $\M_\pm$ according to \eqnref{eq:Q_Phi}, and $\bsl{p}(D_{b,i})$ labels the region given by mapping $D_{b,i}$ to $\M_-$ through the map $\bsl{p}$.
The winding number $\W_{b,i}$ is integer-valued, and does not depend on the specific shape of $D_{b,i}$. (See \appref{app:Euler_Obstructed_Pairing_General} for details.)

The total winding number of all pairing nodes of $\Delta_b$ is determined by the normal-state Euler numbers $\N_\pm$ as
\eq{
\label{eq:W_b_E_b_N}
\sum_{i} \W_{b,i} = 2 \E_b\ ,
}
where $\E_b$ is called the Euler index of $\Delta_b$ since it is determined by the Euler numbers as
\eq{
\label{eq:E_b_N}
\E_b=\frac{1}{2}\left[\N_+ - (-1)^b \N_-\right]\ .
}
To see this, since $\W_{b,i}$ does not care about the specific shape of $D_{b,i}$, we can choose all $D_{b,i}$ such that (i) they have no intersection, and (ii) $\M-\cup_i D_{b,i}$ has zero measure on $\M$.
In this case, Stokes' theorem naturally suggests that the first term in \eqnref{eq:W_b_i} does not contribute to the total winding, while the last two terms give the Euler index.

\eqnref{eq:W_b_E_b_N} is the main result of this section, which relates the total winding number of the pairing nodes of $\Delta_b$ on $\M_+$ to the Euler numbers on $\M_\pm$.
It also indicates that nonzero $\E_b$ requires $\Delta_b$ to have pairing nodes on $\M_+$, because if $\Delta_b$ does not have any pairing nodes, the total winding number of the pairing nodes must be zero.
As a result, \eqnref{eq:E_b_N} suggests that $\Delta_{\shpa}$ always has pairing nodes, while $\Delta_{\perp}$ can be nodeless if $\N_+ = \N_-$.
Compared to the monopole Cooper pairing proposed in \refcite{Li2018WSMObstructedPairing} whose pairing nodes are enforced by the nonzero monopole charge determined by Chern numbers, here the pairing nodes of a Euler obstructed pairing channel are enforced by the nonzero Euler index determined by Euler numbers.
{
Moreover, conceptually speaking, the zeros of $\Delta_{\shpa}$ are the superconducting generalization of the Euler-number-enforced normal-state band touching points discussed in \refcite{Ahn2019TBGFragile}, since the zeros of $\Delta_{\shpa}$ are also enforced by the normal-state Euler numbers.
}

Besides enforcing pairing nodes, nonzero $\E_b$ can also provide obstruction to smooth $\Delta_{b}(\bsl{q})$ in certain gauges.
For example, we can separate $\M_+$ into two hemispheres, and $\M_-$ will also be separated into two patches according to the one-to-one correspondence $\bsl{p}(\bsl{q})$.
Then, given any real oriented gauge for the normal-state basis based on the patch choice, $\Delta_b(\bsl{q})$ with nonzero $\E_b$ is not smooth in $\M_+$, and cannot be expanded in terms of the normal spherical Harmonics. 
Instead, $\Delta_b(\bsl{q})$ should be expanded in terms of the monopole Harmonics~\cite{Wu1976MonopoleHarmonics,Li2018WSMObstructedPairing} with the monopole charge determined by $\E_b$.

Besides the real oriented gauges, we can also choose a Chern gauge for the normal-state basis, for which two normal-state bands with a nonzero Euler number are converted into two $\PT$-related sectors with opposite Chern numbers~\cite{Xie2020TopologyBoundSCTBG,Bouhon2020WeylNonabelian}.
{ 
Specifically, the Chern gauge is defined as the following.
As shown above, we can choose an oriented real gauge for the basis on $\M_{\pm}$ as $\ket{u^{RO}_{\pm,\bsl{q}}}$.
Then, according to \refcite{Xie2020TopologyBoundSCTBG,Bouhon2020WeylNonabelian}, we can transform the real oriented gauge to a complex gauge for the basis on $\M_{\pm}$ as
\eqa{
\label{eq:Chern_gauge_pm}
& \ket{u^{Ch}_{\pm,\bsl{q}}} = \frac{1}{\sqrt{2}}\ket{u^{RO}_{\pm,\bsl{q}}}\mat{1 & 1 \\ \ii & -\ii}\ .
}
$\ket{u^{Ch}_{a,\bsl{q}}}$ have a well-defined Chern number $Ch_a$ as
\eq{
Ch_a =(-1)^{a-1} e_2\ ,
}
and thus $\ket{u^{Ch}_{a,\bsl{q}}}$ is called a Chern gauge.}
With a two-patch Chern gauge, the elements of $\Delta_b$ becomes the pairing between Chern states with well defined Chern numbers, and then the channel $\Delta_b$ with nonzero $\E_b$ can be viewed as a $\PT$-protected double version of the monopole Cooper pairing proposed in \refcite{Li2018WSMObstructedPairing}, which does not have smooth representations. (See \appref{app:Euler_Obstructed_Pairing_General} for details.)
We would like to emphasize that the obstruction to smooth representations can only happen to certain gauges, and the connection between our proposed Euler obstructed Cooper pairing and the monopole Cooper pairing proposed in \refcite{Li2018WSMObstructedPairing} only holds for the Chern gauge, while the above discussion of the enforced pairing nodes is completely gauge-independent.

Now we conclude this section.
Between any two sphere-like FSs with nonzero Euler numbers, the $\PT$-invariant Cooper pairing order parameter with momentum-independent spin part (\eqnref{eq:H_pairing}) can always be split into two channels (\eqnref{eq:spliting_Delta_b_two_channels}).
Each channel has its own Euler index determined by the sum or difference of Euler numbers on the two FSs, and the Euler index determines the total winding number of the pairing nodes of the channel on one FS (\eqnref{eq:W_b_E_b_N}).
A nonzero Euler index requires the corresponding channel to have pairing nodes on the FSs, and provides obstruction to the smooth matrix representation of the channel for certain gauges of the normal-state basis.
We refer to a pairing channel with a nonzero Euler index as being Euler obstructed, and at least one of the two channels in \eqnref{eq:spliting_Delta_b_two_channels} is Euler obstructed.

\section{Euler Obstructed Cooper Pairing in Semimetals with $\TR$ and $\P$}

In this section, we will focus on the TR-invariant centrosymmetric (inversion-invariant) normal-state platforms and discuss the physical consequences of the Euler obstructed Cooper pairing in them.
The TR symmetry requires the normal state to be nonmagnetic (and also free of any other TR-breaking effects).
In particular, we focus on the $\PT$-invariant Cooper pairing order parameter that has zero total momentum.
Therefore, we should consider two inversion-related (or TR-related) sphere-like FSs $\M_\pm$ with nonzero Euler numbers, meaning that $\M_-=-\M_+$, $\bsl{K}_-=-\bsl{K}_+$, and $\bsl{p}=-\bsl{q}$ for \eqnref{eq:H_pairing} and \figref{fig:pairing_FS}.
As a result, the pairing operator in \eqnref{eq:H_pairing} becomes
\eq{
\label{eq:H_pairing_TR}
H_{pairing}=\sum_{\bsl{q}\in \M_{+}} c^{\dagger}_{+,\bsl{q}} \Delta(\bsl{q})\otimes\Pi (c^{\dagger}_{-,-\bsl{q}})^T + h.c.\ ,
}
where the pairing for $c^{\dagger}_{-,\bsl{q},a,s} c^{\dagger}_{+,-\bsl{q},a',s'}$ with $\bsl{q}\in\M_-$ has been included based on the anticommutation relation for fermions. 
According to \eqnref{eq:spliting_Delta_b_two_channels}, the pseudo-spin pairing matrix $\Delta$ can be split into two channels $\Delta_{\perp}$ and $\Delta_{\shpa}$ based on the gauge group, and they have their own Euler indices $\E_{\perp}$ and $\E_{\shpa}$ determined by the Euler numbers $\N_\pm$ through \eqnref{eq:E_b_N}.
Owing to TR symmetry, the Euler numbers on two FSs are the same ($\N_+ = \N_-\neq 0$), resulting in 
\eqa{
\label{eq:E_b_N_TR}
\E_{\perp}& =\frac{1}{2}(\N_+ - \N_-) = 0 \\
\E_{\shpa}& =\frac{1}{2}(\N_+ + \N_-) = \N_+ \neq 0\ .
}
It means that only the $\Delta_{\shpa}$ pairing channel is Euler obstructed and must have pairing nodes on the FSs $\M_\pm$, whereas $\Delta_{\perp}$ is allowed to be nodeless.
(See \appref{app:PandT} for more details.)

\subsection{Nodal Superconductivity}
\label{sec:nodalSC}

In this subsection, we will discuss the nodal superconductivity induced by the Euler obstructed $\Delta_{\shpa}$.
As discussed above, the nonzero Euler index $\E_\shpa$ requires $\Delta_{\shpa}$ to have pairing nodes on the FS $\M_+$.
So if $\Delta_{\shpa}$ exists on the FS by itself (or equivalently $\Delta_\perp=0$ on the FS), the pairing nodes of $\Delta_{\shpa}$ on the FS $\M_+$ become the zero-energy gapless points of the BdG Hamiltonian (called the BdG point nodes), which define the nodal superconductivity.
However, the trivial $\Delta_{\perp}$ is allowed to coexist with the Euler obstructed $\Delta_{\shpa}$ in general, because $\Delta_{\shpa}$ and $\Delta_{\perp}$ are separated based on the gauge group, which is not a physical symmetry of the normal state.
Then, the remaining question becomes whether the BdG point nodes would be gapped out if a perturbatively small $\Delta_{\perp}$ is included.
If not, we can say that the nodal superconductivity can be induced by a sufficiently-dominant Euler obstructed $\Delta_{\shpa}$.

To address this question, we first use the symmetries of the normal state to split the pairing operator into channels that typically cannot coexist.
Since the spin $\SU(2)$ symmetry has been exploited in \eqnref{eq:H_pairing} to separate the spin-singlet channel from the spin-triplet channel, we will use the inversion symmetry to further split the pairing based on the parity.
To exploit the inversion symmetry, we choose the gauge for the normal-state basis such that 
\eqa{
\label{eq:reps_PT_T_P}
& \PT c^{\dagger}_{\pm,\bsl{q}} (\PT)^{-1}= c^{\dagger}_{\pm,\bsl{q}} \tau_0\otimes \ii s_y \\  & \TR c^{\dagger}_{\pm,\bsl{q}} \TR^{-1}= c^{\dagger}_{\mp,-\bsl{q}} \tau_0\otimes \ii s_y \\
& \P c^{\dagger}_{\pm,\bsl{q}} \P^{-1}= c^{\dagger}_{\mp,-\bsl{q}} \tau_0\otimes s_0 \ ;
}
the gauge choice does not lose any generality since we have shown that the enforced pairing nodes for the Euler obstructed channel are gauge-independent.
With this gauge, the expressions of $\Delta_{\perp}$ and $\Delta_{\shpa}$ for different parities and spin channels are shown in \tabref{tab:parity_pairing}, where we have used $\zeta=-1$ for $\bsl{p}(\bsl{q})=-\bsl{q}$. (See \appref{app:PandT} for details.)
Since we care about the Euler obstructed $\Delta_\shpa$, we will only consider the spin-singlet parity-even pairing or the spin-triplet parity-odd pairing, as the other two channels have vanishing $\Delta_\shpa$ according to \tabref{tab:parity_pairing}.
In this case, the pseudo-spin pairing matrix reads
\eq{
\label{eq:ssPeven_stPodd}
\Delta= \Delta_\perp +\Delta_\shpa = d_{0} \tau_0 +\bsl{d}_{\shpa}\cdot\bsl{\tau}_\shpa\ ,
}
where $\Delta_\perp  = d_{0} \tau_0$, $\Delta_\shpa=\bsl{d}_{\shpa}\cdot\bsl{\tau}_\shpa$, $\bsl{d}_{\shpa}=(d_z,d_x)$, and the $\bsl{q}$-dependence of $d_{0}$ and $\bsl{d}_{\shpa}$ is implicit.
The simplified form of the pairing above will be justified near the superconducting transition in \secref{sec:LGE}.

\begin{table}[t]
    \centering
    \begin{tabular}{|c|c|c|}
    \hline
        &  parity-even & parity-odd\\
    \hline
         spin-singlet (\eqnref{eq:Pi_form_spin_singlet})  & $(d_0\tau_0,\bsl{d}_\shpa\cdot\bsl{\tau}_\shpa)$ & $(d_y\ii\tau_y,0)$\\
         \hline
         spin-triplet (\eqnref{eq:Pi_form_spin_triplet})  & $(d_y\ii\tau_y,0)$ & $(d_0\tau_0,\bsl{d}_\shpa\cdot\bsl{\tau}_\shpa)$\\
         \hline
    \end{tabular}
    \caption{The expressions of $(\Delta_\perp,\Delta_{\shpacap})$ for different parities and spin-channels of the pairing operator (\eqnref{eq:H_pairing_TR} with \eqnref{eq:spliting_Delta_b_two_channels}) after choosing the gauge in \eqnref{eq:reps_PT_T_P}.
    The momentum dependence of $\Delta_{\perp,{\shpacap}}$ and $d_{0,x,y,z}$ is implied, and $d_{0,x,y,z}\in \dsR$.
    The normal-state platform is chosen to preserve the TR and inversion symmetries.
    }
    \label{tab:parity_pairing}
\end{table}

With the above simplification of the pairing, we now discuss the BdG nodes in general, without assuming the dominance of the Euler obstructed $\Delta_\shpa$.
In general, the BdG Hamiltonians on $\M_\pm$, labeled by $H_{BdG,\pm}$, are related by the particle-hole symmetry (or particle-hole redundancy), and thus we only need to study the nodes of $H_{BdG,+}$.
$H_{BdG,+}$ has the spin $\SU(2)$ symmetry for spin-singlet pairing (\eqnref{eq:Pi_form_spin_singlet}); for the spin-triplet pairing (\eqnref{eq:Pi_form_spin_triplet}), we can first rotate the spin to make $\Pi=- s_z s_y$, and then $H_{BdG,+}$ is invariant under the spin $\U(1)$ rotation along $z$.
Thus, in both cases, we can block diagonalize $H_{BdG,+}$ into two spin blocks $H_{BdG,+,\uparrow}$ and $H_{BdG,+,\downarrow}$, according to the conserved spin component along $z$.
It turns out the two spin blocks are related by either the spin rotation symmetry or the combined spin-charge rotation symmetry, meaning that we only need to study $H_{BdG,+,\uparrow}$, while the  $H_{BdG,+,\downarrow}$ has the same BdG nodes as $H_{BdG,+,\uparrow}$.
Combined with \eqnref{eq:ssPeven_stPodd}, the most general form of the matrix representation of $H_{BdG,+,\uparrow}$ reads
\eqa{
\label{eq:H_cal}
& \H(\bsl{q}) =\mat{ \epsilon \tau_0 + m \bsl{g}_{\shpa} \cdot\bsl{\tau}_{\shpa} & d_{0} \tau_0+\bsl{d}_{\shpa}\cdot\bsl{\tau}_\shpa \\
d_{0} \tau_0+\bsl{d}_{\shpa}\cdot\bsl{\tau}_\shpa & -[\epsilon \tau_0+ m \bsl{g}_{\shpa} \cdot\bsl{\tau}_{\shpa}] 
} \ ,
}
where the $\bsl{q}$-dependence of $\epsilon$, $\bsl{g}_{\shpa}=(g_z,g_x)$, $d_{0}$, and $\bsl{d}_{\shpa}$ is implicit, $\epsilon(\bsl{q})=E_0(\bsl{q})-\mu$ with $E_0(\bsl{q})$ the average energy of the top-two occupied normal-state bands on $\M_+$, and $m$ measures the FS splitting.
(See \appref{app:PandT} for details.)

The spin-up block $\H$ on $\M_+$ has an effective $\PT$ symmetry as 
\eq{
\label{eq:H_cal_PT}
\H^*(\bsl{k})=\H(\bsl{k})\ ,
}
and also has an effective chiral symmetry as
\eq{
\label{eq:H_cal_chiral}
\rho_y\tau_0\H(\bsl{q})\rho_y\tau_0=-\H(\bsl{q})\ .
}
Since the effective $\PT$ and effective chiral symmetries anticommute with each other, $\H$ belongs to the CI nodal class~\cite{Bzdusek2017AZInversionNodal}, which can support the zero-energy gapless lines of the BdG Hamiltonian (called the BdG line nodes) in the 3D momentum space.
The BdG line nodes are classified by the $\dsZ_2$ monopole charge protected by the effective $\PT$ symmetry.

\begin{figure*}[t]
    \centering
    \includegraphics[width=1.9\columnwidth]{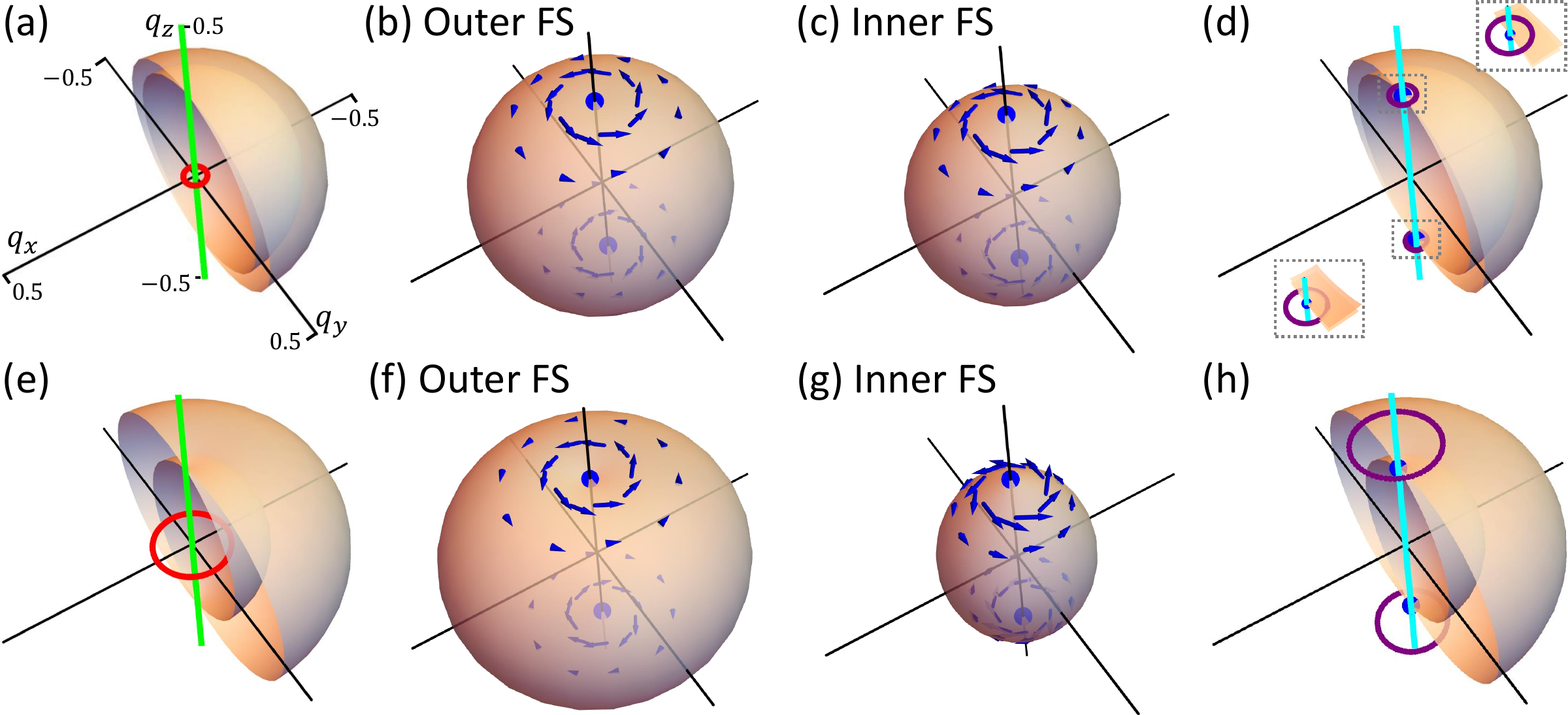}
    \caption{FSs and Euler obstructed Cooper pairing for the effective model \eqnref{eq:H_pm_eff} and \eqnref{eq:pairing_eff_Podd_st} around one valley $\bsl{K}_+$.
    (a-d) are for the parameter set A, and (e-h) are for the parameter set B, where sets A and B are defined in \eqnref{eq:eff_parameter_set}.
    The coordinate systems in all plots are exactly the same, and the orange surfaces in all plots are the normal-state FSs with Euler numbers being 1.
    In (a,e), we plot the normal-state MNL (red) and the normal-state {\NLs} (green), and the {\NLs} exactly aligns with and covers the $q_z$ axis.
    In (b,c,f,g), we plot the pairing nodes (blue dots) and the vector field $\bsl{v}_{\shpacap}$ (blue arrows) of the Euler obstructed $\Delta_{\shpacap}$ on the FSs in (a,e).
    We only plot the projection of $\bsl{v}_{\shpacap}$ on the tangent plane of the FS, since only these components are related to the Euler index, and we omit $\bsl{v}_{\shpacap}$ when its amplitude is too large or too small for clarity.
    In (d,h), we plot the zero-energy BdG MNLs (purple) and the BdG {\NLs} (cyan) for the spin-up block of the BdG Hamiltonian, as well as the pairing nodes (blue dots) of $\Delta_{\shpacap}$ on the FSs.
    The spin-down block has the same nodal structure as (d,h).
    Here the normal-state (BdG) {\NLs} stands for the touching line between the top two occupied bands of the normal-state (BdG) Hamiltonian in each spin subspace. 
    }
    \label{fig:EOP_Nodal}
\end{figure*}

Next, we show the BdG nodes (lines or points) of $\H$ can originate from a sufficiently-dominant Euler obstructed $\Delta_\shpa$.
Let us start with a special case where $m=0$ and $|\Delta_\perp|=|d_0|=0$ on $\M_+$ (and slightly away from $\M_+$).
As mentioned above, in this case, all pairing nodes of $|\Delta_\shpa|=|\bsl{d}_\shpa|$ on $\M_+$, which are required by its nonzero Euler index $\E_\shpa$, become the BdG point nodes of $\H$.
Since the FS surface splitting $m$ is zero, those BdG point nodes are actually zeros of $\H(\bsl{q})$, having four-fold degeneracy.
A direct calculation shows that the BdG point nodes have nonzero $\dsZ_2$ monopole charges. (See \appref{app:PandT} for details.)
Then, even if we include infinitesimal $m$ and $\Delta_\perp$, which preserve the effective $\PT$ and chiral symmetries, the BdG point nodes cannot be gapped out but are expanded into the doubly-degenerate BdG line nodes with nonzero $\dsZ_2$ monopole charge---the zero-energy BdG MNLs.
As we increase $m$, the zero-energy BdG MNLs should typically exist as long as $|m|\ll|\mu|$, since the four-fold degenerate BdG point nodes for $m=0$ are typically scattered in the momentum space with distances on the scale of the Fermi momentum.
Therefore, when the FS splitting is small (typically compared to the chemical potential), a sufficiently-dominant Euler obstructed $\Delta_\shpa$ leads to nodal SC with zero-energy BdG MNLs in each spin subspace.
The zero-energy BdG MNLs should be close to the pairing nodes of $\Delta_\shpa$ on FSs, since the former originate from the latter.

Next, we consider the case where the FS splitting becomes large (\eg, comparable with the chemical potential).
For the special case where $|\Delta_\perp|=|d_0|=0$ on (and slightly away from) $\M_+$, the pairing nodes of $\Delta_\shpa$ on $\M_+$, enforced by $\E_\shpa\neq 0$, again become the BdG point nodes of $\H$ on $\M_+$.
If a BdG point node is given by a pairing node of $\Delta_\shpa$ that coincides with a zero of $\bsl{g}_\shpa$ in \eqnref{eq:H_cal}, it is four-fold degenerate and has nonzero $\dsZ_2$ monopole charges, similar as above.
Otherwise, the BdG point node is doubly degenerate and has a $\pi$ Berry phase along an infinitesimal circle that encloses the node on $\M_+$. 
As a circle with $\pi$ Berry phase in 3D momentum space always encloses a $\PT$-protected nodal line, the doubly-degenerate BdG point node on $\M_+$ should be given by the intersection between a BdG line node and $\M_+$. (See \appref{app:PandT} for details.)
Then, in both cases, the SC is nodal, and remains nodal (with BdG line nodes) even if we add a perturbatively small $\Delta_\perp$ owing to the stable topological invariants of the BdG nodes.
Therefore, even if the FS splitting becomes large, a sufficiently-dominant Euler obstructed $\Delta_\shpa$ on $\M_+$ still leads to nodal SC, though the $\dsZ_2$ monopole charge of the BdG line nodes might be trivial.

The above discussion is done for the BdG Hamiltonian on $\M_\pm$.
In practice, we often care about the full BdG Hamiltonian in the atomic basis, which gives the BdG Hamiltonian on $\M_\pm$ by projection.
It turns out as long as the pairing is weak---the maximum pairing amplitude is much smaller than the gap above and below the top two occupied normal-state bands on $\M_\pm$, all the above BdG line nodes of $\H$ should still exist in the full BdG Hamiltonian, since the effective $\PT$ and effective chiral symmetries of \eqnref{eq:H_cal} still exist for the full BdG Hamiltonian. (See \appref{app:PandT} for details.)
Then, we conclude that for weak pairing, a sufficiently-dominant Euler obstructed $\Delta_\shpa$ on $\M_\pm$ always leads to nodal SC; if the FS splitting is small, the nodal SC contains the zero-energy BdG MNLs in each spin subspace that originate from the pairing nodes of $\Delta_\shpa$.
If the $C_6$ symmetry exists, the zero-energy BdG MNLs may shrink to the zero-energy BdG MSDPs.

To verify the above general discussion, we next study an effective model built from atomic orbitals.
The normal-state MNLs have been predicted to exist in a 3D graphdiyne~\cite{Nomura2018GraphdiyneNLSM,Ahn2018MonopoleNLSM}, in which there are two inversion-related MNLs centered at two valleys $\bsl{K}_\pm$ with $\bsl{K}_-=-\bsl{K}_+$.
We adopt the effective model for the two MNLs in the 3D graphdiyne, which reads
\eq{
\label{eq:H_pm_eff}
H_\pm^{eff}=\sum_{\bsl{q}}\Psi^\dagger_{\pm,\bsl{q}} h_{\pm}^{eff}(\bsl{q})\otimes s_0 \Psi_{\pm,\bsl{q}}\ ,
}
where
\eq{
h_{\pm}^{eff}(\bsl{q})=q_x\tau_x\sigma_x - q_y\tau_x\sigma_z \pm q_z\tau_z\sigma_0 \pm m \tau_y\sigma_y\ .
}
Here we can treat $\Psi^\dagger_{\bsl{k}}=(...,\Psi^\dagger_{\bsl{k},a',s},...)$ as basis derived from the atomic orbitals with $a'=1,2,3,4$ labelling the orbital and sublattice degrees of freedom, and then $\Psi^\dagger_{\pm,\bsl{q}}=\Psi^\dagger_{\bsl{K}_\pm+\bsl{q}}$ with $\bsl{K}_\pm=(0,0,\pm K_0)$.
Furthermore, both $\tau$ and $\sigma$ are now Pauli matrices for the $a'$ index, and $m$ measures the value of the FS splitting.
Compared to the more realistic model in \refcite{Nomura2018GraphdiyneNLSM}, we have omitted the identity term and have chosen a proper basis to make the matrix Hamiltonian real for simplicity.
The model has the TR and inversion symmetries, which are represented as $\TR  \Psi^\dagger_{+,\bsl{q}} (\TR)^{-1} = \Psi^\dagger_{-,-\bsl{q}} U_{\TR}\otimes\ii\s_y$ and $\P  \Psi^\dagger_{+,\bsl{q}} (\P)^{-1} = \Psi^\dagger_{-,-\bsl{q}} U_{\TR}\otimes s_0$, where $U_{\TR}=\tau_z\sigma_0$.
We further include a $\PT$-invariant parity-odd spin-triplet pairing
\eq{
\label{eq:pairing_eff_Podd_st}
H_{pairing}^{eff} =\sum_{\bsl{q}} \Psi_{+,\bsl{q}}^\dagger \Delta_{eff}  (\tau_{y}\sigma_y U_{\TR}^T)  \otimes \ii  s_x (\Psi_{-,-\bsl{q}}^\dagger)^T + h.c.
}
with $\Delta_{eff}\in \dsR$.
The effective BdG Hamiltonian is given by \eqnref{eq:H_pm_eff} and \eqnref{eq:pairing_eff_Podd_st}.

Let us consider two sets of parameter values
\eqa{
\label{eq:eff_parameter_set}
& \text{Set A:  } m=0.03,\ \mu=-0.3,\ \Delta_{eff}=0.03\\
& \text{Set B:  } m=0.1,\ \mu=-0.3,\ \Delta_{eff}=0.2\ .
}
Here we have chosen and will always choose the unit system in which $\hbar=c=k_B=1$ and the momentum cutoff (or the lattice constant) is 1.
For both sets of the parameter values, we have a normal-state MNL together with two sphere-like FSs at each valley (see \figref{fig:EOP_Nodal}(a,e) for $\bsl{K}_+$), and the top two occupied normal-state bands have Euler numbers equal to 1 on each FS, meaning that $\E_\perp=0$ and $\E_\shpa=1$ on each FS around $\bsl{K}_+$ according to \eqnref{eq:E_b_N_TR}.
One physical difference between the two parameter sets is that the FS splitting is much smaller than the chemical potential $|m|\ll |\mu|$ for set A, while the FS splitting is not small for set B.
The other physical difference is that the pairing amplitude $\Delta_{eff}$ is much smaller than the minimum $E_g$ of the gap above the two occupied normal-state bands on FSs for set A as $\Delta_{eff}=0.03\ll E_g=0.48$, while $\Delta_{eff}=0.2=E_g$ for set B.
Thus, we have small FS splitting and weak pairing for set A, but large FS splitting and strong pairing for set B. (See \appref{app:model} for more details.)

According to \tabref{tab:parity_pairing} and \eqnref{eq:E_b_N}, the pairing in \eqnref{eq:pairing_eff_Podd_st} should contain a trivial $\Delta_\perp$ and an Euler obstructed $\Delta_\shpa$ on the FSs with nonzero Euler numbers.
To verify it, we project the pairing onto both FSs around $\bsl{K}_+$, and indeed get nonzero $\Delta_\perp$ and $\Delta_\shpa$, agreeing with \tabref{tab:parity_pairing}.
We further calculate the gauge-invariant vector field $\bsl{v}_\shpa$ for $\Delta_\shpa$ according to \eqnref{eq:v_b}, and find the pairing nodes of the Euler obstructed $\Delta_\shpa$ coincide with the vortices of $\bsl{v}_\shpa$ as shown in \figref{fig:EOP_Nodal}(b-c, f-g).
In particular, each pairing node of $\Delta_\shpa$ in \figref{fig:EOP_Nodal}(b-c, f-g) has winding number (\eqnref{eq:W_b_i}) being 1, and thus the total winding number of pairing nodes is 2 on each FS around $\bsl{K}_+$, agreeing with $\E_\shpa=1$ and \eqnref{eq:W_b_E_b_N}.

To test the dominance of the Euler obstructed $\Delta_\shpa$, we calculate the ratio between the averaged magnitudes of $\Delta_\perp$ and $\Delta_\shpa$ channels as
\eq{
\label{eq:Perp_VS_Para}
r_{\perp\shpa}=\frac{\sum_{\bsl{q}\in \M}|\Delta_\perp(\bsl{q})|}{\sum_{\bsl{q}\in \M}|\Delta_\shpa(\bsl{q})|}
}
with $\M$ being any of the two FSs around $\bsl{K}_+$. 
As a result, for set A (B) in \eqnref{eq:eff_parameter_set}, we have $r_{\perp\shpa}=0.016$ ($r_{\perp\shpa}=0.051$) on the outer FS and $r_{\perp\shpa}=0.019$ ($r_{\perp\shpa}=0.11$) on inner FS, meaning that the Euler obstructed channel dominates on the FSs.

Next we discuss the nodal structure of the BdG Hamiltonian (given by \eqnref{eq:H_pm_eff} and \eqnref{eq:pairing_eff_Podd_st}) around $\bsl{K}_+$, since that around $\bsl{K}_-$ is related by the particle-hole symmetry.
\figref{fig:EOP_Nodal}(d) suggests that the BdG Hamiltonian with set A has zero-energy MNLs in each spin subspace, and each zero-energy BdG MNL is close to one pairing node of the Euler obstructed $\Delta_\shpa$ on each FS.
It agrees with the above general discussion since we have small FS splitting, weak pairing, and a dominant Euler obstructed channel for set A.
Interestingly, even for set B, which is beyond the above general discussion due to the strong pairing, the zero-energy BdG MNLs still exist (\figref{fig:EOP_Nodal}(h)).
The nonzero $\dsZ_2$ monopole charges of the zero-energy BdG MNLs in all these cases are verified by the Wilson-loop calculation and by their odd linking number to the BdG {\NLs}---the touching line between the top two occupied bands of the BdG Hamiltonian in each spin subspace. (See \appref{app:model} for more details.)
Therefore, the effective model verifies the general conclusion about the zero-energy BdG MNLs induced by a sufficiently-dominant Euler obstructed pairing channel for small FS splitting and weak pairing, and shows that the zero-energy BdG MNLs may persist even when the FS splitting becomes large and the pairing becomes strong.

\subsection{Hinge Majorana Modes}
\label{sec:hingeMZM}

\begin{figure}[t]
    \centering
    \includegraphics[width=\columnwidth]{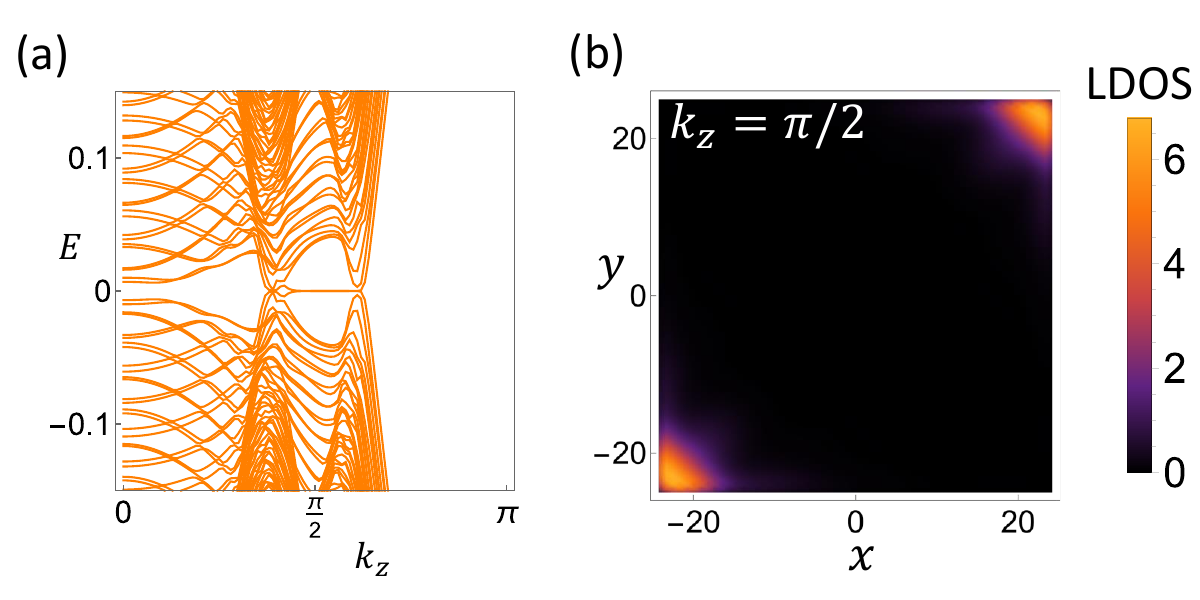}
    \caption{Plots for the spin-up block of the tight-binding BdG Hamiltonian that is finite along the $x$ and $y$ directions and infinite along the $z$ direction.
    All plots are the same for the spin-down block.
    In (a), we plot the BdG band structure for $k_z\geq 0$.
    The bands for $k_z\leq 0$ can be derived from the particle-hole symmetry.
    In (b), we plot zero-energy LDOS in the $x-y$ plane for $k_z=\pi/2$.
    }
    \label{fig:Hinge}
\end{figure}

In the last subsection, we have shown the zero-energy BdG MNLs that originate from a sufficiently-dominant Euler obstructed $\Delta_{\shpa}$ on FSs with small splitting.
In this subsection, we will discuss the hinge MZMs induced by the zero-energy BdG MNLs.

Let us first consider two $q_x-q_y$ planes that enclose the upper zero-energy BdG MNL in \figref{fig:EOP_Nodal}(d) or (h).
In each of the two decoupled spin subspaces, the Wilson loop winding numbers on the two planes must have different parities, owing to the nonzero $\dsZ_2$ monopole charge of the BdG MNL.
Then, one of the two 2D planes must have an odd Wilson loop winding number.
According to \refcite{Wang2019TMDHigherOrder,Ahn2019TBGFragile,Ahn2020HOTSCPT}, if we choose an open boundary condition along the $x$ and $y$ directions for the 2D plane with an odd Wilson loop winding number, there are MZMs localized at the ``corners" in each spin subspace.
These ``corner" MZMs would be extended along the $z$ direction owing to their well-defined $q_z$, and thus they are actually hinge MZMs along $z$.
This argument generally holds for any two parallel planes in 1BZ that enclose a zero-energy BdG MNL, not limited to the $q_x-q_y$ planes for \figref{fig:EOP_Nodal}(d,h).

To verify the general argument, we resort to the effective model (\eqnref{eq:H_pm_eff} and \eqnref{eq:pairing_eff_Podd_st}), and add an extra term to the effective model as
\eq{
\label{eq:H_eff_extra}
H_{extra}^{eff}=\sum_{\bsl{q},\alpha=\pm}\Psi_{\alpha,\bsl{q}}^\dagger m_1 (\tau_0\sigma_x+\tau_0\sigma_z) \otimes s_0\Psi_{\alpha,\bsl{q}}\ ,
}
which preserves TR, inversion, and spin $\SU(2)$ symmetries of the normal state.
We choose a circular open boundary condition in the $x-y$ plane for the effective model, and consider $q_z$ near the upper zero-energy BdG MNL in \figref{fig:EOP_Nodal}(d,h).
We analytically find two (zero) boundary zero modes in each spin subspace for $q_z$ below (above) the zero-energy BdG MNL, coinciding with the fact that an odd Wilson loop winding number only exists on one side of the zero-energy BdG MNL. 
Furthermore, the equation for the two zero modes in each spin subspace reads
\eq{
\label{eq:zero_mode_eq}
[\frac{1}{R}\partial_\theta + m_1 \tau_x (-\cos\theta+ \sin\theta)] V(r,\theta) = 0 \ ,
}
where $(x,y)=r(\cos\theta,\sin\theta)$, $r=R$ on the boundary, and $V(r,\theta)$ is localized at $r=R$. (See \appref{app:model} for more details.)
According to the above equation, the two zero-mode solutions would be two surface modes for $m_1=0$ (\ie, without the extra term \eqnref{eq:H_eff_extra}).
But the surface modes are just an artifact of the special form of the effective model; the symmetry-preserving extra term in \eqnref{eq:H_eff_extra} turns the surface modes into two domain-wall modes at $\theta=\pi/4, 5\pi/4$, where $(-\cos\theta+ \sin\theta)$ changes sign.
The two domain-wall modes are two hinge MZMs in each spin subspace.

To further verify the hinge MZMs, we construct a tight-binding model, which exactly reproduces the effective model (\eqnref{eq:H_pm_eff} and \eqnref{eq:pairing_eff_Podd_st} with the extra term in \eqnref{eq:H_eff_extra}) at low energies.
For all the numerical calculation with the tight-binding model, we choose the parameter values for the tight-binding model such that the tight-binding model matches the effective model with set B in \eqnref{eq:eff_parameter_set}, $\bsl{K}_\pm=(0,0,\pm\pi/2)$, and $m_1=0.05$ at low-energies.

We first consider a square geometry with length 50 in the $x-y$ plane for the tight-binding model, but keep the Bloch momentum along $z$ well-defined. (See \appref{app:model} for the detailed construction of the tight-binding model.)
We plot the band structure of the spin-up block of the tight-binding BdG Hamiltonian in \figref{fig:Hinge}(a), while the spin-down block has the same band structure.
\figref{fig:Hinge}(a) shows a zero-energy flat band, which corresponds to the hinge MZMs, as shown by the zero-energy local density of states (LDOS) plotted in \figref{fig:Hinge}(b).
Since the $k_z$ range of the zero-energy flat band in \figref{fig:Hinge}(a) roughly equals to the $q_z$ distance between two zero-energy BdG MNLs in \figref{fig:EOP_Nodal}(h), the hinge MZMs indeed only appear on one side of each zero-energy BdG MNL.
Furthermore, \figref{fig:Hinge}(b) shows two MZMs localized at the $\theta=\pi/4$ hinge and the $\theta=5\pi/4$ hinge for each spin block of the tight-binding BdG model, which also coincides with the above analysis for the effective model.

\begin{figure}[t]
    \centering
    \includegraphics[width=\columnwidth]{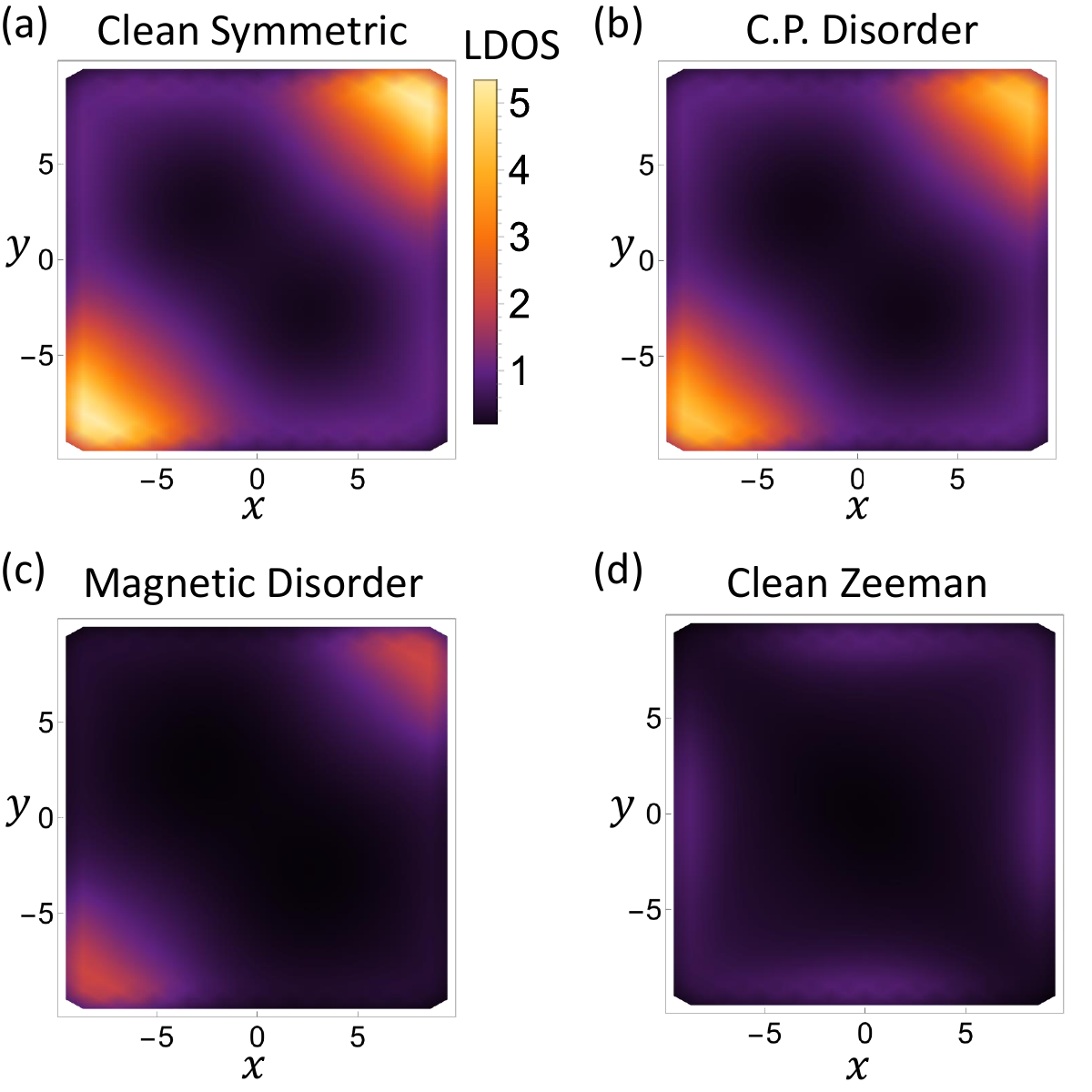}
    \caption{The zero-energy LDOS plots for the tight-binding BdG Hamiltonian in the configuration that is finite along the $x$, $y$, and $z$ directions.
    The zero-energy LDOS are plotted for $k_z=\pi/2$ derived from a Fourier transformation along $z$.
    All plots have the same color bar.
    (a) and (d) are in the clean limit.
    (a) preserves all the symmetries of the finite-size model, while (d) includes a symmetry-breaking Zeeman field. 
    (b) and (c) correspond to the chemical-potential (noted as ``C.P." in the figure) disorder and the magnetic disorder, respectively.
    }
    \label{fig:Hinge_Disorder}
\end{figure}

Next, we study the stability of the hinge MZMs against the chemical-potential disorder and the magnetic disorder.
To introduce the disorder, we need to consider a 3D finite system with an open boundary condition along $x$, $y$ and $z$ directions with the lengths respectively being $L_x=20$, $L_y=20$ and $L_z=10$.
We will include both spin blocks in the following since we will deal with the magnetic disorder.
Although the Bloch momentum along $z$ is not a good quantum number, we can always perform the Fourier transformation~\cite{Wilson2018WSMDisorder} to plot the zero-energy LDOS on the $x-y$ plane with $k_z=\pi/2$.
As a test of this Fourier transformation procedure, we plot the zero-energy LDOS for the clean limit in \figref{fig:Hinge_Disorder}(a), which clearly shows hinge MZMs similar to \figref{fig:Hinge}(b), meaning that the procedure is trustworthy. (See \appref{app:model} for details.)

Now we add the disorder, which is Gaussian with zero mean and standard deviation $W=0.1$.
The zero mean guarantees that the disorder preserves all symmetries on average.
The disorder strength $W=0.1$ is comparable with the bulk gap $(\sim 0.3)$ of the tight-binding BdG model at $k_z=\pi/2$, meaning that the disorder strength is not perturbatively small. (See \appref{app:model} for details.)
According to \figref{fig:Hinge_Disorder}(b-c), the hinge MZMs still exist in the presence of the chemical-potential disorder or the magnetic disorder, though the magnetic disorder has a much stronger effect on the hinge MZMs than the chemical-potential disorder.
As a comparison, we further consider a Zeeman field along $z$ with strength $B=0.02$ in the clean limit, which breaks the effective chiral symmetry in each spin subspace.
Although the strength of the symmetry-breaking field $B=0.02$ is much smaller than the disorder strength $W=0.1$, the zero-energy LDOS is much lower at the hinges for the former (\figref{fig:Hinge_Disorder}(d)) than that for the latter (\figref{fig:Hinge_Disorder}(b-c)).
Therefore, compared to the symmetry-breaking effect, the hinge MZMs are much more robust against the disorder that preserves symmetries on average, even if the disorder strength is comparable with the bulk gap at the same momentum.

At the end of this subsection, we emphasize that although \refcite{Ahn2020HOTSCPT} has discussed the higher-order nodal superconductor with BdG MNLs, the BdG MNLs in \refcite{Ahn2020HOTSCPT} were characterized by the inversion-protected symmetry indicator, and are unlikely to directly come from the Euler obstructed Cooper pairing.
It is because the inversion indicator can only detect an odd number of pairs of MNLs~\cite{Ahn2020HOTSCPT}, while a dominant Euler obstructed pairing channel tends to convert a pair of normal-state MNLs into two pairs of zero-energy BdG MNLs (as shown in \figref{fig:EOP_Nodal}), resulting in an even number of pairs of BdG MNLs in total.
Therefore, without imposing extra symmetries, the higher-order nodal superconductivity discussed in this subsection is beyond the symmetry indicator.
Furthermore, as discussed in the introduction, it features the first class of higher-order nodal superconductivity that originates from the topologically obstructed Cooper pairing.

\subsection{Linearized Gap Equation}
\label{sec:LGE}

The above discussion on the nodal SC and hinge MZMs necessarily requires a $\PT$-invariant pairing with a momentum-independent spin part (\ie, \eqnref{eq:H_pairing_TR}) and a dominant Euler obstructed pairing channel (\ie, the pairing form in \eqnref{eq:ssPeven_stPodd} for the parity-even spin-singlet channel or the parity-odd spin-triplet channel together with a dominant $\Delta_\shpa$).
In this subsection, we will use the linearized gap equation to justify this condition near the superconducting transition.

We first consider the most general symmetry-allowed normal-state Hamiltonian near the FSs, together with the most general symmetry-allowed interaction that accounts for the zero-total-momentum Cooper pairing order parameters, according to the symmetry representation \eqnref{eq:reps_PT_T_P}. (See details in \appref{app:PandT}.)
By solving the linearized gap equation, we find that the highest superconductivity critical temperature can always be achieved by a $\PT$-invariant pairing form with a momentum-independent spin part, justifying \eqnref{eq:H_pairing_TR}.
In particular, we find that when the FS splitting is small, the critical temperature for $\Delta=d_y\ii \tau_y$ in \eqnref{eq:H_pairing_TR}, which corresponds to the spin-singlet parity-odd channel or the spin-triplet parity-even channel according to \tabref{tab:parity_pairing}, is suppressed by the FS splitting $m$ as
\eq{
\label{eq:Tc_dy}
\ln (\frac{T_c}{ T_{c,0} } )= \frac{\left\langle |d_y^{(0)}(\bsl{q})|^2 f_{\bsl{q}}\right\rangle_{\bsl{q}\in FS_0}}{\left\langle |d_y^{(0)}(\bsl{q})|^2\right\rangle_{\bsl{q}\in FS_0}} \leq 0\ ,
}
where $T_{c,0}$ and $T_c$ are the critical temperatures for zero and nonzero $m$, respectively, the superscript ``$(0)$" stands for $m=0$, $\langle ... \rangle_{\bsl{q}\in FS_0}$ means taking the average on the FS for $m=0$, and $ f_{\bsl{q}}$ is a non-positive function.
On the other hand, the critical temperature $T_c$ for $\Delta=d_0 \tau_0+\bsl{d}_\shpa\cdot\bsl{\tau}_\shpa$, which corresponds to the parity-even spin-singlet channel and the pairing-odd spin-triplet channel, reads
\eq{
\label{eq:Tc_d0zx}
\ln (\frac{T_c}{ T_{c,0} } )= \frac{\left\langle [|\bsl{d}_\shpa^{(0)}|^2 - (\bsl{d}_\shpa^{(0)}\cdot\hat{\bsl{g}}_\shpa^{(0)})^2] f_{\bsl{q}}\right\rangle_{\bsl{q}\in FS_0}}{{\left\langle |d_0^{(0)}(\bsl{q})|^2+ \bsl{d}_\shpa^{(0)}(\bsl{q})^2\right\rangle_{\bsl{q}\in FS_0}}} \ ,
}
where $\hat{\bsl{g}}_\shpa=\bsl{g}_\shpa/|\bsl{g}_\shpa|$, and $\bsl{g}_\shpa$ describes the momentum distribution of the FS splitting as defined in \eqnref{eq:H_cal}.
It means that $T_c$ for $d_{0,z,x}$ can be un-suppressed by (i) aligning the nonzero $\bsl{d}_\shpa^{(0)}$ to $\hat{\bsl{g}}_\shpa^{(0)}$ or (ii) having a zero $\bsl{d}_\shpa^{(0)}$, similar to the unsuppressed spin-triplet channel in the presence of noncentrosymmetric SOC discussed in \refcite{Frigeri2004NCSC}.
(See more details in \appref{app:PandT}.)
Given that the FS splitting should generally exist, the $d_y$ channel tends to have lower $T_c$ than $d_{0,z,x}$.
Therefore, the pairing form in \eqnref{eq:ssPeven_stPodd} for the parity-even spin-singlet channel or the pairing-odd spin-triplet channel is justified for small FS splitting.

\begin{figure}[t]
    \centering
    \includegraphics[width=0.6\columnwidth]{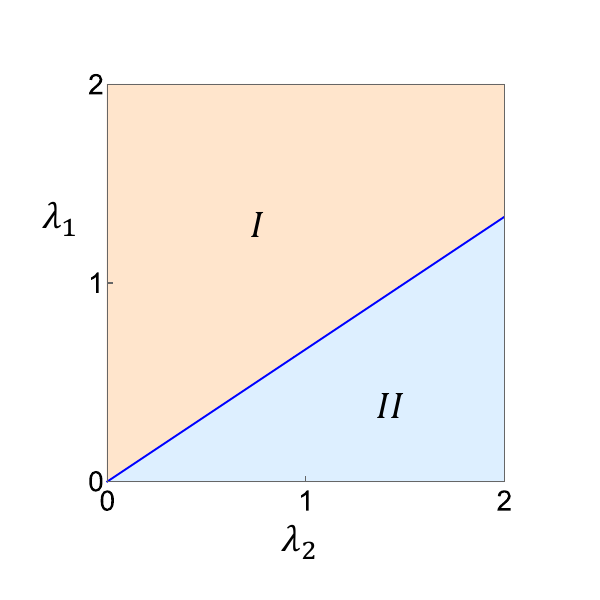}
    \caption{The phase diagram given by the linearized gap equation for the effective model.
    The trivial channel (which leads to a dominant trivial $\Delta_\perp$ on the FSs) has a higher critical temperature $T_c$ in phase $I$, while the pairing channel in \eqnref{eq:pairing_eff_Podd_st} (which leads to a dominant Euler obstructed $\Delta_{\shpacap}$ on the FSs) has a higher $T_c$ in phase $II$.
    The phase boundary is at $\lambda_1=2 \lambda_2 /3$, and $\lambda_{1,2}$ correspond to the absolute strengths of two interaction channels.
    }
    \label{fig:LGE}
\end{figure}

Nevertheless, it is still not clear whether the Euler obstructed $\Delta_\shpa$ dominates within \eqnref{eq:ssPeven_stPodd}.
To address this question, we resort to the effective model (\eqnref{eq:H_pm_eff} and \eqnref{eq:pairing_eff_Podd_st}), and focus on the case where the FS splitting $m$ is small.
We find that the pairing channel in \eqnref{eq:pairing_eff_Podd_st} for the effective model, which gives a dominant Euler obstructed $\Delta_{\shpa}$ on the FSs, is not suppressed by the FS splitting $m$.
To test how promising the pairing is, we introduce a competing spin-singlet parity-even channel, which has a dominant trivial $\Delta_{\perp}$ on the FSs and is not suppressed by the FS splitting $m$.
As shown in \figref{fig:LGE}, we find that the pairing channel in \eqnref{eq:pairing_eff_Podd_st} has a higher $T_c$ if $\lambda_2>3 \lambda_1 /2$, where the positive $\lambda_1$ and $\lambda_2$ measure the strengths of the attractive interactions for the competing trivial channel and the nontrivial \eqnref{eq:pairing_eff_Podd_st}, respectively. (See more details in \appref{app:model}.)
Therefore, the pairing channel in \eqnref{eq:pairing_eff_Podd_st} dominates and then gives a dominant Euler obstructed $\Delta_{\shpa}$ on FSs in a large range of parameter values when the FS splitting is small.

\section{Conclusion and Discussion}
\label{sec:discussion}

In conclusion, we have theoretically demonstrated the existence of the Euler obstructed Cooper pairing between any two sphere-like FSs with nonzero Euler numbers, and have established the nodal superconductivity and hinge MZMs as the physical consequences arising from a sufficiently-dominant Euler obstructed pairing channel.

Conceptually, the notion of the Euler obstructed Cooper pairing that we proposed is the first Cooper pairing order parameter obstructed by the normal-state band topology beyond the conventional AZ ten-fold way.
We emphasize that the notion of the obstructed Cooper pairing in this work, as well as the previously related works~\cite{Li2018WSMObstructedPairing,Murakami2003BerryPhaseMSC,Sun2020Z2PairingObstruction}, is fundamentally different from the obstruction in the study of the nodeless topological superconductors~\cite{Schindler2020PairingObstruction}.
Specifically, the Cooper pairing order parameter is obstructed on the FSs in the former case; in the latter case, the obstruction occurs in the ground state wavefunction of the BdG Hamiltonian, while the Cooper pairing order parameter is allowed to be smooth on FSs for all gauges.
To avoid any confusion, we emphasize that the ``obstruction" means very different concepts in these two distinct contexts, and one does not necessarily imply the other and vice versa.

Our work has physical implications for superconductivity: The dominant Euler obstructed Cooper pairing provides a natural topological mechanism for nodal superconductivity and hinge MZMs to appear in (monopole-charged) nodal-line (or spinless-Dirac) semimetals.
Compared to the normal-state platforms for the previously discussed (\ie, before our work) obstructed Cooper pairings~\cite{Li2018WSMObstructedPairing,Sun2020Z2PairingObstruction}, which are the type-I Weyl/Dirac semimetals, the nodal-line semimetals used in our work have different behaviors of the density of states---typically linear in energy for the nodal-line semimetals in our work and quadratic in energy for the type-I Weyl/Dirac semimetals~\cite{Burkov2011NLSM}. 
Compared to the nodal superconductors given by the previously discussed obstructed Cooper pairings~\cite{Li2018WSMObstructedPairing,Sun2020Z2PairingObstruction}, which have surface Majorana modes, the nodal superconductor given by the Euler obstructed Cooper pairing of our work can have hinge MZMs. 
Experimentally, the density of states can be measured in transport experiments~\cite{Neumaier2009MeasureDOS,Liu2021BaMnSb2}, the nodal superconductivity can be detected by measuring the magnetic penetration depth~\cite{Kim2016YPtBiSCj=3/2}, and signatures of the hinge MZMs can be detected by using scanning tunneling microscopy~\cite{Jack2019MZMHinge}.  
Thus, in addition to introducing the conceptual principle of ‘Euler obstructed Cooper pairing’, our work should help the experimental realization of topological nodal superconductivity (with associated MZMs) in the nodal-line semimetals and related materials, which are of intrinsic interest in themselves.

One future direction is to predict the materials that may realize our predicted Euler obstructed Cooper pairing, which can set the stage for the experimental discovery of the Euler obstructed Cooper pairing and the associated Majorana MZMs.
The first step is to look for the normal-state platforms---the monopole-charged spinless-Dirac/nodal-line semimetals.
The monopole-charged nodal-line semimetal phase has been predicted to exist in 3D graphdiyne~\cite{Nomura2018GraphdiyneNLSM,Ahn2018MonopoleNLSM}, and was also discussed in 3D transition metal dichalcogenides~\cite{Wang2019TMDHigherOrder}. 
For the monopole-charged spinless-Dirac semimetals, one can search the topological materials database~\cite{Bernevig2019TopoMat,Fang2019TopoMat,Wan2019TopoMat} for candidates.
One way is to look for the nonmagnetic materials that (i) have both inversion and $C_6$ symmetries and (ii) only involve the elements in the first three rows of the periodic table to ensure negligible SOC near the Fermi level.
{According to the symmetry data in the database of \refcite{Bradlyn2017TQC,Bernevig2019TopoMat,Vergniory2021TopoDatabase,BCServer,TopoMatDatabase}, we believe that C$_{12}$Li with space group No.\,191 hosts two inversion-related MSDPs along $\Gamma-A$ near the Fermi energy.
Explicit confirmation of our statement requires more detailed first-principle calculations and the experiments.
}

Given a suitable normal-state material, the next step is to predict the possible Cooper pairing order parameters that have a dominant Euler obstructed pairing channel.
To do so, it is useful to relate the pairing symmetry to the dominant Euler obstructed pairing channel.
For example, let us consider a weak zero-total-momentum spin-singlet pairing in the monopole-charged spinless-Dirac semimetals, such as C$_{12}$Li.
In this case, if the pairing is parity-even and $C_6$-odd, it must have a dominant Euler obstructed pairing channel on small sphere-like FSs that enclose normal-state MSDPs, and must lead to zero-energy BdG MSDPs. (See \appref{app:PandT} for details.)
It is also worth utilizing the symmetry indicator~\cite{Po2017SymIndi,Bradlyn2017TQC,Kruthoff2017TCI}, which recently has been generalized to SCs~\cite{Ono2018SITSC,Shiozaki2019SITSC,Ono2019SITSC,Ono2020SITSC,Skurativska2020SITSC,Geier2020SITSC,Ahn2020HOTSCPT,Huang2021SITSC,Ono2021SITSC1651,Ono2021SITSCNodal,Tang2021SITSC}.
Specifically, although the inversion indicator typically cannot capture the zero-energy BdG MNLs arising from an Euler obstructed pairing channel as discussed in the last section, it is useful to explore whether other symmetry indicators work in the presence of extra symmetries in addition to the TR and inversion symmetries. 

{As mentioned in the introduction, nonzero normal-state Euler numbers can also exist in 2D systems like MATBG, besides the 3D systems studied in this work.
Then, our theory can be generalized to the Cooper pairing order parameters in the MATBG~\cite{Yu2021EOCPTBG}.
As we only consider the isolated sets of two spin-doubly degenerate bands near on the FSs, another future direction would be to consider the cases with more than two spin-doubly degenerate bands, perhaps using the second Stiefel-Whitney class.}
We anticipate that our introduction of the concept of the Euler obstructed Cooper pairing in nodal superconductivity along with the emergent hinge MZMs should be a new direction for the active subject of topology in condensed matter phenomena.

\emph{Note Added}: During the finalizing stages of the preparation of this manuscript, an arXiv preprint \refcite{Po20212DNodalSCBy1DAIII} appeared, studying the nodal superconductivity enforced by the $\dsZ$-valued additive topological invariants (\eg, Chern numbers and the 1D chiral-symmetry-protected AIII invariant) of the normal states. 
In contrast, the $\dsZ$-valued Euler number used in our work is not additive (and is unstable), and the corresponding additive invariant is the $\dsZ_2$-valued second Stiefel-Whitney class (when first Stiefel-Whitney classes vanish).
Therefore, the Euler obstructed Cooper pairing proposed in our work is beyond the formalism presented in \refcite{Po20212DNodalSCBy1DAIII}. 

\section{Acknowledgement}
J.Y. thanks Yang-Zhi Chou, Shao-Kai Jian, Zhi-Da Song, Ming Xie, and Rui-Xing Zhang for helpful discussions.
This work is supported by the Laboratory for Physical Sciences and Microsoft Corporation.  
Y.-A.C. acknowledges the support from the JQI Postdoctoral Fellowship.


%


\appendix


\section{Brief Review on MNLs and Euler Class}

\label{app:review_NLSM_Euler}

In the section, we briefly review the $P\TR$-invariant monopole-charged nodal-line semimetal and Euler class, following \refcite{Bzdusek2017AZInversionNodal,Ahn2018MonopoleNLSM,Ahn2019TBGFragile,Bouhon2020WeylNonabelian}.
More details can be found in those references.
Throughout this section, we always consider 3D spinless noninteracting crystals with $\PT$ symmetry, where ``spinless" here means that we always impose spin $\SU(2)$ symmetry (which requires the SOC to be negligible) and suppress the spin index.

\subsection{$P\TR$-invariant Nodal Lines}

Owing to the $\PT$ symmetry, 3D systems can have stable nodal lines.
To verify this statement, we first need to introduce the so-called real gauge---the gauge where $\PT$ is represented by complex conjugate, \ie, $\PT\dot{=}\cc$.
As shown in \refcite{Bouhon2020WeylNonabelian}, the real gauge always exists for any spinless 3D $\PT$-invariant crystal, which we review below.
For any isolated set of $N$ bands, we label the Bloch bases of the set as $\ket{\psi_{\bsl{k}}}=(\cdots,\ket{\psi_{\bsl{k},a}},\cdots)$ with $a=1,2,\cdots,N$.
Here $\ket{\psi_{\bsl{k}}}$ are not required to be the eigenstates, and are in general given by the $U(N)$ transformation of the eigenstates.
Thus,  $\ket{\psi_{\bsl{k}}}$, as well as its periodic part $\ket{u_{\bsl{k}}}=e^{-\ii\bsl{k}\cdot\bsl{r}} \ket{\psi_{\bsl{k}}}$, has a $U(N)$ gauge freedom described by 
\eqa{
\label{eq:UN_gauge}
& \ket{\psi_{\bsl{k}}}\rightarrow \ket{\psi_{\bsl{k}}} R_{\bsl{k}}\\
& \ket{u_{\bsl{k}}}\rightarrow \ket{u_{\bsl{k}}} R_{\bsl{k}} \ .
}
If we choose $\ket{\psi_{\bsl{k}}}$ as the basis for all bands, then $\ket{\psi_{\bsl{k}}}$ become the Bloch basis of the Hamiltonian, which we label as $\ket{\Psi_{\bsl{k}}}$.

The symmetry representation of $\PT$ symmetry is 
\eq{
\PT \ket{\psi_{\bsl{k}}} = \ket{\psi_{\bsl{k}}}U(\bsl{k})\ .
}
Owing to $(\PT)^2=1$, we have 
\eqa{
& U(\bsl{k}) U(\bsl{k})^* = 1 \\
& U(\bsl{k}) U(\bsl{k})^\dagger = 1\ .
}
According to the above equation, we can always split $U(\bsl{k})=U^{\text{Re}}(\bsl{k})+\ii U^{\text{Im}}(\bsl{k})$, where both $U^{\text{Re}}(\bsl{k})$ and $U^{\text{Im}}(\bsl{k})$ are both real and symmetric, $[U^{\text{Re}}(\bsl{k}),U^{\text{Im}}(\bsl{k})] =0$, and $[U^{\text{Re}}(\bsl{k})]^2+[U^{\text{Im}}(\bsl{k})]^2 = 1 $.
It means that $U^{\text{Re}}(\bsl{k})$ and $U^{\text{Im}}(\bsl{k})$ can be simultaneously diagonalized by an orthogonal real matrix $V(\bsl{k})$, resulting in
\eq{
[V(\bsl{k})]^T U(\bsl{k}) V(\bsl{k}) = \mat{
\ddots &  & \\
 & e^{\ii \phi_i(\bsl{k})}  & \\
 &  & \ddots 
}\ .
}
By defining a unitary matrix 
\eq{
U_\phi(\bsl{k})= V(\bsl{k}) \mat{
\ddots &  & \\
 & e^{\ii \phi_i(\bsl{k})/2}  & \\
 &  & \ddots 
}\ ,
}
we have 
\eq{
U_\phi^\dagger (\bsl{k}) U(\bsl{k}) U_\phi^*(\bsl{k}) = 1\ ,
}
meaning that we can define
\eq{
\ket{\widetilde{\psi}_{\bsl{k}}}= \ket{\psi_{\bsl{k}}} U_\phi(\bsl{k})
}
with
\eq{
\PT \ket{\widetilde{\psi}_{\bsl{k}}} = \ket{\widetilde{\psi}_{\bsl{k}}}\ .
}
Then, $\ket{\widetilde{\psi}_{\bsl{k}}}$ is a real gauge.
Note that the real gauge is not unique, since any transformation of $\ket{\widetilde{\psi}_{\bsl{k}}}$ by a real orthogonal matrix still gives us a real gauge.

Based on the existence of the real gauge, \refcite{Bzdusek2017AZInversionNodal} presented a simple argument for the existence of stable phases with nodal lines, which we review in the following.
Given any two bands in a region of the 3D 1BZ, the effective matrix Hamiltonian for them may be expressed as 
\eq{
h_{eff}(\bsl{k})=\tau_0 \epsilon(\bsl{k})+\bsl{\tau}\cdot\bsl{g}(\bsl{k})\ ,
}
where $\bsl{\tau}=(\tau_x,\tau_y,\tau_z)$, and $\tau$'s are Pauli matrices.
After choosing a real gauge,  {\ie} $\PT\dot{=}\cc$, $\PT$ symmetry requires $g_y(\bsl{k})=0$ in $h_{eff}(\bsl{k})$, simplifying the band touching condition for the two bands to
\eq{ 
g_x(\bsl{k})=g_z(\bsl{k})=0\ ,
}
which may have an 1D solution manifold for $\bsl{k}$, representing the possible stable line nodes.

In the following, we present an alternative argument for the stable nodal lines.
For a 3D $\PT$-invariant crystal, the parameters of its Hamiltonian includes three independent components of the Bloch momentum $\bsl{k}$ and $n$ other independent parameters noted by a $n$-dimensional vector $\bsl{\xi}$.
By choosing a Bloch basis for the Hamiltonian, the Hamiltonian can be represented as a Hermitian matrix with $n+3$ parameters---the matrix Hamiltonian is in a $(n+3)$-dimensional parameter manifold $\Lambda$.
Typically, the Bloch basis for the Hamiltonian is built from local atomic orbitals and thus the representation of $\PT$ is a constant matrix. 
As a result, it is possible to choose a real gauge with a gauge transformation that does not depend on any parameters, and thus such gauge transformation cannot affect the dimension of the parameter manifold.
Therefore, in the following, we choose such a real gauge to study the dimension of the band touching.
With a real gauge, the matrix Hamiltonian is real symmetric.
According to \refcite{Keller2008DegCodim}, the parameter manifold, in which a certain eigenvalue of real symmetric matrices become doubly degenerate, has codimension 2.
Then,
there is a $(n+1)$-dimensional parameter submanifold $\Lambda_{nodal}\subset\Lambda$; at each point of $\Lambda_{nodal}$, two specified energies of the Hamiltonian coincide.
Without fine-tuning, the projection of $\Lambda_{nodal}$ to the $n$-dimensional $\bsl{\xi}$ manifold is a $n$-dimensional $\bsl{\xi}$ submanifold $\Lambda_{nodal,\bsl{\xi}}$.
A point of $\Lambda_{nodal,\bsl{\xi}}$ should correspond to a one-dimensional fiber in $\Lambda_{nodal}$, which stands for a touching line between the two specified bands in the first Brillouin zone.
Therefore, if the two specified bands are conduction and valence bands, $\Lambda_{nodal,\bsl{\xi}}$ stands for a stable phase where the system has nodal lines.

\subsection{Topological Indices of Nodal Lines}

In this part, we review the topological indices of the $\PT$-protected nodal lines.
We will focus on the nodal lines between two specified bands.
We refer to the higher band as the conduction band, and the lower band as the valence band.
We further refer to all bands that are not lower than the conduction band as empty bands, and refer to all bands that are not higher than the valence bands as occupied bands.
The conduction band is the lowest empty band, and the valence band is the highest occupied band.

There are two topological indices of the nodal rings in $\PT$-invariant systems~\cite{Bzdusek2017AZInversionNodal}.
To introduce them, let us first review the concept of the Wilson loop.
The Wilson loop is defined for a set of bands that are isolated on a close loop $\gamma$ in the 1BZ torus.
Recall that we label the Bloch basis for an isolated set of bands as $\ket{\psi_{\bsl{k}}}=(\cdots,\ket{\psi_{\bsl{k},a}},\cdots)$, where $a=1,\cdots,N$ with the band number $N$, and its periodic part as $\ket{u_{\bsl{k}}}=e^{-\ii\bsl{k}\cdot\bsl{r}} \ket{\psi_{\bsl{k}}}$; they both have a $U(N)$ gauge freedom as shown in \eqnref{eq:UN_gauge}.
Then, the Wilson loop~\cite{Dai2011Z2WilsonLoop} is defined as
\eq{
\label{eq:WL}
W_{\gamma}(\bsl{k}_0)=\lim_{L\rightarrow\infty}\bra{u_{\bsl{k}_0}} P_{\bsl{k}_1} P_{\bsl{k}_2} ...P_{\bsl{k}_L} \ket{u_{\bsl{k}_0}}\ ,
}
where $P_{\bsl{k}}=\ket{u_{\bsl{k}}} \bra{u_{\bsl{k}}}$ is the projection operator constructed from $\ket{u_{\bsl{k}}}$ rather than $\ket{\psi_{\bsl{k}}}$.
In particular, $\bsl{k}_0$ is the initial point of $\gamma$, and $\bsl{k}_1$, ..., $\bsl{k}_L$ are chosen on the $\gamma$ sequentially with the distance between $\bsl{k}_{l}$ and $\bsl{k}_{l-1}$ on $\gamma$ being $|\gamma|/L$, where $|\gamma|$ is the length of $\gamma$.
\eqnref{eq:WL} suggests that $W_{\gamma}(\bsl{k}_0)$ is unitary.
Since the $U(N)$ gauge transformation and the change of the initial point $\bsl{k}_0$ on $\gamma$ can only change $W_{\gamma}(\bsl{k}_0)$ by a unitary transformation, the spectrum of Wilson loop, as well as its determinant, is gauge invariant.

\begin{figure}[t]
    \centering
    \includegraphics[width=\columnwidth]{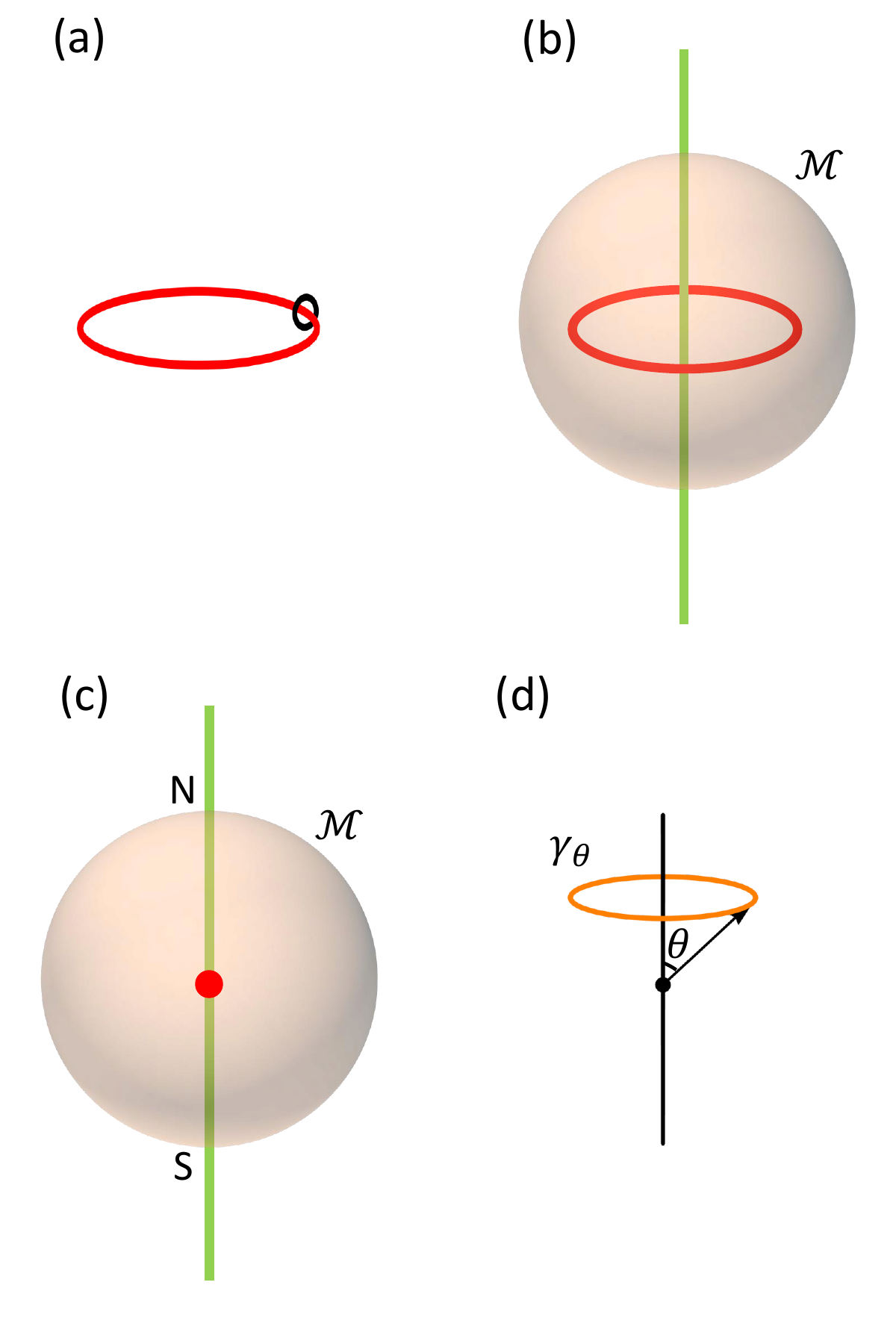}
    \caption{
    This figures schematically shows the nodal structure in 3D spinless $\PT$-invariant noninteracting crystals.
    (a) The red line is a $\PT$ protected nodal ring.
    The determinant of the Wilson loop along the black circle is $-1$.
    (b) The red line is a $\PT$ protected nodal ring with nontrivial $\dsZ_2$ monopole charge evaluated on the (orange) sphere $\M$ enclosing it.
    The green line is a {\NLs}---the degenerate line between the top two occupied bands---which has nontrivial linking to the nodal ring.
    (c)  The red point is a spinless Dirac point given by the crossing between a $\Lambda_1$ band and a $\Lambda_2$ band on a $C_6$-invariant axis.
    The green line is the degenerate line between the top two occupied bands, which is along $C_6$-invariant axis.
    N and S, respectively standing for the north and south poles, mark the intersection between the $C_6$-invariant axis and the (orange) sphere $\M$  that encloses the spinless Dirac point.
    (d) The orange line $\gamma_\theta$ shows a path along the circle with a fixed value of the polar angle $\theta$.
    }
    \label{fig:SpinlessDirac_NLSM}
\end{figure}

Owing to the $\PT$-invariant, the Wilson loop for any isolated set of bands along any close loop $\gamma$ satisfies $W_{\gamma}(\bsl{k}_0)\sim W_{\gamma}^*(\bsl{k}_0)$, where $\sim$ stands for being related by a unitary transformation. 
Then, the gauge invariant $\det(W_{\gamma})$ obeys $\det(W_{\gamma})=\det(W_{\gamma})^*=\pm 1$.
This $\dsZ_2$ index gives the first topological index for the nodal ring.
In particular, we consider $\det(W_{\gamma})$ for the set of all occupied bands on $\gamma$, then $\det(W_{\gamma})=-1$ if $\gamma$ is a circle around a nodal ring~\cite{Bzdusek2017AZInversionNodal}, as shown in \figref{fig:SpinlessDirac_NLSM}(a).

The second topological index is constructed by the Wilson loop for the isolated set of all occupied bands on a sphere that encloses a nodal ring~\cite{Bzdusek2017AZInversionNodal}, as shown in \figref{fig:SpinlessDirac_NLSM}(b).
We can evaluate the Wilson loop $W_{\gamma_\theta}$ along the circle $\gamma_\theta$ with a fixed value of the polar angle $\theta$ (\figref{fig:SpinlessDirac_NLSM}(d)), and plot the phases of the spectrum as a function of $\theta$ from $0$ to $\pi$.
Then, we can count the number of the linear crossings in the spectrum at $\pi$, also called the Wilson loop winding number.
The winding number itself is not the stable topological invariant, since two linear crossings can merge and gap each other for a large band number $N$.
The stable topological invariant is the parity of winding number (\ie, winding number modulo $2$), which serve as a $\dsZ_2$ monopole charge of the nodal ring.
If the monopole charge of the nodal ring is 0, it is called a trivial nodal line; the monopole charge is 1 for a MNL.
MNLs always appear in pairs since the total $\dsZ_2$ monopole charge in the 1BZ must be zero.
Interestingly, each MNL must enclose the degenerate lines between the top two occupied bands, noted as {\NLs}s, as exemplified in \figref{fig:SpinlessDirac_NLSM}(b).
It is because the $\dsZ_2$ monopole charge equals the linking number between the MNL and {\NLs}s modulo 2~\cite{Ahn2018MonopoleNLSM}.
Mathematically, the two topological indices of the nodal rings correspond to the first and second Stiefel-Whitney classes of real vector bundles (evaluated on the fundamental classes of the manifolds)~\cite{Ahn2018MonopoleNLSM}.

\subsection{Euler Class}

In this part, we focus on the situation where we can choose the sphere $\M$ that encloses a nodal ring such that the top two occupied bands are gapped from the lower occupied bands on and in $\M$.
Within this case, we review the concept of the Euler class following \refcite{Ahn2019TBGFragile}.

Since the top two occupied bands are gapped from the lower occupied bands, the lower occupied bands would only deform adiabatically as shrinking the sphere to a point.
Then, the lower bands must have zero $\dsZ_2$ monopole charge, and the $\dsZ_2$ monopole charge of the nodal ring is solely determined by the top two occupied bands.
For the top two occupied bands, the Wilson loop winding becomes $\dsN$-classified~\cite{Ahn2019TBGFragile}, since two linear crossings at $\pi$ can not be merged if they are separated by one linear crossing at $0$.
Then, the meaningful winding number is the number of the $\pi$ linear crossings after merging all $0$ and $\pi$ linear crossings that are allowed to be merged.
In the following, we always mean the meaningful winding number when we refer to the Wilson loop winding number of any isolated set of two bands.

Mathematically, the $\dsN$ winding number corresponds to the Euler class of the real vector bundle~\cite{Ahn2019TBGFragile}.
For any real gauge, the top two occupied bands correspond to a rank-2 real vector bundle on the sphere.
An explicit way to evaluate the Euler class of the rank-2 real vector bundle is given in \refcite{Zhao2017PTRealCN,Ahn2019TBGFragile}, which we review in the following. 
First, we split $\M$ into several open patches, labeled by $\alpha=I,II,...$, and then pick an oriented real patch-wise smooth (PWS) gauge as $\ket{\widetilde{u}^{\alpha}_{\bsl{k}}}$, where being oriented means that the transition function for any $\bsl{k}\in\alpha\cap\alpha'$ is in $\SO(2)$ or has determinant $+1$, \ie, $\ket{\widetilde{u}^{\alpha}_{\bsl{k}}}=\ket{\widetilde{u}^{\alpha'}_{\bsl{k}}}\widetilde{R}_{\bsl{k}}^{\alpha,\alpha'}$ with $\widetilde{R}_{\bsl{k}}^{\alpha,\alpha'}\in \SO(2)$.
Then, define~\cite{Zhao2017PTRealCN,Ahn2019TBGFragile}
\eq{
\bsl{a}_{\bsl{k}}^{\alpha}=  \pf{\bra{\widetilde{u}_{\bsl{k}}^{\alpha}} \nabla_{\bsl{k}} \ket{\widetilde{u}^{\alpha}_{\bsl{k}}}}
=
\frac{1}{2}\Tr[-\ii\tau_y\bra{\widetilde{u}_{\bsl{k}}^{\alpha}} \nabla_{\bsl{k}} \ket{\widetilde{u}^{\alpha}_{\bsl{k}}}]
}
and 
\eq{
\bsl{f}_{\bsl{k}}^{\alpha}= \nabla_{\bsl{k}} \times \bsl{a}_{\bsl{k}}^{\alpha}\ .
}
Here, we actually choose $\M$ to have an infinitesimal thickness, and therefore $\nabla_{\bsl{k}}$ is well defined on $\M$ in all three directions.
Nevertheless, only the component of $\bsl{f}_{\bsl{k}}^{\alpha}$ along the normal direction of $\M$ is meaningful for the Euler class as shown below.
As $\bsl{f}_{\bsl{k}}^{\alpha}=\bsl{f}_{\bsl{k}}^{\alpha'}$ for all $\bsl{k}\in\alpha\cap\alpha'$, we can drop the patch index in $\bsl{f}_{\bsl{k}}$, and $\bsl{f}_{\bsl{k}}$ is smooth everywhere in $\M$.
The Euler class of the two bands is given by the surface integral of $\bsl{f}_{\bsl{k}}$ on the sphere $\M$ as \eqnref{eq:e_2_original}~\cite{Zhao2017PTRealCN,Ahn2019TBGFragile}.
Changing the real oriented PWS gauge would change the sign of $e_2$, since $\ket{\widetilde{u}_{\bsl{k}}}\rightarrow \ket{\widetilde{u}_{\bsl{k}}} \tau_z$ gives us another real oriented PWS gauge and this gauge transformation leads to $e_2\rightarrow -e_2$.
Therefore, we use $|e_2|$ instead of $e_2$ to capture the Euler class in this work because $|e_2|$ is invariant under the change of the real oriented gauges.
$|e_2|$ evaluated for any real oriented gauge for the top two occupied bands is equal to the Wilson loop winding for the two bands, meaning that the $\dsZ_2$ monopole charge of the enclosed nodal ring is given by $|e_2|\ \mod\ 2$, which agrees with the theorem: the Euler class mod 2 is the top Stiefel-Whitney class of the vector bundle \cite{Milnor2016}.

For MNLs, the intersection between {\NLs} and the sphere $\M$ can be understood from the Euler class.
Specifically, the Wilson loop winding on $\M$ for the top two occupied bands must be odd, and thus Euler class must be odd $|e_2| \ \mod\ 2 = 1$ for any real oriented gauge.
Because of the nonzero $|e_2|$, the top two isolated bands must be connected; if they are disconnected, the rank-2 real vector bundle on the sphere is the Whitney sum of two rank-1 real vector bundles, which are trivial (since the sphere is simply-connected), and thus it must have zero $|e_2|$.
In general, the band touching occurs at isolated points on the sphere $\M$.
In particular, as demonstrated in \refcite{Ahn2019TBGFragile}, each isolated band touching point can be assigned a winding number $\widetilde{w}_i$, and the total winding number is related to $e_2$ in \eqnref{eq:e_2_original} (not $|e_2|$) according to a modified Nielsen-Ninomiya theorem.
In the following, we review the two derivations for the modified Nielsen-Ninomiya theorem in \refcite{Ahn2019TBGFragile}.

For the first derivation, let us choose two patches for the sphere $\M$, noted as $I$ and $II$ (two hemispheres overlapping on a thin strip), and then make sure the oriented real PWS $\ket{\widetilde{u}_{\bsl{k}}}$ is smooth in each patch.
Furthermore, we choose
\eq{
\label{eq:oriented_gauge}
\ket{\widetilde{u}^{II}_{\bsl{k}}}=\ket{\widetilde{u}^{I}_{\bsl{k}}} R_{\bsl{k}}
} 
with $R_{\bsl{k}}=e^{\ii \tau_y \Gamma_{\bsl{k}}}\in \SO(2)$ for any $\bsl{k}\in I\cap II$, meaning that this is an oriented real gauge.
The Euler number in this oriented real gauge can be evaluated as
\eq{
\label{eq:e2_real_gauge}
e_2=-\int_{\partial I} d\bsl{k}\cdot\nabla_{\bsl{k}}\Gamma\ ,
}
where $\partial I$ is the boundary of $I$.
Then, we expand the Bloch Hamiltonian as
\eq{
\label{eq:projected_normal_H_real_gauge}
H=\sum_{\bsl{k}\in \M}\ket{\widetilde{\psi}_{\bsl{k}}}\widetilde{h}(\bsl{k}) \bra{\widetilde{\psi}_{\bsl{k}}} + ...\ ,
}
where 
\eq{
\widetilde{h}(\bsl{k})=E_{0,\bsl{k}}\tau_0+g_{\bsl{k}} \cos(\theta_{\bsl{k}})\tau_z +g_{\bsl{k}} \sin(\theta_{\bsl{k}}) \tau_x\ .
}
Since $\ket{\widetilde{\psi}_{\bsl{k}}}$ is given by the unitary rotation of the eigenbases, the energy bands of the two band are $\epsilon_{\bsl{k}}\pm g_{\bsl{k}}$, meaning that the band touching is determined by $g_{\bsl{k}}=0$.
In either patch $\alpha=I$ or $II$, both $\ket{\widetilde{u}^{\alpha}_{\bsl{k}}}$ and $\widetilde{h}_{\alpha}(\bsl{k})$ are smooth; yet, $(\cos(\theta_{\bsl{k}}^{\alpha}),\sin(\theta_{\bsl{k}}^{\alpha}))$ is not always smooth---singular $(\cos(\theta_{\bsl{k}}^{\alpha}),\sin(\theta_{\bsl{k}}^{\alpha}))$ infers $g_{\bsl{k}}= 0$ since  $(\cos(\theta_{\bsl{k}}^{\alpha}),\sin(\theta_{\bsl{k}}^{\alpha}))$ is smooth if $g_{\bsl{k}}\neq 0$.
Then, given a gapless point at $\bsl{k}_{i}$ in patch $\alpha$, we can define the winding of $\bsl{k}_{i}$ as
\eq{
\label{eq:w_tilde_patch}
\widetilde{w}_i=\frac{1}{2\pi}\int_{\gamma_i} d\bsl{k}\cdot\nabla_{\bsl{k}}\theta^{\alpha}\ ,
}
where $\gamma_i$ is a small loop surrounding only $\bsl{k}_i$.
If a band touching point $\bsl{k}_i$ is in the overlap of two patches, {\ie} $\bsl{k}_i\in I\cap II$, we choose $\alpha=I$ for $\widetilde{w}_i$ in the above equation.
Then, applying Stokes' theorem (on the region outside these loops $\gamma_i$), we have
\eq{
\sum_i \widetilde{w}_i=\frac{1}{2\pi} \int_{\partial I} d\bsl{k}\cdot\nabla_{\bsl{k}}(\theta^{I}-\theta^{II}),\
}
where the minus sign is due to the different orientations of the boundaries $\partial I$ and $\partial II$.
According to \eqnref{eq:projected_normal_H_real_gauge} and \eqnref{eq:oriented_gauge}, we have
\eq{
\nabla_{\bsl{k}}(\theta^{I}_{\bsl{k}}-\theta^{II}_{\bsl{k}})=-2\nabla_{\bsl{k}}\Gamma_{\bsl{k}}\ .
}
Combined with \eqnref{eq:e2_real_gauge}, we arrive at the modified Nielsen-Ninomiya theorem
\eq{
\label{eq:nodes_e2_rel_gauge_dep}
\sum_i \widetilde{w}_i= 2 e_2\ .
}
It is different from the conventional Nielsen-Ninomiya theorem in the sense that the total winding numbers of the nodes \eqnref{eq:nodes_e2_rel_gauge_dep} sum to a topological index instead of zero.

The above derivation of \eqnref{eq:nodes_e2_rel_gauge_dep}, relies on the choice of real oriented PWS gauges.
The second derivation for \eqnref{eq:nodes_e2_rel_gauge_dep} exploits energy eigenstates.
Specifically, we choose the two energy eigenstates $\ket{\psi^{eig}_{\bsl{k}}}=(\ket{\psi^{eig}_{1,\bsl{k}}},\ket{\psi^{eig}_{2,\bsl{k}}})$ for the two bands everywhere on the sphere $\M$ except the band touching points.
Since $\ket{\psi^{eig}_{\bsl{k}}}$ are not required to be real, $\ket{u^{eig}_{\bsl{k}}}$ can be made smooth on $\M$ except at the band touching points, by choosing the proper $U(1)$ factors for both eigenvectors, since they have different energies. 
With this gauge choice, $\PT$ is represented as 
\eq{
\PT \ket{u^{eig}_{\bsl{k}}} = \ket{u^{eig}_{\bsl{k}}} \mat{ e^{\ii \Gamma_1(\bsl{k})} & \\ & e^{\ii \Gamma_2(\bsl{k})}}\ .
}
Then, they defined a vector based on the off-diagonal part of the Berry connection as
\eq{
\bsl{b} (\bsl{k}) =  -\ii \bra{u^{eig}_{1,\bsl{k}}}\nabla_{\bsl{k}}\ket{u^{eig}_{2,\bsl{k}}}  \sqrt{e^{-\ii [\Gamma_1(\bsl{k})-\Gamma_2(\bsl{k})]}}\ ,
}
where a global sign has been chosen for the square root.
Eventually, they demonstrated \eqnref{eq:nodes_e2_rel_gauge_dep} by redefining the winding $\widetilde{w}_i$  as 
\eq{
\widetilde{w}_i=\frac{1}{2\pi}\int_{\gamma_i} d\bsl{k}\cdot \bsl{b} (\bsl{k})
}
with $\gamma_i$ infinitesimal.
Nevertheless, both derivations in \refcite{Ahn2019TBGFragile} relies on certain gauge choices.
We will show a generalization of \eqnref{eq:nodes_e2_rel_gauge_dep} to all gauges in \appref{app:reformulate_Euler}, thus demonstrating the modified Nielsen Ninomiya theorem in all gauges.

\section{Review on the Monopole-Charged Spinless Dirac Semimetals}
\label{app:spinless_DSM}

In the section, we review the Monopole-Charged spinless Dirac semimetal proposed in \refcite{Lenggenhager2021TPBBC}.
Throughout this section, we consider 3D spinless noninteracting crystals with $\PT$ symmetry and a six-fold rotation $C_6$ symmetry, where ``spinless" means that we always impose spin $\SU(2)$ symmetry (which requires the SOC to be negligible) and suppress the spin index.


The group spanned by $\PT$ and $C_6$ has four spinless ICRs~\cite{Bradley2009MathSSP,Bernevig2020MagneticTQC,Lenggenhager2021TPMNL,Lenggenhager2021TPBBC}, noted as $A$, $B$, $\Lambda_1$ and $\Lambda_2$.
Specifically, $\Lambda_1$ and $\Lambda_2$ are 2D ICRs, and $A$ and $B$ are 1D ICRs.
Their explicit symmetry representations are 
\eq{
\label{eq:A_B_Lambda12}
A:\ \left\{\begin{array}{ll}
& C_6\dot{=} 1 \\
& \PT \dot{=} \cc
\end{array}
\right. \ ,
}
\eq{
B:\ \left\{\begin{array}{ll}
& C_6\dot{=} -1 \\
& \PT \dot{=} \cc
\end{array}
\right. \ ,
}
\eq{
\Lambda_1:\ \left\{\begin{array}{ll}
& C_6\dot{=} \exp[-\ii \tau_y \frac{2\pi}{6}] \\
& \PT \dot{=}  \cc
\end{array}
\right. \ ,
} 
\eq{
\Lambda_2:\ \left\{\begin{array}{ll}
& C_6\dot{=} \exp[-\ii \tau_y \frac{4\pi}{6}] \\
& \PT \dot{=}  \cc
\end{array}
\right. \ .
}
The above 4 ICRs can be simply derived as follows.
First, we have $C_6^6=1$ owing to the spinless condition, meaning that the $C_6$ eigenvalues are $e^{\pm \ii 2\pi/6}$, $e^{\pm \ii 4\pi/6}$ and $\pm 1$.
Then, given a $C_6$ eigenstate $\ket{\lambda}$ with eigenvalue $\lambda$, $\PT \ket{\lambda}$ is also a $C_6$ eigenstate with eigenvalue $\lambda^*$, owing to $[\PT,C_6]=0$.
If $\lambda\neq\lambda^*$ for $\lambda=e^{\pm \ii 2\pi/6},\  e^{\pm \ii 4\pi/6}$, then $\PT \ket{\lambda}$ and $\ket{\lambda}$ must be different, resulting in 2D ICRs.
Otherwise for $\lambda=\pm$, we cannot distinguish $\PT \ket{\lambda}$ from $\ket{\lambda}$ based on symmetries, and thus they correspond to 1D ICRs.

At a $C_6$-invariant momentum $\bsl{k}$, which obeys $C_6 \bsl{k} = \bsl{k}+\bsl{G}$ with $\bsl{G}$ a reciprocal lattice vector, the Bloch energy eigenstates can be classified by the ICRs that they furnish.
Then, along the $C_6$-invariant axis in the 1BZ, we can have doubly degenerate bands for both $\Lambda_1$ and $\Lambda_2$, while the bands correspond to $A$ and $B$ are single bands unless fine-tuned.
According to \refcite{Lenggenhager2021TPBBC}, the spinless Dirac point given by the band crossing between a $\Lambda_1$ band and a $\Lambda_2$ band.
The spinless Dirac point must be penetrated by a degenerate line between the top two occupied band along the $C_6$-invariant axis, as shown in \figref{fig:SpinlessDirac_NLSM}(c).
In \refcite{Lenggenhager2021TPBBC}, a symmetry argument has been provided to show that the spinless Dirac points given by the crossing between $\Lambda_1$ and $\Lambda_2$ bands must have nonzero $\dsZ_2$ monopole charge protected by $\PT$ symmetry.
In the following, we will review this.

Consider a sphere $\M$ that encloses a spinless Dirac point.
As shown in \figref{fig:SpinlessDirac_NLSM}(c), we can always make $\M$ small enough such that the top two occupied bands are gapped from the lower bands on and in $\M$.
Then, as discussed in \appref{app:review_NLSM_Euler}, the $\dsZ_2$ monopole charge of the spinless Dirac point equals the parity of the Euler class of the top two occupied bands on $\M$, since the lower bands on $\M$ must have trivial $\dsZ_2$ monopole charge.

To calculate the Euler class of the top two occupied bands on $\M$, we perform a stereographic projection to project the sphere $\M$ to a 2D plane.
Specifically, we choose the gapless Dirac point as the origin, and $\M$ is defined by a fixed $|\bsl{k}|$.
Without loss of generality, we choose the $C_6$-invariant axis as the $z$ axis, and the parametrize $\bsl{k}\in\M$ as
\eq{
\bsl{k}=|\bsl{k}|\left(\frac{2 x}{1+x^2 + y^2}, \frac{2 y}{1+x^2 + y^2}, \frac{-1 + x^2 + y^2}{1+x^2 + y^2}\right)
}
with $(x,y)\in\dsR^2$.
With this parametrization, the north pole is at $x=y=0$, while the south pole is at $|x^2+y^2|\rightarrow \infty$.
Recall that computing the Euler class with \eqnref{eq:e_2_original} requires a real oriented PWS gauge for the basis $\ket{\widetilde{u}^{\alpha}_{\bsl{k}}}=(\ket{\widetilde{u}^{\alpha}_{1,\bsl{k}}},\ket{\widetilde{u}^{\alpha}_{2,\bsl{k}}})$ of the top two occupied bands on the sphere $\M$.
To do so, we split the 2D $x-y$ plane into two patches, $I:x>0^-$ and $II:x<0^+$, and we choose a real smooth gauge of $\ket{u_{x,y}}$ in $I$, noted as $\ket{\widetilde{u}^{I}_{x,y}}$.
Note that $\ket{\widetilde{u}^{I}_{x,y}}$ is constant for $|x^2+y^2|\rightarrow \infty$.
Owing to the two-fold rotation $C_2=C_6^3$ symmetry, we can define 
\eq{
\label{eq:e2_real_gauge_C2}
\ket{\widetilde{u}^{II}_{x,y}}=C_2\ket{\widetilde{u}^{I}_{-x,-y}}\ .
}
%
At $x=0$, the transition matrix between the two patches is 
\eqa{
& \braket{\widetilde{u}^{II}_{0,y}}{\widetilde{u}_{0,y}^{I}}=\bra{\widetilde{u}^{I}_{0,-y}}C_2^\dagger\ket{\widetilde{u}_{0,y}^{I}}\\
& =\bra{\widetilde{u}^{I}_{0,-y}}C_2\ket{\widetilde{u}_{0,y}^{I}}=U_{C_2}(0,y)\ ,
}
where $C_2^2=1\Rightarrow C_2=C_2^\dagger$ is used.
Owing to the $\PT$ symmetry, $U_{C_2}(0,y)$ is a real matrix, and thus $U_{C_2}(0,y)\in O(2)$.
At $(x,y)=(0,0)$, the corresponding $\bsl{k}$ is at the north pole, which is a $C_6$-invariant momentum.
Thus, we know $\ket{\widetilde{u}_{0,0}^{I}}$ furnish a $\Lambda_1$ or $\Lambda_2$ ICR, meaning that $C_2 \ket{\widetilde{u}_{0,0}^{I}}=\pm \ket{\widetilde{u}_{0,0}^{I}}\Rightarrow\det[U_{C_2}(0,0)]=1$.
Since $U_{C_2}(0,y)$ is smooth, $\det[U_{C_2}(0,y)]=1$ for any $y$, meaning that $\ket{u_{x,y}^{I/II,R}}$ is a oriented gauge.

With this oriented gauge, the Euler number reads
\eq{
 |e_2| = \left|\frac{1}{2\pi}\int_{I} dx dy\ f_I(x,y)+\frac{1}{2\pi}\int_{II} dx dy\ f_{II}(x,y)\right|\ ,
}
where 
\eq{
f_\alpha(x,y)=\epsilon^{i j}\partial_i \pf[\bra{\widetilde{u}^{I}_{x,y}} \partial_j \ket{\widetilde{u}^{I}_{x,y}}]
}
with $i,j=x,y$.
Owing to \eqnref{eq:e2_real_gauge_C2}, we have $f_{II}(x,y)=f_{I}(-x,-y)$, resulting in 
\eqa{
& |e_2| =\left|\frac{1}{\pi} \int_{I} dx dy\ f_I(x,y)\right| \\
& =\left|\frac{1}{\pi} \int_{\partial I} \left\{dx\ \pf[\bra{\widetilde{u}^{I}_{x,y}} \partial_x \ket{\widetilde{u}^{I}_{x,y}}] + dy\ \pf[\bra{\widetilde{u}^{I}_{x,y}} \partial_y \ket{\widetilde{u}^{I}_{x,y}}]\right\}\right| \\
& =\left|\frac{1}{\pi} \int_{-\infty}^{\infty}  dy\ \pf[\bra{\widetilde{u}^{I}_{0,y}} \partial_y \ket{\widetilde{u}^{I}_{0,y}}]\right|
\ ,
}
where the second equality uses Stokes' theorem, and the third equality is derived from the fact that $\ket{\widetilde{u}^{I}_{x,y}}$ is constant for $|x^2+y^2|\rightarrow \infty$. 
Combined with $C_2 \ket{\widetilde{u}^{I}_{0,y}}=\ket{\widetilde{u}^{I}_{0,-y}} U_{C_2}(0,y)$, we get
\eqa{
& |e_2|=\left|\frac{1}{\pi} \int_{0}^{\infty}  dy\ \left\{ \pf[\bra{\widetilde{u}^{I}_{0,y}} \partial_y \ket{\widetilde{u}^{I}_{0,y}}] -\pf[\bra{\widetilde{u}^{I}_{0,-y}} \partial_y \ket{\widetilde{u}^{I}_{0,-y}}] \right\} \right|\\
& =\left|-\frac{1}{\pi} \int_{0}^{\infty}  dy\ \pf[ U_{C_2}(0,y)\partial_y U_{C_2}^\dagger(0,y)]\right|\ .
}
As $U_{C_2}(0,y)\in \SO(2)$, it should have the form $U_{C_2}(0,y)=e^{\ii \tau_y \Gamma(0,y)}$.
Moreover, as shown in \figref{fig:FS}(a), the top two occupied band should correspond to different 2D ICRs at the north and south poles, meaning that $U_{C_2}(0,0)=-U_{C_2}(0,y)=\pm 1\Rightarrow \Gamma(0,0)-\Gamma(0,\infty)=\pi\ \mod\ 2\pi$.
Then, we eventually arrive at 
\eq{
 |e_2|=\left|\frac{1}{\pi}[-\Gamma(0,0)+\Gamma(0,\infty)]\right|=1\ \mod\ 2\ .
}
It means that the top two occupied bands have an odd $|e_2|$, and thus the spinless Dirac point given by the crossing between one $\Lambda_1$ band and one $\Lambda_2$ band has a nonzero $\dsZ_2$ monopole charge, which we refer to as MSDP.

This nonzero monopole charge infers that even if the $C_6$ symmetry is broken, the MSDP will split to a MNL instead of disappearing, as shown in \figref{fig:FS}(b,c).

\section{Gauge-Invariant Euler number}
\label{app:reformulate_Euler}

In the section, we present more details on the gauge-invariant Euler number that captures the topology characterized by the Euler class $|e_2|$.
Again, throughout this section, we always consider 3D spinless noninteracting crystals with $\PT$ symmetry, unless specified otherwise.

Before going into details of the gauge-invariant formalism, we emphasize again what do we mean by gauge.
Consider a generic isolated set of bands on a manifold $\M$ with basis $\ket{u_{\bsl{k}}}$, where $\ket{u_{\bsl{k}}}$ is the periodic part of the Bloch basis as discussed above \eqnref{eq:UN_gauge}, and  
\eq{
\PT\ket{u_{\bsl{k}}}=\ket{u_{\bsl{k}}}U_{\bsl{k}}\ .
}
$\ket{u_{\bsl{k}}}$ is a single-valued function on $\M$, and choosing a gauge for $\ket{u_{\bsl{k}}}$ is nothing but specifying the form of the function.
We can have many different function forms for $\ket{u_{\bsl{k}}}$, and different forms are related by the gauge transformation in \eqnref{eq:UN_gauge}.
One example is the smooth gauge, which just means that we choose a form of $\ket{u_{\bsl{k}}}$ that is smooth in $\M$.

Now we discuss the real gauges of $\ket{u_{\bsl{k}}}$.
In the last two sections, we used the real PWS gauges, which are technically not gauges of $\ket{u_{\bsl{k}}}$, since they are not single-valued at some momenta.
Nevertheless, the real PWS gauges are closed related to the real gauges of $\ket{u_{\bsl{k}}}$ that we will use.
A real (oriented) gauge refers to $\ket{u_{\bsl{k}}}$ that satisfies that (i) $\PT\ket{u_{\bsl{k}}}=T\ket{u_{\bsl{k}}}$ and (ii) there exists a real (oriented) PWS $\ket{\widetilde{u}^\alpha_{\bsl{k}}}$ such that $\ket{u_{\bsl{k}}}=\ket{\widetilde{u}^{\alpha_{\bsl{k}}}_{\bsl{k}}}$ with $\alpha$ labelling the patches and $\bsl{k}\in\alpha_{\bsl{k}}$.
In other words, a real (oriented) gauge of $\ket{u_{\bsl{k}}}$ is given by assigning a special patch $\alpha_{\bsl{k}}$ to each $\bsl{k}$ in a real (oriented) PWS $\ket{\widetilde{u}^\alpha_{\bsl{k}}}$.
The real curvature and the Euler class for a real oriented gauge of $\ket{u_{\bsl{k}}}$ can be unambiguously defined by the corresponding real oriented PWS $\ket{\widetilde{u}^\alpha_{\bsl{k}}}$, though $\ket{\widetilde{u}^\alpha_{\bsl{k}}}$ is not unique for one $\ket{u_{\bsl{k}}}$.
It is because different real oriented PWS gauges for one real oriented gauge of $\ket{u_{\bsl{k}}}$ always have the same real curvature and same Euler class.
In some cases, we want to exploit the properties of the PWS gauge, and then we can choose a gauge of $\ket{u_{\bsl{k}}}$ such that it is equal to a PWS gauge $\ket{\widetilde{u}^\alpha_{\bsl{k}}}$ within a certain patch $\alpha$, \ie, $\ket{u_{\bsl{k}}}=\ket{\widetilde{u}^\alpha_{\bsl{k}}}$ for $\bsl{k}\in\alpha$.
The above discussion of the real gauges of $\ket{u_{\bsl{k}}}$ is useful for the discussion in the following.

In the rest of this section, we will first review the path-dependent parallel transport in general, then define the $\eta$ factor, and finally provide a gauge-invariant generalization of $|e_2|$, which we call the Euler number $\N$.
We also present a generalized version of \eqnref{eq:nodes_e2_rel_gauge_dep} for all gauges or a demonstration of the modified Nielsen-Ninomiya theorem for all gauges.
Although the Wilson loop winding number can be viewed as one gauge-invariant generalization of $|e_2|$, it does not provide the $\bsl{k}$-resolved information of the band topology, which is required for deriving a generalized version of \eqnref{eq:nodes_e2_rel_gauge_dep}.

\subsection{Review of Parallel Transport}

In this part, we first review the parallel transport, which was studied in \refcite{Soluyanov2011WannierZ2,Soluyanov2012SmoothGaugeZ2,Li2020TIHOTIGaugeInvLines}.
Throughout this part, we consider a set of $N$ bands that are gapped from other bands along certain 1D paths in the 1BZ.
Throughout this work, all 1D paths are chosen to be continuous and can be given by gluing smooth parts.
We label the Bloch basis of $N$ bands by $\ket{\psi_{\bsl{k}}}=(\ket{\psi_{1,\bsl{k}}},...,\ket{\psi_{a,\bsl{k}}},...,\ket{\psi_{N,\bsl{k}}})$, and use $\ket{u_{\bsl{k}}}=e^{-\ii \bsl{k}\cdot\bsl{r}}\ket{\psi_{\bsl{k}}}$ to label the periodic part of the Bloch basis.
We do not require $\ket{\psi_{\bsl{k}}}$ to be the eigenstates of the $N$ bands, and thus $\ket{\psi_{\bsl{k}}}$ has $\U(N)$ gauge freedom (\eqnref{eq:UN_gauge}).

In \appref{app:review_NLSM_Euler}, we reviewed the Wilson loop \eqnref{eq:WL} for a set of bands that are isolated along a closed loop $\gamma$.
Now let us consider a non-closed $\gamma$ in the 1BZ, and label the initial point and the final point of $\gamma$ as $\bsl{k}_i$ and $\bsl{k}_f$, respectively.
Then, \eqnref{eq:WL} is changed to \eqnref{eq:parallel_transport} with $(\bsl{k}_0,\bsl{k})$ replaced by $(\bsl{k}_i,\bsl{k}_f)$, which corresponds to the parallel transport~\cite{Soluyanov2011WannierZ2,Soluyanov2012SmoothGaugeZ2,Li2020TIHOTIGaugeInvLines}.
\eqnref{eq:parallel_transport} was also called the Wilson line in the literature~\cite{Li2020TIHOTIGaugeInvLines}.

Of particular interest is the determinant of $W(\bsl{k}_i\xrightarrow[]{\gamma} \bsl{k}_f)$, labeled as $\det[W(\bsl{k}_i\xrightarrow[]{\gamma} \bsl{k}_f)]$ (with the determinant operation followed by the limitation).
The following three known properties of $\det[W(\bsl{k}_i\xrightarrow[]{\gamma} \bsl{k}_f)]$ are relevant to our later discussion~\cite{Soluyanov2011WannierZ2,Soluyanov2012SmoothGaugeZ2,Li2020TIHOTIGaugeInvLines}.
First, unlike the determinant of Wilson loop, $\det[W(\bsl{k}_i\xrightarrow[]{\gamma} \bsl{k}_f)]$ is gauge dependent---under $U(N)$ gauge transformation \eqnref{eq:UN_gauge}, we have
\eq{
\det[W(\bsl{k}_i\xrightarrow[]{\gamma} \bsl{k}_f)]\rightarrow \det[W(\bsl{k}_i\xrightarrow[]{\gamma} \bsl{k}_f)] \frac{\det[R_{\bsl{k}_f}]}{\det[R_{\bsl{k}_i}]}\ .
}
Therefore, $\det[W(\bsl{k}_i\xrightarrow[]{\gamma} \bsl{k}_f)]$ not only depends on the path $\gamma$ but also depends on the gauge choice for the basis at $\bsl{k}_{i}$ and $\bsl{k}_{f}$.

Second, based on \eqnref{eq:parallel_transport}, it is straightforward to get the following composite rule
\eqa{
& \det[W(\bsl{k}_i\xrightarrow{\gamma_1} \bsl{k}')]\det[W(\bsl{k}'\xrightarrow{\gamma_2} \bsl{k}_f)]\\
& =\det[W(\bsl{k}_i\xrightarrow{\gamma_1+\gamma_2} \bsl{k}_f)]\ ,
}
where  the end of $\gamma_1$ and the start of the $\gamma_2$ are both $\bsl{k}'$, and $\gamma_1+\gamma_2$ stands for the path given by connecting the end of $\gamma_1$ to the start of the $\gamma_2$.

Third, $\det[W(\bsl{k}_i\xrightarrow[]{\gamma} \bsl{k}_f)]$ is a $U(1)$ number.
To see this, let us note that when $\gamma$ is smooth and for a smooth gauge of $\ket{u_{\bsl{k}}}$ along $\gamma$, we have
\eq{
\label{eq:eta_Berry_phase}
\det[W(\bsl{k}_i\xrightarrow[]{\gamma} \bsl{k}_f)]= \exp\left[\ii  \int_{\bsl{k}_i\xrightarrow{\gamma} \bsl{k}_f} d\bsl{k}\cdot \Tr(\bsl{A}(\bsl{k})) \right]\ ,
}
where $\bsl{A}(\bsl{k})=-\ii \bra{u_{\bsl{k}}} \nabla_{\bsl{k}} \ket{u_{\bsl{k}}}$ is the non-abelian Berry connection.
The above equation means that $\det[W(\bsl{k}_i\xrightarrow[]{\gamma} \bsl{k}_f)]\in \U(1)$ holds for any smooth gauge of $\ket{u_{\bsl{k}}}$ along smooth $\gamma$, and since the gauge transformation can only change $\det[W(\bsl{k}_i\xrightarrow[]{\gamma} \bsl{k}_f)]$ by a $\U(1)$ factor, we know $\det[W(\bsl{k}_i\xrightarrow[]{\gamma} \bsl{k}_f)]\in \U(1)$ holds for any gauge along smooth $\gamma$.
The result holds even if $\gamma$ is not smooth, since $\gamma$ can always be split into smooth parts.
Then, combined with \eqnref{eq:parallel_transport}, we have
\eqa{
\det[W(\bsl{k}_i\xrightarrow[]{\gamma} \bsl{k}_f)]^*& =\frac{1}{\det[W(\bsl{k}_i\xrightarrow[]{\gamma} \bsl{k}_f)]} \\
& =\det[W(\bsl{k}_f\xrightarrow{\overline{\gamma}} \bsl{k}_i)]\ ,
}
where $\overline{\gamma}$ is the reverse of $\gamma$.

$\det[W(\bsl{k}_i\xrightarrow[]{\gamma} \bsl{k}_f)]$ has been applied to various cases, such as (spinful) TR-protected topological insulators~\cite{Soluyanov2011WannierZ2,Soluyanov2012SmoothGaugeZ2,Li2020TIHOTIGaugeInvLines} and 3D higher-order topological insulators protected by the combination of time-reversal and rotation symmetries~\cite{Li2020TIHOTIGaugeInvLines}.
In the presence of spinless $\PT$ symmetry (or the combination of two-fold rotation and TR symmetry for 2D systems), the Wilson loop or parallel transport was studied with smooth gauges along the path~\cite{Ahn2018MonopoleNLSM,Ahn2019SymRepC2T} .
In the following, we will construct a path-independent $\eta$ factor from the parallel transport expression \eqnref{eq:parallel_transport}.

\subsection{Path-Independent $\eta$ Factor}

In this part, we first discuss how $\det[W(\bsl{k}_i\xrightarrow[]{\gamma} \bsl{k}_f)]$ becomes path-independent on a closed 2D sphere-like (or equivalently diffeomorphic to a sphere) manifold in the presence of spinless $\PT$ symmetry.

Let us consider a generic 2D closed submanifold $\M$ in the 3D 1BZ, which we assume is sphere-like.
Suppose there exists $N$ bands that are gapped from all other bands at every point on $\M$.
According to \appref{app:review_NLSM_Euler}, the determinant of the Wilson loop $\det(W_{\gamma})$ for the $N$ bands takes values $\pm 1$ for any closed loop $\gamma$ on $\M$.
Furthermore, since $\M$ is sphere-like, every closed loop on $\M$ can be smoothly contracted to a point, meaning that 
\eq{
\label{eq:detW_1_orientable}
\det(W_{\gamma})=1\ \forall \text{ closed path }\gamma\subset\M\ .
}
%
%
\eqnref{eq:detW_1_orientable} suggests that for any two points $\bsl{k}_0$ and $\bsl{k}$ on $\M$, $\det[W(\bsl{k}_0\xrightarrow{\gamma} \bsl{k})]$ is independent of the path $\gamma$.
To see this, consider two paths from $\bsl{k}_0$ to $\bsl{k}$, labelled as $\gamma$ and $\gamma'$.
Then, we have
\eqa{
& \frac{\det[W(\bsl{k}_0\xrightarrow{\gamma} \bsl{k})]}{\det[W(\bsl{k}_0\xrightarrow{\gamma'} \bsl{k})]}=\det[W(\bsl{k}_0\xrightarrow{\gamma} \bsl{k})]\det[W(\bsl{k}\xrightarrow{\overline{\gamma}'} \bsl{k}_0)]\\
&=\det[W(\bsl{k}_0\xrightarrow{\gamma+\overline{\gamma}'} \bsl{k}_0)]=\det(W_{\gamma+\overline{\gamma}'})=1\\
&
\Rightarrow  \det[W(\bsl{k}_0\xrightarrow{\gamma} \bsl{k})]=\det[W(\bsl{k}_0\xrightarrow{\gamma'} \bsl{k})]\ .
}

As $\det[W(\bsl{k}_0\xrightarrow{\gamma} \bsl{k})]$ is independent of the path, we can construct the path-independent $\eta$ factor from it as \eqnref{eq:eta_path_indep}.
Note that the path-independent nature of $\eta_{\bsl{k}_0}(\bsl{k})$ relies on three things: (i) the spinless $\PT$, (ii) $\M$ being sphere-like, and (iii) the $N$ bands being isolated on $\M$.

Based on the three properties of $\det[W(\bsl{k}_0\xrightarrow{\gamma} \bsl{k})]$ discussed in the last part, we have the following four properties of the $\eta_{\bsl{k}_0}(\bsl{k})$.
First, \eq{
\label{eq:eta_PT_based_point_dep}
\eta_{\bsl{k}_0}(\bsl{k})=\eta_{\bsl{k}_0}^*(\bsl{k})\det[U_{\bsl{k}}]\ ,
}
where $U_{\bsl{k}}=\braket{u_{\bsl{k}}}{u_{\bsl{k}}^{\PT}}$.
%
Second, $\eta_{\bsl{k}_0}(\bsl{k})\in \U(1)$ since $\det[\braket{u^{\PT}_{\bsl{k}_0}}{u_{\bsl{k}_0}}]=\det[U_{\bsl{k}_0}]^*\in \U(1)$.
Third, under $\U(N)$ gauge transformation \eqnref{eq:UN_gauge}, we have 
\eq{
\eta_{\bsl{k}_0}(\bsl{k})\rightarrow \eta_{\bsl{k}_0}(\bsl{k}) \chi(R_{\bsl{k}_0})  \det[R_{\bsl{k}}]\ ,
}
where 
\eq{
\chi(R_{\bsl{k}_0})=\frac{\sqrt{\det[\braket{u^{\PT}_{\bsl{k}_0}}{u_{\bsl{k}_0}}]\det[R_{\bsl{k}_0}]^2} }{\sqrt{\det[\braket{u^{\PT}_{\bsl{k}_0}}{u_{\bsl{k}_0}}]} \det[R_{\bsl{k}_0}] } =\pm 1\ .
}
Fourth, for $\bsl{k}_0'\in\M$,
\eq{
\label{eq:eta_base_point_change}
\eta_{\bsl{k}_0'}(\bsl{k})=\chi_{\bsl{k}_0',\bsl{k}_0}\eta_{\bsl{k}_0}(\bsl{k})\ ,
}
where 
\eq{
\label{eq:eta_base_point_change_factor}
\chi_{\bsl{k}_0',\bsl{k}_0}=\eta_{\bsl{k}_0'}(\bsl{k}_0)\frac{1}{\sqrt{\det[\braket{u^{\PT}_{\bsl{k}_0}}{u_{\bsl{k}_0}}]}}=\pm 1\ .
}
Here $\chi_{\bsl{k}_0',\bsl{k}_0}=\pm 1$ is derived from $\chi_{\bsl{k}_0',\bsl{k}_0}\in \dsR$ and $\chi_{\bsl{k}_0',\bsl{k}_0}\in \U(1)$.
%
In the following, we treat $\bsl{k}_0$ as a base point, and then $\eta_{\bsl{k}_0}$ is a function from $\M$ to $U(1)$ after fixing a gauge for $\ket{u_{\bsl{k}}}$.
With this convention, \eqnref{eq:eta_base_point_change} should be viewed as a change of the base point.

Interestingly, $\eta_{\bsl{k}_0}(\bsl{k})$ can remember the orientation of basis in any real gauge for $\ket{u_{\bsl{k}}}$.
Consider a generic real gauge (not necessarily oriented) of $\ket{u_{\bsl{k}}}$ with corresponding real PWS $\ket{\widetilde{u}_{\bsl{k}}^{\alpha}}$, where $\alpha=I,II,...$ labels the patches.
Then, 
\eq{
\eta_{\bsl{k}_0}(\bsl{k}) = \widetilde{\eta}_{\alpha_{\bsl{k}_0},\bsl{k}_0}(\alpha_{\bsl{k}},\bsl{k})
}
with
\eq{
\widetilde{\eta}_{\alpha_0,\bsl{k}_0}(\alpha,\bsl{k})= \lim_{L\rightarrow\infty} \det\left[ \bra{\widetilde{u}_{\bsl{k}_0}^{\alpha_0}} P_{\bsl{k}_1} P_{\bsl{k}_2} ...P_{\bsl{k}_L} \ket{\widetilde{u}_{\bsl{k}}^{\alpha}} \right]\ ,
}
which takes values $\pm 1$.
Here we have used $\PT\ket{\widetilde{u}_{\bsl{k}_0}^{\alpha_0}}=\ket{\widetilde{u}_{\bsl{k}_0}^{\alpha_0}}$.
Now we show that for all $\bsl{k}_0\in \alpha_0$ and $\bsl{k}\in \alpha$,
we have 
\eq{
\label{eq:eta_R_bar}
\widetilde{\eta}_{\alpha_0,\bsl{k}_0}(\alpha,\bsl{k})=\overline{\eta}_{\alpha_0}(\alpha)
}
with $\overline{\eta}_{\alpha_0}(\alpha)$ only depending on the patches $\alpha_0$ and $\alpha$.
To show this, let us note that  
\eqa{
 \widetilde{\eta}_{\alpha,\bsl{k}_0}(\alpha,\bsl{k})=\exp\left[\ii  \int_{\bsl{k}_0\rightarrow \bsl{k}} d\bsl{l}\cdot \Tr[\widetilde{\bsl{A}}^{\alpha}(\bsl{l})] \right]= 1\ ,
}
where $\Tr[\widetilde{\bsl{A}}^{\alpha}(\bsl{k})]=-\ii \Tr[\bra{\widetilde{u}^{\alpha}_{\bsl{k}}} \nabla_{\bsl{k}} \ket{\widetilde{u}^{\alpha}_{\bsl{k}}}]=0$ and \eqnref{eq:eta_Berry_phase} are used. 
Thus, for all $\bsl{k}_0,\bsl{k}_0'\in\alpha_0$ and $\bsl{k},\bsl{k}'\in \alpha$, we have
\eqa{
& \widetilde{\eta}_{\alpha_0,\bsl{k}_0'}(\alpha,\bsl{k}')= \widetilde{\eta}_{\alpha_0,\bsl{k}_0'}(\alpha_0,\bsl{k}_0) \widetilde{\eta}_{\alpha_0,\bsl{k}_0}(\alpha,\bsl{k})  \widetilde{\eta}_{\alpha,\bsl{k}}(\alpha,\bsl{k}') \\
& = \widetilde{\eta}_{\alpha_0,\bsl{k}_0}(\alpha,\bsl{k})\ ,
}
proving \eqnref{eq:eta_R_bar}.

In fact, $\overline{\eta}_{\alpha_0}(\alpha)=1$ indicates that the orientation of the basis in $\alpha$ is the same as that in $\alpha_0$; otherwise, opposite.
To see this, let us choose a path from $\bsl{k}_0$ in $\alpha_0$ to $\bsl{k}$ in $\alpha$, and sequentially find the patches that the path passes through, noted as $\alpha_{0,1,2,...,m}$ with $\alpha_m=\alpha$.
Then, select one momentum for each intersection of two neighboring patches according to the path as $\bsl{k}_{i}\in \alpha_i\cap \alpha_{i-1}$ for $i=1,...,m$, and relabel $\bsl{k}_{m+1}=\bsl{k}$, which allow us to split the $\eta$ as
\eqa{
& \overline{\eta}_{\alpha_0}(\alpha)=\widetilde{\eta}_{\alpha_0,\bsl{k}_0}(\alpha,\bsl{k}) 
 = \prod_{i=1}^{m}\det\left( \braket{\widetilde{u}_{\bsl{k}_{i}}^{\alpha_{i-1}}}{\widetilde{u}_{\bsl{k}_{i}}^{\alpha_i}}\right)\ .
}
Since $\braket{\widetilde{u}_{\bsl{k}_{i}}^{\alpha_{i-1}}}{\widetilde{u}_{\bsl{k}_{i}}^{\alpha_i}}$ is nothing but the transition function at $\bsl{k}_{i}$, $\det\left(\braket{\widetilde{u}_{\bsl{k}_{i}}^{\alpha_{i-1}}}{\widetilde{u}_{\bsl{k}_{i}}^{\alpha_i}}\right)$ is $1$ ($-1$) if patches $\alpha_{i-1}$ and $\alpha_{i}$ has the same (opposite) orientations.
Then, $\overline{\eta}_{\alpha_0}(\alpha)=\widetilde{\eta}_{\alpha_0,\bsl{k}_0}(\alpha,\bsl{k})$ is equal to $1$ ($-1$) if the orientation is flipped by an even (odd) times as the path goes from $\bsl{k}_0$ to $\bsl{k}$.
As $\widetilde{\eta}_{\alpha_0,\bsl{k}_0}(\alpha,\bsl{k})$ is path independent, whether the orientation is flipped by even or odd times is path independent, which should indicate whether $\ket{\widetilde{u}_{\bsl{k}_0}^{\alpha_0}}$ and $\ket{\widetilde{u}_{\bsl{k}}^{\alpha}}$ have the same or opposite orientations. 
Eventually, we have that $\overline{\eta}_{\alpha_0}(\alpha)$ equals to $1$ ($-1$) if the real gauge has the same (opposite) orientations in $\alpha$ and $\alpha_0$.
Clearly, the real gauge is oriented iff $\overline{\eta}_{\alpha_0}(\alpha)=1$ for a certain $\alpha_0$ and all $\alpha$.
Then, for a generic real gauge, we may always pick a $\alpha_0$ and flip the orientation for all $\alpha$ with $\overline{\eta}_{\alpha_0}(\alpha)=-1$ to get an oriented gauge.
Therefore, the $\widetilde{\eta}_{\alpha_0,\bsl{k}_0}(\alpha,\bsl{k})$ factor can help select out an orientation for a real gauge, which can be helpful for calculating the Euler class using \eqnref{eq:e_2_original}.

Nevertheless, this is not what we will do, since we still need to pick a real gauge to use \eqnref{eq:e_2_original}.
What we ultimately want is the gauge-invariant Euler number $\N$.
Before moving on to the gauge-invariant formalism for $\N$, we will discuss the smoothness property of $\eta_{\bsl{k}_0}(\bsl{k})$ based on \eqnref{eq:eta_R_bar}.

The first statement that we want to make is that for any $\bsl{k}\in\M$ and any open neighborhood $D_{\bsl{k}}\in\M$ of $\bsl{k}$, we can always choose a gauge $\ket{u_{\bsl{k}}}$ such that $\ket{u_{\bsl{k}}}$ and $\eta_{\bsl{k}_0}(\bsl{k})$ is smooth in $D_{\bsl{k}}$.
To prove the above statement, note that we can always choose $\ket{u_{\bsl{k}}}$ to be smooth everywhere in $\M$ since $\PT$ symmetry requires the Chern number of $\ket{u_{\bsl{k}}}$ to be zero on $\M$~\cite{Brouder2007Wannier}.
Then, we can prove the above statement if we demonstrate that for any gauge $\ket{u_{\bsl{k}}}$ that is everywhere smooth in $\M$, $\eta_{\bsl{k}_0}(\bsl{k})$ is everywhere smooth in $\M$.
To prove the stronger statement, let us first choose a real PWS $\ket{\widetilde{u}_{\bsl{k}}^{\alpha}}$, and we know $\widetilde{\eta}_{\alpha_0,\bsl{k}_0}(\alpha,\bsl{k})=\overline{\eta}_{\alpha_0}(\alpha)$ that is constant in $\bsl{k}$ for all $\alpha$.
Then, for any $\bsl{k}\in\M$, there exists a patch $\alpha$ such that $\bsl{k}\in \alpha$.
From $\ket{\widetilde{u}_{\alpha,\bsl{k}}}R_{\alpha,\bsl{k}}=\ket{u_{\bsl{k}}}$, we know 
\eq{
\label{eq:eta_eta_R}
\eta_{\bsl{k}_0}(\bsl{k})=\frac{\sqrt{\det^2[R_{\alpha_0,\bsl{k}_0}]}}{\det[R_{\alpha_0,\bsl{k}_0}]}\widetilde{\eta}_{\alpha_0,\bsl{k}_0}(\alpha,\bsl{k})\det[R_{\alpha,\bsl{k}}] \text{ for }\bsl{k}\in \alpha\ .
}
Since $\ket{\widetilde{u}_{\alpha,\bsl{k}}}$ is smooth in $\alpha$, $R_{\alpha,\bsl{k}}$ and $\det[R_{\alpha,\bsl{k}}]$ are smooth in $\alpha$, meaning that $\eta_{\bsl{k}_0}(\bsl{k})$ is smooth in $\alpha$.
Thus, $\eta_{\bsl{k}_0}(\bsl{k})$ is smooth at $\bsl{k}$ for any $\bsl{k}$ in $\M$, meaning that $\eta_{\bsl{k}_0}(\bsl{k})$ is smooth in $\M$ for any gauge $\ket{u_{\bsl{k}}}$ that is everywhere smooth in $\M$.

The second statement that we make is that for any gauge $\ket{u_{\bsl{k}}}$, if $\ket{u_{\bsl{k}}}$ is smooth in an open $D\subset\M$, $\eta_{\bsl{k}_0}(\bsl{k})$ is smooth in $D$.
To see that, we label a globally smooth gauge as $\ket{u_{\bsl{k}}^{GS}}$, and then $\eta_{\bsl{k}_0}^{GS}(\bsl{k})$ is smooth in $D$.
Then, as $\ket{u_{\bsl{k}}}=\ket{u_{\bsl{k}}^{GS}} R_{\bsl{k}}$ with $R_{\bsl{k}}$ smooth in $D$, we have $\eta_{\bsl{k}_0}(\bsl{k})=\chi[R_{\bsl{k}_0}]\eta_{\bsl{k}_0}^{GS}(\bsl{k}) \det[R_{\bsl{k}}]$ smooth in $D$.

\subsection{Gauge-Invariant Euler Number}
With the $\eta$ factor, we will provide an alternative gauge-invariant expression for topology characterized by the Euler class in this part.
Throughout this part, we again again consider a generic 2D sphere-like submanifold $\M$ in the 3D 1BZ.
Unlike the last part, we focus on the case where there exists two bands that are gapped from all other bands at every point on $\M$ in this part.
So the Bloch basis of the two bands in the 3D 1BZ are $\ket{\psi_{\bsl{k}}}=(\ket{\psi_{1,\bsl{k}}},\ket{\psi_{2,\bsl{k}}})$, and the periodic parts of the Bloch basis are $\ket{u_{\bsl{k}}}=e^{-\ii \bsl{k}\cdot\bsl{r}}\ket{\psi_{\bsl{k}}}$.
Specifically, we only require the orthonormal $\ket{\psi_{a,\bsl{k}}}$ to be linear combinations of the eigenstates of the two bands at $\bsl{k}$, and thus $\ket{u_{\bsl{k}}}$ has $\U(2)$ gauge freedom (\eqnref{eq:U2_gauge}).

To get a gauge-invariant formalism, we need the projection operator for the two bands $P_{\bsl{k}}= \ket{u_{\bsl{k}}}\bra{u_{\bsl{k}}}$.
Besides $P_{\bsl{k}}$, the other quantity that we use for the reformulation is the Q operation \eqnref{eq:Q_k0}.
As discussed in the main text, the $U(2)$ gauge transformation of $\ket{u_{\bsl{k}}}$ (\eqnref{eq:UN_gauge} with $N=2$ or equivalently \eqnref{eq:U2_gauge}) and changing the base point only change the $Q_{\bsl{k}_0}(\bsl{k})$ by a $\bsl{k}$-independent sign factor.
Owing to the $\bsl{k}$-independent gauge freedom, $Q_{\bsl{k}_0}$ is always a smooth function of $\bsl{k}\in \M$ for any gauge of $\ket{u_{\bsl{k}}}$ and for any base point $\bsl{k}_0$.
It is because $Q_{\bsl{k}_0}$ must be a smooth function of $\bsl{k}\in\M$ for a globally smooth gauge, and the gauge transformation and base point change can only change $Q_{\bsl{k}_0}$ by a $\bsl{k}$-independent sign factor.

With \eqnref{eq:Q_k0} and $P_{\bsl{k}}$, we define $\bsl{\Phi}_{\bsl{k}_0}(\bsl{k})$ as \eqnref{eq:Phi}, which is smooth in $\M$ for any gauge.
We further define $\N_{\bsl{k}_0}$ as \eqnref{eq:N_k0}.
Under gauge transformations or bases point changes, $\bsl{\Phi}_{\bsl{k}_0}(\bsl{k})$ and $\N_{\bsl{k}_0}$ only changes by a sign factor as $Q_{\bsl{k}_0}(\bsl{k})$.
Then, we define the gauge-invariant quantity $\N$ as \eqnref{eq:N}.
Importantly, we have $\bsl{\Phi}_{\bsl{k}_0}(\bsl{k})$ is real, $\N_{\bsl{k}_0}\in \dsZ$ and thus $|\N|\in \dsN$ for any gauges, and 
\eqa{
\label{eq:N_e2abs}
\N=|e_2|\text{ for any real oriented gauge}\ .
}
To see this, let us choose a generic real oriented gauge of $\ket{u_{\bsl{k}}}$ with corresponding PWS $\ket{\widetilde{u}^\alpha_{\bsl{k}}}$.
Owing to \eqnref{eq:eta_R_bar}, $Q_{\bsl{k}_0}(\bsl{k})$ can be simplified to
\eq{ 
\label{eq:Q_k0_real_gauge}
Q_{\bsl{k}_0}(\bsl{k})=-\frac{1}{\sqrt{2}}\ket{\widetilde{u}_{\bsl{k}}^{\alpha}}\ii \tau_y\bra{\widetilde{u}_{\bsl{k}}^{\alpha}}\ .
}
\eqnref{eq:Q_k0_real_gauge} represents the orientation of the gauge in each patch, since it is invariant under an orientation-preserving real gauge transformation but changes sign under an orientation-flipping real gauge transformation.
As a result, we have
\eqa{
& \bsl{\Phi}_{\bsl{k}_0}(\bsl{k}) =  \bra{\nabla_{\bsl{k}}\widetilde{u}_{\bsl{k},1}^{\alpha}}\times \ket{\nabla_{\bsl{k}}\widetilde{u}_{\bsl{k},2}^{\alpha}} = \bsl{f}_{\bsl{k}}\ .
}
Then, $\bsl{\Phi}_{\bsl{k}_0}(\bsl{k})$ is real and
\eq{
\N_{\bsl{k}_0} =  e_2 \in \dsZ \Rightarrow \N=|e_2|
}
hold for any real oriented gauges, proving \eqnref{eq:N_e2abs}.
Therefore, $\N$ is a gauge-invariant expression for topology characterized by the Euler class, which we call the Euler number.
Furthermore, since base point change and the gauge transformation only changes $\bsl{\Phi}_{\bsl{k}_0}(\bsl{k})$ and $\N_{\bsl{k}_0}$ by $\pm 1$, we have $\bsl{\Phi}_{\bsl{k}_0}(\bsl{k})$ is real and $\N_{\bsl{k}_0}\in\dsZ$ for all gauges and base point choices.

When $\N$ is nonzero, we can construct based-point-independent $\eta$ factor, $\bsl{\Phi}$, and $Q$, which are very useful for the derivation of Euler obstructed Cooper pairing in the next section.
The nonzero $\N$ allows us to define the base-point-independent $\eta( \bsl{k} )$ as
\eq{
\label{eq:eta_path_indep_no_base_point}
\eta( \bsl{k} )=\frac{\N_{\bsl{k}_0}}{\N} \eta_{\bsl{k}_0}( \bsl{k} )\ .
}
We have 
\eq{
\eta(\bsl{k}) \rightarrow \eta(\bsl{k}) \det[R_{\bsl{k}}]
}
under the $\U(2)$ gauge transformation \eqnref{eq:U2_gauge}, and 
\eq{
\label{eq:eta_PT}
\eta^*( \bsl{k} ) = \eta( \bsl{k} ) \det[U_{\bsl{k}}]
}
based on \eqnref{eq:eta_PT_based_point_dep}.
Further, it gives us a gauge-invariant and base-point-independent $Q$ operator and $\bsl{\Phi}$ as \eqnref{eq:Q_Phi}.
The gauge invariant forms can also be understood as fixing the gauge at $\bsl{k}_0$ by making the nonzero $\N_{\bsl{k}_0}$ positive.

At the end of this part, we emphasize that although \eqnref{eq:N_e2abs} infers that $|\N_{\bsl{k}_0}|=\N$ is equal to the Wilson loop winding number for all gauges since both $\N$ and the Wilson loop winding number are gauge invariant, $\N_{\bsl{k}_0}$ has the advantage of containing the $\bsl{k}$-resolved information of the band topology, which allows us to prove the modified Nielsen-Ninomiya theorem for all gauges (or equivalently generalize  \eqnref{eq:nodes_e2_rel_gauge_dep} to all gauges), as we show in the next part.

\subsection{Demonstration of Modified Nielsen-Ninomiya theorem For All Gauges}

In this part, we prove the modified Nielsen-Ninomiya theorem proposed in \refcite{Ahn2019TBGFragile} for all gauges (or equivalently generalize  \eqnref{eq:nodes_e2_rel_gauge_dep} to all gauges) by using \eqnref{eq:Q_k0} and \eqnref{eq:N_k0}.
Similar to the last part, we again consider a generic 2D sphere-like submanifold $\M$ in the 3D 1BZ, and we focus on the case where there exists two bands that are gapped from all other bands at every point on $\M$.

Besides \eqnref{eq:Q_k0} and \eqnref{eq:N_k0}, another quantity that we need is the gauge-invariant projection of the Bloch Hamiltonian as
\eq{
\label{eq:PHP_normal}
P_{\bsl{k}}H_{\bsl{k}} P_{\bsl{k}} = \ket{u_{\bsl{k}}} \left( \epsilon_{\bsl{k}}\tau_0 + \bsl{g}_{\bsl{k}}\cdot \bsl{\tau} \right) \bra{u_{\bsl{k}}}\ ,
}
where $P_{\bsl{k}}= \ket{u_{\bsl{k}}}\bra{u_{\bsl{k}}}$ and $H_{\bsl{k}}=e^{-\ii \bsl{k}\cdot\bsl{r}} H e^{\ii \bsl{k}\cdot\bsl{r}}$.
Since $\ket{\psi_{\bsl{k}}}$ are given by a $\U(2)$ gauge transformation of eigenstates of the two bands, the two bands should read $\epsilon_{\bsl{k}}\pm |\bsl{g}_{\bsl{k}}|$, meaning that the two bands touches at $\bsl{k}$ with $|\bsl{g}_{\bsl{k}}|=0$.
As $\epsilon_{\bsl{k}} P_{\bsl{k}}$ is gauge-invariant, we know the following quantity is also gauge invariant
\eq{
P_g(\bsl{k}) \equiv P_{\bsl{k}}H_{\bsl{k}} P_{\bsl{k}}-\epsilon_{\bsl{k}} P_{\bsl{k}}= \ket{u_{\bsl{k}}}  \bsl{g}_{\bsl{k}}\cdot \bsl{\tau} \bra{u_{\bsl{k}}}\ .
}
Of particular interest is 
\eq{
\label{eq:P_g_hat}
P_{\hat{g}}(\bsl{k})\equiv \frac{P_g(\bsl{k})}{|\bsl{g}_{\bsl{k}}|}\ ,
}
which is also gauge invariant.
Since the two bands are isolated in $\M$, $P_{\bsl{k}}H_{\bsl{k}} P_{\bsl{k}}$ and $\epsilon_{\bsl{k}} P_{\bsl{k}}$ are smooth in $\M$, meaning that $P_g(\bsl{k})$ is smooth in $\M$.
Combined with the fact that $|\bsl{g}_{\bsl{k}}|$ is smooth at $\bsl{k}$ if $|\bsl{g}_{\bsl{k}}|\neq 0$, $P_{\hat{g}}(\bsl{k})$ is smooth at $\bsl{k}\in\M$ if $|\bsl{g}_{\bsl{k}}|\neq 0$.
Since $|\bsl{g}_{\bsl{k}}|=0$ infers bands touching points and the two bands in general only touch at isolated points in $\M$, we know the singular behavior of $P_{\hat{g}}(\bsl{k})$ only occurs at certain (not necessarily all) isolated band touching points in $\M$.

Combining \eqnref{eq:Q_k0} with \eqnref{eq:P_g_hat}, we can define the following vector field on $\M$
\eq{
\label{eq:v_k0}
\bsl{v}_{g,\bsl{k}_0}(\bsl{k})=\frac{1}{\sqrt{2}}\Tr\left[Q_{\bsl{k}_0}(\bsl{k}) P_{\hat{g}}(\bsl{k})\nabla_{\bsl{k}} P_{\hat{g}}^\dagger(\bsl{k}) \right]\ .
}
The gauge freedom of $\bsl{v}_{g,\bsl{k}_0}$ is determined by $Q_{\bsl{k}_0}$: the $\U(2)$ gauge transformation and the base point change can only change $\bsl{v}_{g,\bsl{k}_0}$ by a $\bsl{k}$-independent sign factor.
On the other hand, the smoothness of $\bsl{v}_{g,\bsl{k}_0}(\bsl{k})$ is determined by $P_{\hat{g}}(\bsl{k})$: for any gauge and base point choice, $\bsl{v}_{g,\bsl{k}_0}(\bsl{k})$ is smooth at $\bsl{k}$ if $|\bsl{g}_{\bsl{k}}|\neq 0$.
As a result, we can claim that for any gauge and base point choice, singularities of $\bsl{v}_{g,\bsl{k}_0}$ infer band touching points of the two bands.

As discussed above, the two bands only touch at several isolated points in $\M$, which have zero measure, without fine-tuning.
Then, we can define winding number of $\bsl{v}_{g,\bsl{k}_0}$ around each band touching point as
\eq{
w_{i,\bsl{k}_0}=\frac{1}{2\pi}\int_{\partial D_i} d\bsl{k}\cdot \bsl{v}_{g,\bsl{k}_0}(\bsl{k})\ ,
}
where $D_i\in\M$ is an infinitesimal open neighborhood of the $i$th band touching point.
In the following, we will show that
\eq{
\label{eq:winding_e2_rel}
\sum_{i} w_{i,\bsl{k}_0}=2 \N_{\bsl{k}_0}\ ,
}
and $w_{i,\bsl{k}_0}\in\dsZ$ hold for any gauge and any base point, and thus \eqnref{eq:winding_e2_rel} provides a generalization of the modified Nielson-Ninomiya theorem in \eqnref{eq:nodes_e2_rel_gauge_dep} to all gauges.

Let us first choose an oriented real PWS $\ket{\widetilde{u}^{\alpha}_{\bsl{k}}}$.
For $\bsl{k}\in\alpha-\cup_i D_i$, \eqnref{eq:eta_R_bar} indicates that
\eqa{
& P_{\hat{g}}(\bsl{k})= \ket{\widetilde{u}_{\bsl{k}}^{\alpha}}  ( \cos(\theta^{\alpha}_{\bsl{k}}) \tau_z+ \sin(\theta^{\alpha}_{\bsl{k}})\tau_x ) \bra{\widetilde{u}_{\bsl{k}}^{\alpha}} \\
& 
Q_{\bsl{k}_0}(\bsl{k})=-\frac{1}{\sqrt{2}}\ket{\widetilde{u}_{\bsl{k}}^{\alpha}}\ii \tau_y\bra{\widetilde{u}_{\bsl{k}}^{\alpha}}\ ,
}
resulting in
\eq{
\label{eq:v_k0_real_ortiented}
\bsl{v}_{g,\bsl{k}_0}(\bsl{k})=\nabla_{\bsl{k}}\theta^{\alpha}_{\bsl{k}}+ \Tr[\ii\tau_y  \bra{\widetilde{u}_{\bsl{k}}^{\alpha}}\nabla_{\bsl{k}}\ket{\widetilde{u}_{\bsl{k}}^{\alpha}}] = \nabla_{\bsl{k}}\theta^{\alpha}_{\bsl{k}} - 2\widetilde{\bsl{a}}_{\bsl{k}}^{\alpha}\ .
}
The above equations can be derived by choosing a real oriented gauge $\ket{u_{\bsl{k}}}$ such that $\ket{u_{\bsl{k}}}=\ket{\widetilde{u}^{\alpha}_{\bsl{k}}}$ for $\bsl{k}\in \alpha$.
The chosen real oriented gauge may vary for different patches $\alpha$.
Then, for $\bsl{k}\in\alpha-\cup_i D_i$
\eq{
\nabla_{\bsl{k}}\times \bsl{v}_{g,\bsl{k}_0}(\bsl{k})=- 2 \bsl{f}_{\bsl{k}}\ .
}
Thus, using Stokes' theorem, we have
\eqa{
\sum_i w_{i,\bsl{k}_0} & = - \frac{1}{2\pi}\int_{\M-\cup_i D_i} d\bsl{S}\cdot \nabla_{\bsl{k}}\times \bsl{v}_{g,\bsl{k}_0}\\
&=2 e_2=2 \N_{\bsl{k}_0}\ ,
}
where we use 
\eq{
\int_{\cup_i D_i} d\bsl{S}\cdot \bsl{f}_{\bsl{k}}=0
}
as $D_i$'s are infinitesimal and $\bsl{f}_{\bsl{k}}$ is smooth in $\M$.
Furthermore, for each $i$th band touching point, there always exists patch $\alpha_i$ such that $D_i\in\alpha_i$.
As $\partial D_i\subset \alpha_i-\cup_{i'} D_{i'}$, \eqnref{eq:v_k0_real_ortiented} suggests that 
\eqa{
& w_{i,\bsl{k}_0}=\frac{1}{2\pi}\int_{\partial D_i} d\bsl{k}\cdot [\nabla_{\bsl{k}}\theta^{\alpha_i}_{\bsl{k}} - 2\widetilde{\bsl{a}}_{\bsl{k}}^{\alpha_i}] \\
& = \widetilde{w}_{i}-2\frac{1}{2\pi}\int_{ D_i} d\bsl{S}\cdot \bsl{f}_{\bsl{k}} = \widetilde{w}_{i}\in\dsZ\ ,
}
where \eqnref{eq:w_tilde_patch} is used.
Then, we know \eqnref{eq:winding_e2_rel} returns to \eqnref{eq:nodes_e2_rel_gauge_dep} for any real oriented gauge, and thus holds for all real oriented gauges and all base-point choice. 

Moreover, the gauge transformation of $\N_{\bsl{k}_0}$ and $w_{i,\bsl{k}_0}$ are the same.
Specifically, under the $\U(2)$ gauge transformation of $\ket{u_{\bsl{k}}}$ (\eqnref{eq:UN_gauge} with $N=2$), we have 
\eqa{
& \N_{\bsl{k}_0}\rightarrow \N_{\bsl{k}_0}\chi( R_{\bsl{k}_0} ) \\
& w_{i,\bsl{k}_0}\rightarrow w_{i,\bsl{k}_0}\chi( R_{\bsl{k}_0} )\ .
}
Thus, \eqnref{eq:winding_e2_rel} holds for all gauges and all base points, serving as a generalization of \eqnref{eq:nodes_e2_rel_gauge_dep} to all gauges, proving that the modified Nielson-Ninomiya theorem proposed in \refcite{Ahn2019TBGFragile} holds for all gauges.

\section{More Details on Euler Obstructed Cooper Pairing}
\label{app:Euler_Obstructed_Pairing_General}

In this section, we present more details on the Euler obstructed Cooper pairing in general.
We start with the general setup.
Then, we give a gauge-invariant discussion for the $\PT$-protected Euler  obstructed Cooper pairing, and show that the total winding number of its pairing nodes is determined by Euler numbers on the FSs.
Finally, we choose a special type of complex gauges for the normal-state basis, called Chern gauges~\cite{Xie2020TopologyBoundSCTBG}, and show that for Chern gauges, the Euler obstructed Cooper pairing can be viewed as a $\PT$-protected double version of the monopole Cooper pairing proposed in \refcite{Li2018WSMObstructedPairing,Murakami2003BerryPhaseMSC}.

\subsection{General Setup}
\label{app:Euler_Obstructed_Pairing_General-General Setup}

Throughout this section, we choose the normal states to be 3D $\PT$-invariant (effectively) noninteracting crystals with spin $\SU(2)$ symmetry.
As discussed in \appref{app:review_NLSM_Euler} and \appref{app:spinless_DSM}, we can have sphere-like FSs with nonzero Euler numbers in such systems, when the MNLs or MSDPs are near the Fermi energy.
We focus on two sphere-like FSs with nonzero Euler numbers.
We will not suppress the spin index, and thus the nonzero Euler number on each FS is carried by the top two spin-doubly-degenerate bands.
Recall that the FSs might not be manifold but we will always use two manifolds $\M_\pm$ to approximate them.
For the convenience of later discussion, we use $\bsl{K}_\pm$ to label the ``center" of the two sphere-like FSs, and choose $\bsl{K}_\pm$ as the origins of $\M_{\pm}$.
In other words, the actual Bloch momenta on two sphere-like FSs are in $\bsl{K}_\pm+\M_{\pm}$.

On $\M_{\pm}$, both the top two occupied bands should be included for the consideration of pairing, as disucssed in the main text.
Owing to the spin $\SU(2)$ symmetry in the normal phase, the eigenstates can be expressed as a tensor product of a spinless part and a spin part.
Then, we label the Bloch basis for the top two occupied bands as $\ket{\psi_{\pm,\bsl{q},a}}\otimes\ket{s}$ for $\bsl{q}\in\M_{\pm}$, where $s=\uparrow,\downarrow$ is the spin index and we call $a=1,2$ the pseudo-spin index.
Explicitly, $\ket{\psi_{\pm,\bsl{q},a}}$ are the spinless Bloch orthonormal basis for the two top occupied bands on $\M_{\pm}$, which are linear combinations of the spinless eigenstates for the two bands.
As $\ket{\psi_{\pm,\bsl{q},a}}$ are not required to be the eigenstates for the two bands, $\ket{\psi_{\pm,\bsl{q}}}=(\ket{\psi_{\pm,\bsl{q},1}},\ket{\psi_{\pm,\bsl{q},2}})$ has the $\U(2)$ gauge freedom described by the following transformation
\eq{
\label{eq:U2_gauge_psi_pm}
\ket{\psi_{\pm,\bsl{q}}}\rightarrow \ket{\psi_{\pm,\bsl{q}}} R_{\pm}(\bsl{q})\ .
}
The periodic parts of the Bloch basis are $\ket{u_{\pm,\bsl{q}}}=e^{-\ii (\bsl{K}_\pm+\bsl{q})\cdot\bsl{r}}\ket{\psi_{\pm,\bsl{q}}}$, and the creation operators for $\ket{\psi_{\pm,\bsl{q},a}}\otimes\ket{s}$ read $c^{\dagger}_{\pm,\bsl{q},a,s}$, whose $\U(2)$ gauge transformation reads \eqnref{eq:U2_gauge_u_pm}.

The general Cooper pairing operator that pairs one electron at $\bsl{q}\in\M_{+}$ to another electron at $\bsl{p}=\bsl{p}(\bsl{q})\in\M_{-}$ is given in \eqnref{eq:H_pairing}.
As shown in \eqnref{eq:Pi_form_spin_singlet}-\eqref{eq:Pi_form_spin_triplet}, we choose $\Pi$ in \eqnref{eq:H_pairing} to be $\bsl{q}$-independent since one main origin of a $\bsl{q}$-dependent spin-channel---SOC---is negligible~\cite{Frigeri2004NCSC}.

Furthermore, we consider the case where the pairing does not spontaneously break the $\PT$ symmetry, \ie
\eq{
\label{eq:PT_invariant_pairing}
[\PT, H_{pairing}]=0\ .
}
%
%
Owing to
\eq{
\PT c^{\dagger}_{\pm,\bsl{q}} (\PT)^{-1} = c^{\dagger}_{\pm,\bsl{q}}U_{\pm}(\bsl{q})\otimes \ii s_y\ ,
}
\eqnref{eq:PT_invariant_pairing} indicates \eqnref{eq:Delta_PT}.
In the following, we will discuss how the $\PT$-invariant pairing \eqnref{eq:H_pairing} is obstructed by the odd Euler numbers on $\M_\pm$.

\subsection{Euler Obstructed Cooper Pairing}
\label{app:Euler_Obstructed_Pairing_General-Euler}

In this part, we present more details on the gauge invariant formalism of the Euler obstructed Cooper pairing.
All the key quantities defined and constructed in this part are gauge invariant, and therefore the formalism is gauge invariant.
Nevertheless, it does not mean that we will complete the whole derivation without choosing any gauges.
When proving certain gauge-invariant statements, we will choose certain convenient gauges, since the gauge-invariant nature requires the statement to be true for all gauges if it is true for one gauge.

In the main text, we have present the gauge invariant $P_{\Delta}(\bsl{q})$ in \eqnref{eq:P_Delta} constructed for the pairing matrix.
Besides being gauge invariant, $P_{\Delta}(\bsl{q})$ is also smooth everywhere in $\M_+$.
Since $P_{\Delta}(\bsl{q})$ is gauge-invariant, $P_{\Delta}(\bsl{q})$ must be smooth for all gauges as long as we can show it is smooth for one gauge.
The gauge that will choose is a complex gauge for $\ket{u_{\pm,\bsl{q}}}$ that is everywhere smooth in $\M_\pm$,  which is allowed by the $\PT$ symmetry, as discussed in \appref{app:review_NLSM_Euler}.
For this smooth gauge, if we can show $\Delta(\bsl{q})$ is also smooth, we have a smooth $P_{\Delta}(\bsl{q})$ in $\M_\pm$.

To show $\Delta(\bsl{q})$ is smooth for this gauge, we need to impose two constraints on the pairing matrix $\Delta$, which are true for most physical situations.
The first physical constraint is that the full Hamiltonian in the normal state is constructed from atomic orbitals within the tight-binding approximation.
The Bloch basis for the full normal-state Hamiltonian is $\ket{\Psi_{\bsl{k},a'}}\otimes\ket{s}$, where $\ket{\Psi_{\bsl{k},a'}}$ is the spinless part that is generated from the Fourier transformation of atomic orbitals.
Here $\bsl{k}$ takes values in the 1BZ, and $a'$ labels the degrees of freedom other than Bloch momenta and spin.
The top two occupied bands on $\M_\pm$ are just two bands of the full normal-state Hamiltonian on $\M_\pm$, and $\ket{\psi_{\pm,\bsl{q},a}}$ are given by the linear combinations of $\ket{\Psi_{\bsl{K}_{\pm}+\bsl{q},a'}}$.
Explicitly, with $\ket{\Psi_{\bsl{k}}} =( ..., \ket{\Psi_{\bsl{k},a'}}, ... )$, the spinless basis for the top two occupied bands on $\M_\pm$ can be expressed as $\ket{\psi_{\pm,\bsl{q},a}}= \ket{\Psi_{\bsl{K}_{\pm}+\bsl{q}}} \xi_{a}(\bsl{K}_{\pm}+\bsl{q})$, where $\xi_{a}(\bsl{K}_{\pm}+\bsl{q})$'s are othonormal vectors.
If we label the creation operator of $\ket{\Psi_{\bsl{K}_{\pm}+\bsl{q},a'}}\otimes\ket{s}$ as  $\Psi^{\dagger}_{\bsl{K}_{\pm}+\bsl{q},a',s}$ and define $\xi(\bsl{K}_{\pm}+\bsl{q})=(\xi_{1}(\bsl{K}_{\pm}+\bsl{q}),\xi_{2}(\bsl{K}_{\pm}+\bsl{q}))$, then we have $c^{\dagger}_{\pm,\bsl{q},s} = \Psi^{\dagger}_{\bsl{K}_{\pm}+\bsl{q},s}\xi(\bsl{K}_{\pm}+\bsl{q})$.

The second constraint is that the pairing operator should be expressible in terms of $\ket{\Psi_{\bsl{K}_{\pm}+\bsl{q},a'}}\otimes\ket{s}$ as 
\eq{
\label{eq:pairing_operator_full}
\widetilde{H}_{pairing}=\sum_{\bsl{q}\in \M_{+}} \Psi^{\dagger}_{\bsl{K}_+ + \bsl{q}} \widetilde{\Delta}(\bsl{q})\otimes\Pi (\Psi^{\dagger}_{\bsl{K}_-+\bsl{p}})^T + h.c.\ .
}
In the case, the pairing matrix $\Delta(\bsl{q})$ in \eqnref{eq:H_pairing} can be re-expressed as projection of the $\widetilde{\Delta}(\bsl{q})$
\eq{
\label{eq:pairing_mat_project}
\Delta(\bsl{q})= \xi^\dagger(\bsl{K}_{+}+\bsl{q})\widetilde{\Delta}(\bsl{q}) \xi^*(\bsl{K}_{-}+\bsl{p})\ .
}
With the two constraints, we know $\Delta(\bsl{q})$ is smooth in $\M_+$ for the chosen smooth gauge, since the pairing matrix $\widetilde{\Delta}(\bsl{k})$ is a smooth function of $\bsl{k}$, and $\xi^\dagger(\bsl{K}_{\pm}+\bsl{q})$ is smooth in $\M_\pm$ owing to the chosen smooth gauge.
As a result, we know $P_{\Delta}(\bsl{q})$ is smooth in $\M_+$ for the chosen smooth gauge.
Since $P_{\Delta}(\bsl{q})$ is gauge invariant, $P_{\Delta}(\bsl{q})$ should be smooth in $\M_+$ for any gauge.

Next we split $\Delta(\bsl{q})$ into channels.
We want each channel to have the same gauge transformation rule \eqnref{eq:Delta_U2U2} and $\PT$-invariant constraint \eqnref{eq:Delta_PT} as $\Delta$. 
To do so, we need the path-independent $\eta$ factor and the gauge-invariant $\bsl{Q}$ operator defined in \appref{app:reformulate_Euler}.
Since the top two occupied bands are isolated on $\M_{\pm}$, we can define the path-independent $\eta_{\pm,\bsl{q}_0^\pm}(\bsl{q})$  factor by substituting $\ket{u_{\pm,\bsl{q}}}$ into \eqnref{eq:eta_path_indep} with the base point $\bsl{q}_0^\pm\in\M_{\pm}$.
Then, we can calculate $\N_{\bsl{q}_0^\pm}^\pm$ for the top two occupied bands on $\M_{\pm}$ according to \eqnref{eq:N_k0}.
$\N_{\pm,\bsl{q}_0^\pm}$ must be nonzero, since the Euler numbers $\N_\pm= |\N_{\pm,\bsl{q}_0^\pm}|$ are nonzero.
Then, according to \eqnref{eq:eta_path_indep_no_base_point}, we can define base-point-independent
\eq{
\label{eq:eta_pm}
\eta_{\pm}(\bsl{q})=\frac{\N_{\pm,\bsl{q}_0^\pm}}{\N_\pm}\eta_{\pm,\bsl{q}_0^\pm}(\bsl{q})\ .
}
Combined with \eqnref{eq:Q_Phi}, we get the gauge-invariant and base-point-independent $Q_{\pm}(\bsl{q})$ and $\bsl{\Phi}_\pm(\bsl{q})$ as
\eqa{
\label{eq:Q_pm}
& Q_{\pm}(\bsl{q})=-\frac{\eta^*_\pm(\bsl{q})}{\sqrt{2}}\ket{u_{\pm,\bsl{q}}}\ii \tau_y\bra{u_{\pm,\bsl{q}}^{\PT}} \\
& \bsl{\Phi}_{\pm}(\bsl{q})=\frac{1}{\sqrt{2}}\Tr[Q_{\pm}(\bsl{q})\nabla_{\bsl{q}}P_\pm(\bsl{q})\times\nabla_{\bsl{q}}P_\pm(\bsl{q})]
}
with $P_\pm(\bsl{q})=\ket{u_{\pm,\bsl{q}}}\bra{u_{\pm,\bsl{q}}}$.
As discussed in \appref{app:reformulate_Euler}, $Q_{\pm}(\bsl{q})$ is smooth in $\M_\pm$.

To proceed, let us discuss the $\zeta$ factor that characterizes the compatibility between the normal directions and the one-to-one correspondence $\bsl{p}(\bsl{q})$.
Consider a generic parametrization of $\M_+$, $\bsl{q}=\bsl{q}(x,y)$.
We require $\bsl{q}(x,y)$ to respect the chosen normal direction $\hat{\bsl{n}}_{+}$ of $\M_+$, \ie, the normal direction can be determined by
\eq{
\hat{\bsl{n}}_{+,\bsl{q}}=\frac{\partial_x \bsl{q} \times \partial_y \bsl{q}}{|\partial_x \bsl{q} \times \partial_y \bsl{q}|}\ . 
}
Note that $\hat{\bsl{n}}_{+}$ is independent of the parametrization, and the above equation just means that the parametrization is compatible with the chosen $\hat{\bsl{n}}_{+}$.
The parametrization of $\M_+$ always provides a parametrization of $\M_-$ as $\bsl{p}=\bsl{p}(x,y)=\bsl{p}(\bsl{q}(x,y))$.
The parametrization of $\M_-$ can give a normal direction of $\M_-$, labeled as $\widetilde{\bsl{n}}_-$, as 
\eq{
 \widetilde{\bsl{n}}_{-,\bsl{p}}=\frac{\partial_x \bsl{p} \times \partial_y \bsl{p}}{|\partial_x \bsl{p} \times \partial_y \bsl{p}|} = \frac{S^{-1}_{\bsl{q}}\hat{\bsl{n}}_{+,\bsl{q}}}{|S^{-1}_{\bsl{q}}\hat{\bsl{n}}_{+,\bsl{q}}|}\sgn{\det[S_{\bsl{q}}]}\ ,
}
where $[S_{\bsl{q}}]_{ij}=\partial p_j/\partial q_i$.
Clearly, $\widetilde{\bsl{n}}_{-,\bsl{p}}$ is the unit vector given by mapping $\hat{\bsl{n}}_{+}$ to $\M_-$ through $\bsl{p}(\bsl{q})$, which is also independent of the parametrization of $\M_+$.
$\widetilde{\bsl{n}}_{-,\bsl{p}}$ is either the same or opposite to the chosen normal vector $\hat{\bsl{n}}_{-,\bsl{q}}$ of $\M_-$, meaning that 
\eq{
\label{eq:zeta_definition}
\widetilde{\bsl{n}}_{-,\bsl{p}}=\zeta \hat{\bsl{n}}_{-,\bsl{p}}
}
with $\zeta=\pm$.
Since both $\widetilde{\bsl{n}}_{-,\bsl{p}}$ and $\hat{\bsl{n}}_{-,\bsl{p}}$ are smooth, $\zeta$ should be independent of $\bsl{q}$.

With the $\zeta$ factor and based on \eqnref{eq:Q_pm}, we split $P_{\Delta}(\bsl{q})$ into two channels as \eqnref{eq:P_Delta_b_two_channels} with
\eqa{
\label{eq:Delta_b_two_channels}
\Delta_b(\bsl{q}) & = \frac{1}{2}\left[\Delta(\bsl{q})+\zeta (-1)^b \tau_y \Delta^*(\bsl{q})\tau_y \eta^*_{+,\bsl{q}}\eta^*_{-,\bsl{p}}\right]\\
& =\bsl{\delta}_b(\bsl{q}) \cdot\bsl{\Lambda}_b
\mat{1 & \\ & \zeta\eta_{+,\bsl{q}}^*\eta_{-,\bsl{p}}^*}\ ,
}
$\bsl{\delta}_b\in\dsR^4$, and
\eqa{
& \bsl{\Lambda}_\perp=(\tau_0,\ii \bsl{\tau}) \\
& \bsl{\Lambda}_\shpa=(\ii \tau_0,\bsl{\tau}) \ .
}
The $P_{\Delta_b}$ of each channel $b$ is gauge-invariant and smooth in $\M_+$.
The gauge invariant \eqnref{eq:P_Delta_b_two_channels} also suggests that $\Delta_b(\bsl{q})$ has the same $\U(2)\times\U(2)$ gauge transformation rule as
\eq{
\label{eq:Delta_b_U2U2}
\Delta_b(\bsl{q})\rightarrow R_+^\dagger(\bsl{q})\Delta_b(\bsl{q}) R_-^*(\bsl{p})\ .
}
As a result, we can define the gauge-invariant pairing operator for $\Delta_b$ as \eqnref{eq:H_b_pairing}.
Combining \eqnref{eq:Delta_b_two_channels} with \eqnref{eq:Delta_PT} and \eqnref{eq:eta_PT}, we further have
\eq{
\label{eq:Delta_b_PT}
U_{+}(\bsl{q})\Delta^*_b(\bsl{q}) U_{-}^T(\bsl{p}) = \Delta_b(\bsl{q})\ ,
}
meaning that 
\eq{
\label{eq:H_b_PT}
[\PT, H_{b,pairing}]=0\ .
}
Therefore, we have split the pairing matrix into two channels that have the same gauge transformation rule and $\PT$-invariant, and we can define the gauge-invariant $\PT$-invariant pairing operator for the two channels.
The channel splitting, especially the $P_{b}$, allows us to study the effect of normal-state band topology on the Cooper pairing order parameters, as discussed in the main text.

Now we explicitly show that the winding number $\W_{b,i}$ defined in \eqnref{eq:W_b_i} is integer valued ($\W_{b,i}\in \dsZ$).
%
%
Let us choose an oriented real gauge of $\ket{u_{+,\bsl{q}}}$ with the corresponding PWS $\ket{\widetilde{u}_{+,\bsl{q}}^\alpha}$ ($\alpha=I,II,...$ labels the patches).
We can map each patch $\alpha$ on $\M_+$ to the patch $\beta=\bsl{p}(\alpha)$ on $\M_-$, and then we have an open cover of $\M_-$ with patches $\beta=\bsl{p}(I),\bsl{p}(II),...$.
Then, we can choose an oriented real gauge for $\ket{u_{-,\bsl{p}}}$ with the corresponding PWS $\ket{\widetilde{u}_{-,\bsl{p}}^\beta}$.
In particular, we choose the real oriented gauges such that (i) there exist a patch $\alpha_0$ that contains $D_{b,i}$ and (ii) $\ket{u_{+,\bsl{q}}}=\ket{\widetilde{u}_{+,\bsl{q}}^{\alpha_0}}$ and $\ket{u_{-,\bsl{p}}}=\ket{\widetilde{u}_{-,\bsl{p}}^{\bsl{p}(\alpha_0)}}$ for all $\bsl{q}\in \alpha_0$.

With this gauge choice, we have $\eta_\pm(\bsl{q})=\sgn{e_{2,\pm}}$, where $e_{2,\pm}$ are Euler class for the top two occupied bands on $\M_\pm$ according to \eqnref{eq:e_2_original}, and have
\eqa{
\Delta_b^{\alpha_0}(\bsl{q}) = |\Delta_b|  e^{\ii (-1)^b \theta_b^{\alpha_0} \tau_y} \mat{1 & \\ & \zeta (-1)^b \sgn{e_{2,+} e_{2,-}}}
}
for all $\bsl{q}\in\alpha_0$, where $\theta_b^{\alpha_0}$ is real.
Plugging the expression in $\hat{P}_b$ gives
\eq{
\hat{P}_b(\bsl{q})= \ket{\widetilde{u}_{+,\bsl{q}}^{\alpha_0}} \Delta_b^{\alpha_0}(\bsl{q})\bra{\widetilde{u}_{-,\bsl{p}}^{\beta_0}}\ ,
}
where $\beta_0=\bsl{p}(\alpha_0)$.
Combined with 
\eq{
Q_{+}(\bsl{q})=\sgn{e_{2,+}}\frac{-\ii}{\sqrt{2}}\ket{\widetilde{u}_{+,\bsl{q}}^{\alpha_0}}\tau_y\bra{\widetilde{u}_{+,\bsl{q}}^{\alpha_0}}\ ,
}
we can obtain the expression of the $\bsl{v}_b$ vector as 
\eqa{
\label{eq:v_b_explicit_real_oriented}
\bsl{v}_b & =\frac{1}{\sqrt{2}}\Tr[Q_{+,\bsl{q}} \hat{P}_b(\bsl{q}) \nabla_{\bsl{q}} \hat{P}_b^\dagger(\bsl{q}) ]\\
& = -\sgn{e_{2,+}}(-1)^b \nabla_{\bsl{q}} \theta_b^{\alpha_0}(\bsl{q}) - \sgn{e_{2,+}} \bsl{a}_+^{\alpha_0}(\bsl{q})\\
&\quad +(-1)^b \zeta \sgn{e_{2,-}}\pf[\bra{\widetilde{u}_{-,\bsl{p}}^{\beta_0}}\nabla_{\bsl{q}}\ket{\widetilde{u}_{-,\bsl{p}}^{\beta_0}}]
}
for $\bsl{q}\in\alpha_0$ (but away from the pairing nodes).
Eventually combined with $\bsl{\Phi}_{\pm,\bsl{q}}=\sgn{e_{2,\pm}}\bsl{f}_{\pm,\bsl{q}}$ and the discussion above \eqnref{eq:zeta_definition} on the $\zeta$ factor, 
we arrive at
\eqa{
 \W_{b,i}& = -\sgn{e_{2,+}} (-1)^b \frac{1}{2\pi} \int_{\partial D_{b,i}} d\bsl{q}\cdot  \nabla_{\bsl{q}} \theta_b^{\alpha_0}(\bsl{q})\in \dsZ \ .
}
Thus, we have demonstrate $\W_{b,i}\in \dsZ$ for the chosen real oriented gauge.
Since $\W_{b,i}\in \dsZ$ is gauge invariant, it should hold for any gauge.

At the end of this part, we compare \eqnref{eq:W_b_E_b_N} to \eqnref{eq:winding_e2_rel}.
Although both \eqnref{eq:W_b_E_b_N} for the Cooper pairing order parameter and the modified Nielson-Ninomiya theorem in \eqnref{eq:winding_e2_rel} for the normal-state bands have the form of Stokes' theorem, the winding number and topological index in the two expressions have different physical meaning.
The winding number and topological index in \eqnref{eq:W_b_E_b_N} are defined for describing the Cooper pairing order parameters, while the winding number and topological index in \eqnref{eq:winding_e2_rel} are defined solely for normal-state band structure.
The key physical difference is that the Cooper pairing order parameter spontaneously breaks charge-$\U(1)$ symmetry while the normal state preserves it, resulting in different gauge groups for the two cases---$\U(2)\times\U(2)$ for the Cooper pairing order parameter but $\U(2)$ for the normal-state bands.
Such difference further leads to a different definitions of $\hat{P}_b$ for the Cooper pairing order parameter (\eqnref{eq:P_Delta_b_two_channels}) and $P_{\hat{g}}$ for the normal-state bands (\eqnref{eq:P_g_hat}).
As a result, the winding number in \eqnref{eq:W_b_E_b_N} is defined for the pairing nodes of the pairing matrix (\ie, the zeros of the pairing matrix), while the winding number in \eqnref{eq:winding_e2_rel} is defined for the band touching point between two normal-state bands.
Furthermore, the Euler index of inter-FS Cooper pairing involves the Euler numbers of two sets of normal-state bands, while the topological index in \eqnref{eq:winding_e2_rel} is always determined by the Euler number of the only set of normal-state bands in consideration.

\subsection{Obstruction to Smooth Representations For Real Oriented Gauges}

In this part, we discuss the obstruction to smooth representations for Euler obstructed pairing channels.
The discussion in this part mirrors the one in \refcite{Li2018WSMObstructedPairing}.

We focus on the real oriented PWS $\ket{\widetilde{u}_{+,\bsl{q}}^\alpha}$ and $\ket{\widetilde{u}_{-,\bsl{p}}^{\bsl{p}(\alpha)}}$ with Euler classes respectively being $e_{2,+}$ and $e_{2,-}$.
Specifically, let us parametrize $\bsl{q}\in \M_+$ by its direction $\bsl{q}/|\bsl{q}|=(\sin(\theta)\cos(\phi),\sin(\theta)\sin(\phi),\cos(\theta))$, \ie, $\bsl{q}=\bsl{q}(\theta,\phi)$.
Such a parametrization is allowed since $\M_{+}$ is sphere-like.
Then, we can use the same $(\theta,\phi)$ to parameterize $\bsl{p}\in\M_-$ via $\bsl{p}=\bsl{p}(\bsl{q}(\theta,\phi))$.
Based on the parametrization, we can simultaneously choose two patches for $\M_{\pm}$ as the north hemisphere $N: \theta<\pi/2+0^+$ and south hemisphere $S: \theta>\pi/2+0^-$.
Then, the oriented real PWS gauges are $\ket{\widetilde{u}^{\alpha}_{\pm,\theta,\phi}}$ with $\alpha=N/S$.

With this gauge, the $\bsl{\delta}$ vector in the pairing matrix (\eqnref{eq:Delta_b_two_channels}) has the form
\eqa{
\label{eq:d_vec_real_oriented_gauge}
& \bsl{\delta}_\perp^\alpha=(d_0^\alpha, 0, d_y^\alpha, 0) =|\Delta_\perp|(\cos(\theta_\perp^\alpha),0,\sin(\theta_\perp^\alpha),0)\\
& \bsl{\delta}_\shpa^\alpha=(0, d_x^\alpha, 0, d_z^\alpha) = |\Delta_\shpa|(0,\sin(\theta_\shpa^\alpha),0,\cos(\theta_\shpa^\alpha))\ .
}
In particular, we can always choose the transition function for the normal-state bases along the equator as
\eqa{
\label{eq:N_S_real_transition_special}
& \ket{\widetilde{u}^{S}_{+,\frac{\pi}{2},\phi}}=\ket{\widetilde{u}^{N}_{+,\frac{\pi}{2},\phi}}e^{-\ii \phi \tau_y e_{2,+} }\\
& \ket{\widetilde{u}^{S}_{-,\frac{\pi}{2},\phi}}=\ket{\widetilde{u}^{N}_{-,\frac{\pi}{2},\phi}}e^{-\ii \tau_y  \phi \zeta e_{2,-}}\ .
}
Then, the transition functions for the pairing matrices are
\eqa{
\label{eq:complexified_Euler_obstructed_pairing}
& [d^N_0+ \ii d^N_y](\frac{\pi}{2},\phi) = [d^S_0+ \ii d^S_y](\frac{\pi}{2},\phi) e^{-\ii \phi 2\sgn{e_{2,+}} \E_\perp} \\ 
& [d^N_z+ \ii d^N_x](\frac{\pi}{2},\phi) = [d^S_z+ \ii d^S_x](\frac{\pi}{2},\phi) e^{\ii  \phi 2\sgn{e_{2,+}} \E_\shpa} \ . 
}
Since $\bsl{\delta}_b^\alpha$ is smooth in each path $\alpha$, the above form of the transition function means that if $\E_b$ is nonzero, $\bsl{\delta}_b$ is not smooth in $\M_+$ for real oriented gauges that correspond to the two-patch PWS $\ket{\widetilde{u}^{\alpha}_{\pm,\theta,\phi}}$.
Then, the two-patch expressions of $(d_0+ \ii d_y)$ and $(d_z+ \ii d_x)$ should be expanded in series of monopole Harmonics with monopole charges $-\sgn{e_{2,+}} \E_\perp$ and $\sgn{e_{2,+}} \E_\shpa$, respectively, according to the convention used in \refcite{Wu1976MonopoleHarmonics}.
Thus, the nonzero Euler index provides obstruction to smooth representation of the pairing channel in the chosen real oriented gauges, and when that happens, the pairing matrix should be expanded in terms of the monopole Harmonics with monopole charge determined by the Euler index.

\subsection{Double Monopole Cooper Pairing}
\label{app:Euler_Obstructed_Pairing_General-Double_Monopole}

In the last part, we presented the obstruction to smooth representations for Euler obstructed pairing channels with a special real oriented gauge, and pointed out that we should use monopole Harmonics to expand the pairing matrix $\Delta_b$ in that gauge when its Euler index is nonzero.
\refcite{Li2018WSMObstructedPairing,Murakami2003BerryPhaseMSC} has shown that the monopole Harmonics should be used to expand the monopole Cooper pairing, and thus the discussion in the last part indicates there is a relation between between the Euler obstructed Cooper pairing and the monopole Cooper pairing.
In this part, we demonstrate this relation: for a special gauge called Chern gauge, the Euler obstructed Cooper pairing can be viewed as a $\PT$-protected double version of the monopole Cooper pairing proposed in \refcite{Li2018WSMObstructedPairing}.

To show this, let us first review the Chern gauge in 3D $\PT$-invariant spinless noninteracting crystals in the normal phase following \refcite{Xie2020TopologyBoundSCTBG,Bouhon2020WeylNonabelian}.
Consider a generic 2D closed sphere-like manifold $\M$ in the 3D 1BZ, and consider an isolated set of two bands on $\M$ with  $\ket{u_{\bsl{k}}}=(\ket{u_{1,\bsl{k}}},\ket{u_{2,\bsl{k}}})$ the periodic part of the Bloch basis.
Let us choose a real oriented gauge of $\ket{u_{\bsl{k}}}$ on $\M$ with corresponding PWS $\ket{\widetilde{u}^\alpha_{\bsl{k}}}$ where $\alpha=I,II,...$ labels the patches for $\M$, and we label the Euler class of the two bands for this gauge as $e_2$.
Then, based on $\ket{\widetilde{u}^\alpha_{\bsl{k}}}$, we can define a complex PWS $\ket{u^{Ch,\alpha}_{\bsl{k}}}$ with
\eq{
\label{eq:Chern_gauge}
\ket{u^{Ch,\alpha}_{\bsl{k}}}=\frac{1}{\sqrt{2}}\ket{\widetilde{u}^\alpha_{\bsl{k}}}\mat{1 & 1 \\ \ii & -\ii}\ .
}
For any two patches $\alpha$ and $\alpha'$ that overlap, the transition function for the oriented real PWS gauge at $\bsl{k}\in\alpha\cap\alpha'$ is $\ket{\widetilde{u}^{\alpha}_{\bsl{k}}}=\ket{\widetilde{u}^{\alpha'}_{\bsl{k}}}e^{\ii \tau_y \Gamma_{\bsl{k}}^{\alpha,\alpha'}}$, resulting in
\eq{
\ket{u^{Ch,\alpha}_{\bsl{k}}}=\ket{u^{Ch,\alpha'}_{\bsl{k}}}e^{\ii \tau_z \Gamma_{\bsl{k}}^{\alpha,\alpha'}}\ .
}
The above expression suggests that the transition function for each component of $\ket{u^{Ch,\alpha}_{\bsl{k}}}$ is a $\U(1)$ factor, meaning that each component $\ket{u^{Ch,\alpha}_{a,\bsl{k}}}$ is a PWS section of a rank-1 complex vector bundle.
Then, $\ket{u^{Ch,\alpha}_{a,\bsl{k}}}$ (with $a=1,2$) must have a well-defined patch-independent Berry curvature $\bsl{F}_a$ as
\eqa{
& \bsl{F}_a(\bsl{k})=\nabla_{\bsl{k}}\times (-\ii \bra{u^{Ch,\alpha}_{a,\bsl{k}}}\nabla_{\bsl{k}} \ket{u^{Ch,\alpha}_{a,\bsl{k}}})\\
& = (-1)^{a-1}\bsl{f}_{\bsl{k}}\ ,
}
where $\bsl{f}_{\bsl{k}}$ is the real curvature.
Eventually, we know $\ket{u^{Ch,\alpha}_{a,\bsl{k}}}$ must have a well-defined Chern number $Ch_a$
\eq{
Ch_a=\frac{1}{2\pi}\int_{\M}d\bsl{S}\cdot \bsl{F}_a(\bsl{k}) =(-1)^{a-1} e_2\ .
}
Here $\bsl{F}_1(\bsl{k})=-\bsl{F}_2(\bsl{k})$ and $Ch_1=-Ch_2$ are protected by the $\PT$ symmetry, owing to
\eq{
\label{eq:Chern_gauge_PT}
\PT\ket{u^{Ch,\alpha}_{\bsl{k}}}=\ket{u^{Ch,\alpha}_{\bsl{k}}}\tau_x\ .
}
In sum, by complexifying a generic oriented real PWS gauge with \eqnref{eq:Chern_gauge}, we get a special complex gauge, for which two components of the basis are $\PT$-related two sectors with opposite Chern numbers.
Therefore, we define Chern gauges as $\ket{u^{Ch}_{\bsl{k}}}=\ket{u^{Ch,\alpha_{\bsl{k}}}_{\bsl{k}}}$. 
%

Next, we show that when choosing Chern gauges, the Euler obstructed Cooper pairing can be viewed as a double version of the monopole Cooper pairing proposed in \refcite{Li2018WSMObstructedPairing}.
As shown in the last part, we can choose an oriented real gauge for the basis on $\M_{\pm}$ as $\ket{u^{RO}_{+,\bsl{q}}}$ and $\ket{u^{RO}_{-,\bsl{p}}}$ with ``$RO$" indicating the real oriented gauge.
Then, combined with \eqnref{eq:Chern_gauge}, we can transform the oriented real gauge to a Chern gauge for the basis on $\M_{\pm}$ as
\eqa{
\label{eq:Chern_gauge_pm}
& \ket{u^{Ch}_{+,\bsl{q}}} = \frac{1}{\sqrt{2}}\ket{u^{RO}_{+,\bsl{q}}}\mat{1 & 1 \\ \ii & -\ii}\\
& \ket{u^{Ch}_{-,\bsl{p}}}= \frac{1}{\sqrt{2}}\ket{u^{RO}_{-,\bsl{p}}}\mat{1 & 1 \\ \ii & -\ii}\ .
}
With the Chern gauge \eqnref{eq:Chern_gauge_pm}, the $\eta$ factor becomes
\eq{
\eta_\pm(\bsl{q})=-\ii\ \sgn{e_{2,\pm}}\ ,
}
and then $\Delta_b$ reads
\eqa{
\Delta_\perp^{Ch}(\bsl{q}) & = |\Delta_\perp(\bsl{q})| \mat{e^{\ii \theta_{\perp}(\bsl{q})} & \\ & e^{-\ii \theta_{\perp}(\bsl{q})}}\delta_{\zeta\sgn{e_{2,+}e_{2,-}},-1}\\
& \quad + |\Delta_\perp(\bsl{q})| \mat{ & e^{\ii \theta_{\perp}(\bsl{q})}\\ e^{-\ii \theta_{\perp}(\bsl{q})} & }\delta_{\zeta\sgn{e_{2,+}e_{2,-}},1}\\
\Delta_\shpa^{Ch}(\bsl{q}) &  = |\Delta_\shpa(\bsl{q})| \mat{ & e^{-\ii \theta_{\shpa}(\bsl{q})}\\ e^{\ii \theta_{\shpa}(\bsl{q})} & }\delta_{\zeta\sgn{e_{2,+}e_{2,-}},-1}\\
& \quad  + |\Delta_\shpa(\bsl{q})| \mat{e^{-\ii \theta_{\shpa}(\bsl{q})} & \\ & e^{\ii \theta_{\shpa}(\bsl{q})}}\delta_{\zeta\sgn{e_{2,+}e_{2,-}},1}\ .
}

From the above expression, we can see that regardless of the value of $\zeta\sgn{e_{2,+}e_{2,-}}$, $\Delta_{b}^{Ch}$ only has two nonzero elements, and each element features the pairing between two specific Chern sectors.
As the monopole Cooper pairing appears as the pairing between Chern sectors \cite{Li2018WSMObstructedPairing}, both elements of $\Delta_{b}^{Ch}$ might be monopole Cooper pairings.
Moreover, the two elements of $\Delta_{b}^{Ch}$ are related by the $\PT$ symmetry according to \eqnref{eq:Chern_gauge_PT}, and thus $\Delta_{b}^{Ch}$ might be a $\PT$-protected double version of the monopole Cooper pairing.

To demonstrate it, consider $\bsl{p}=-\bsl{q}$ (\ie, $\zeta=-1$) and $e_{2,+}=-e_{2,-}=1$, and we have \eq{
\Delta_\shpa^{Ch}(\bsl{q})= \mat{e^{-\ii \theta_{\shpa}(\bsl{q})} & \\ & e^{\ii \theta_{\shpa}(\bsl{q})}}\ .
}
$|\Delta_\shpa(\bsl{q})| e^{-\ii \theta_{\shpa}(\bsl{q})}$ is the complex pairing between the Chern sector $\ket{u^{Ch}_{+,1,\bsl{q}}}$ and the Chern sector $\ket{u^{Ch}_{-,1,\bsl{p}}}$, and owing to the $\PT$-symmetry, $|\Delta_\shpa(\bsl{q})| e^{\ii \theta_{\shpa}(\bsl{q})}$ is the complex pairing between the Chern sector $\ket{u^{Ch}_{+,2,\bsl{q}}}$ and the Chern sector $\ket{u^{Ch}_{-,2,\bsl{p}}}$.
Then, both $|\Delta_\shpa(\bsl{q})| e^{\pm \ii \theta_{\shpa}(\bsl{q})}$ are monopole Cooper pairings with nonzero monopole charges $q_\pm=\mp [e_{2,+}-e_{2,-}]/2$, according to \refcite{Li2018WSMObstructedPairing}.
The monopole charges are related to the Euler index of $\Delta_\shpa$ as $q_\pm=\mp \sgn{e_{2,+}}\E_\shpa$, and the opposite monopole charges of $|\Delta_\shpa(\bsl{q})| e^{\pm \ii \theta_{\shpa}(\bsl{q})}$ are required by the $\PT$ symmetries.
Thus, $\Delta_\shpa^{Ch}(\bsl{q})$ is a $\PT$-protected double version of the monopole Cooper pairing with nonzero monopole charges, and thus must have pairing nodes, coinciding with the fact that $\Delta_\shpa(\bsl{q})$ is Euler obstructed.

We emphasize that since the Chern sectors reply on the Chern gauges, the relation between the Euler obstructed Cooper pairing and the monopole Cooper pairing only holds for the Chern gauges. 
Yet, the general equation \eqnref{eq:W_b_E_b_N} for the Euler obstructed Cooper pairing is gauge-independent.

\section{More Details on Euler Obstructed Cooper Pairing for TR-Invariant Centrosymmetric Normal States}
\label{app:PandT}

In \appref{app:Euler_Obstructed_Pairing_General}, we have discussed the Euler obstructed Cooper pairing in a general way.
In this section, we will discuss the Euler obstructed Cooper pairing given normal states with TR, inversion and spin $\SU(2)$ symmetries, which usually occur in nonmagnetic centrosymmetric systems with negligible SOC.
Throughout this section, we focus on the $\PT$-invariant Cooper pairing that (i) has zero total momentum and (ii) occurs between electrons on (or near) inversion-related sphere-like FSs with nonzero Euler numbers.

\subsection{More Details on Zero-Total-Momentum Euler Obstructed Cooper Pairing}
\label{app:PandT_EOP}

We start from the Euler obstructed Cooper pairing in this case.
As we focus on the zero-total-momentum Cooper pairing, we should consider two sphere-like FSs $\M_\pm$ that are related by inversion symmetry, meaning that $\M_-=-\M_+$, $\bsl{K}_-=-\bsl{K}_+$, and $\bsl{p}=-\bsl{q}$ for \eqnref{eq:H_pairing} and \figref{fig:pairing_FS}.
As a result, the pairing operator for the Cooper pairing between the two TR-related pairing-relevant manifolds becomes \eqnref{eq:H_pairing_TR}.
We require $H_{pairing}$ to preserve $\PT$ symmetry as \eqnref{eq:PT_invariant_pairing}, and require $\Pi$ in $H_{pairing}$ to be momentum-independent as in \eqnref{eq:Pi_form_spin_singlet} and \eqnref{eq:Pi_form_spin_triplet}.
As $\bsl{p}=-\bsl{q}$ from $\M_+$ to $\M_-$ is incompatible with the orientation of $\M_\pm$, we have $\zeta=-1$ in \eqnref{eq:zeta_definition}.

As discussed in \appref{app:Euler_Obstructed_Pairing_General-Euler}, the pseudo-spin part $\Delta(\bsl{q})$ can be split into two channels whose Euler indices are determined by the sum or difference of the Euler numbers on $\M_\pm$.
Owing to TR symmetry, we have
\eq{
\TR  c^{\dagger}_{+,\bsl{q}} \TR^{-1} =  c^{\dagger}_{-,-\bsl{q}}V_{\TR}(\bsl{q})\otimes \ii s_y\ ,
}
resulting in
\eq{
\TR \ket{u_{+,\bsl{q}}} =\ket{u_{-,-\bsl{q}}}  V_{\TR}(\bsl{q})\ .
}
Here $V_{\TR}(\bsl{q})$ is a $\U(2)$ matrix, and $\ket{u_{\pm,\bsl{q}}}=(\ket{u_{\pm,\bsl{q},1}},\ket{u_{\pm,\bsl{q},2}})$ are the spinless periodic parts of the Bloch basis created by $c^{\dagger}_{\pm,\bsl{q}}$.
Owing to the TR symmetry, the base-point dependent $\eta_{\pm,\pm\bsl{q}_0}(\bsl{q})$ satisfies
\eq{
 \eta_{+,\bsl{q}_0}(\bsl{q}) = \chi_{\TR}(\bsl{q}_0) \det[ V_{\TR}(\bsl{q})]^* \eta_{-,-\bsl{q}_0}^*(-\bsl{q})\ ,
}
where we use \eqnref{eq:eta_path_indep} and
\eq{
\chi_{\TR}(\bsl{q}_0)=\frac{\sqrt{\det[\braket{u^{\PT}_{-,-\bsl{q}_0}}{u_{-,-\bsl{q}_0}}]^*(\det[V_{\TR}(\bsl{q}_0)]^*)^2} }{\left(\sqrt{\det[\braket{u^{\PT}_{-,-\bsl{q}_0}}{u_{-,-\bsl{q}_0}}]}\right)^* \det[V_{\TR}(\bsl{q}_0)]^* } =\pm 1\ .
}
Then, the $Q$ operators (\eqnref{eq:Q_k0}) are related as 
\eqa{
\TR Q_{+,\bsl{q}_0}(\bsl{q}) \TR^{\dagger}= \chi_{\TR}(\bsl{q}_0) Q_{-,-\bsl{q}_0}(-\bsl{q})\ .
}
Combined with $P_{\pm}(\bsl{q})=\ket{u_{\pm,\bsl{q}}}\bra{u_{\pm,\bsl{q}}}$ and $\TR P_{+}(\bsl{q}) \TR^{\dagger} = P_{-}(-\bsl{q})$, we have
\eq{
\bsl{\Phi}_{+,\bsl{q}_0}(\bsl{q})=\chi_{\TR}(\bsl{q}_0) \bsl{\Phi}_{-,-\bsl{q}_0}(-\bsl{q})\ ,
}
resulting in
\eqa{
\N_{+,\bsl{q}_0} = - \chi_{\TR}(\bsl{q}_0) \N_{-,-\bsl{q}_0}\ ,
}
where we have used \eqnref{eq:Phi}, \eqnref{eq:N_k0}, and the fact that 
the map $\bsl{p}=-\bsl{q}$ from $\M_+$ to $\M_-$ is incompatible with the orientation.
Then, the Euler numbers of the top-two occupied bands on $\M_\pm$ are the same
\eq{
\label{eq:N_TR}
\N_{+}= |\N_{+,\bsl{q}_0}| = |\N_{-,-\bsl{q}_0}| = \N_{-}\ .
}
Owing to \eqnref{eq:N_TR}, the Euler indices (\eqnref{eq:E_b_N}) of the two channels of $\Delta$ (\eqnref{eq:Delta_b_two_channels}) are
\eqa{
& \E_{\perp}=\frac{1}{2}(\N_+ - \N_-) = 0 \\
& \E_{\shpa}=\frac{1}{2}(\N_+ + \N_-) = \N_+ \neq 0\ ,
}
meaning that only the $\Delta_{\shpa}$ pairing channel is Euler obstructed and must have nodes, whereas $\Delta_{\perp}$ is not Euler obstructed and is allowed to be fulled gapped.

Since the Euler obstruction and the pairing nodes do not rely on the gauge choice, and neither does the position of the pairing nodes, we, in the rest of this section, will choose the gauges that correspond to \eqnref{eq:reps_PT_T_P}.
With this gauge, we have $\chi_{\TR}(\bsl{q}_0)=1$ and $\det(V_{\TR}(\bsl{q}))=1$, resulting in $\N_{+,\bsl{q}_0}=-\N_{-,-\bsl{q}_0}$, $\eta_{+,\bsl{q}_0}(\bsl{q})=\eta_{-,-\bsl{q}_0}^*(-\bsl{q})$, and
\eq{
\label{eq:eta_TR}
\eta_{+}(\bsl{q})=-\eta_{-}^*(-\bsl{q})\ .
}

With \eqnref{eq:reps_PT_T_P} and \eqnref{eq:eta_TR}, the $\PT$ symmetry requires the pairing matrix $\Delta$ to be real, and the two channels (\eqnref{eq:spliting_Delta_b_two_channels}) of the pairing matrix $\Delta$ have the form
\eqa{
& \Delta_{\perp}(\bsl{q}) = d_0(\bsl{q}) \tau_0 + d_y(\bsl{q}) \ii \tau_y \\ 
& \Delta_{\shpa}(\bsl{q}) = \bsl{d}_{\shpa}(\bsl{q})\cdot\bsl{\tau}_{\shpa}\ ,
}
where $\bsl{\tau}_{\shpa}=(\tau_z,\tau_x)$, and $\bsl{d}_{\shpa}=(d_z,d_x)$ as defined in the main text.
Then, $|\Delta_{\perp}(\bsl{q})|=\sqrt{ d_0^2(\bsl{q}) + d_y^2(\bsl{q})}$ and $|\Delta_{\shpa}(\bsl{q})|=| \bsl{d}_{\shpa}(\bsl{q})|$.

The separation of the $\Delta_\perp$ and $\Delta_\shpa$ channels is based on the gauge group, which is convenient for the discussion of the Euler obstruction. 
However, such separation cannot forbid the mixing between the two channels as the gauge group is not a physical symmetry.
Based on the physical spin $\SU(2)$ symmetry and the inversion symmetry of the normal phase, we can split the pairing physically into four channels: spin-singlet parity-even, spin-singlet parity-odd, spin-triplet parity-even, and spin-triplet parity-odd.
The pairing mixing between these four channels is forbidden by the spin $\SU(2)$ and inversion symmetries of the normal state, which will be discussed in \appref{app:PandT_LGE}.
Therefore, we only need to consider each of them separately.
Note that under inversion symmetry, the projected pairing matrix transforms as
\eq{
\Delta(\bsl{q})\otimes\Pi\rightarrow -[\Delta(\bsl{q})\otimes\Pi]^T\ ,
}
resulting that $d_y$ has opposite parity to $d_0$ and $\bsl{d}_\shpa$ as listed \tabref{tab:parity_pairing}.
Thus, the $\Delta_\shpa$ only appears in $\Delta$ for spin-singlet parity-even pairing or spin-triplet parity-odd pairing corresponding to \eqnref{eq:ssPeven_stPodd}.

In the following, we will consider spin-singlet parity-even pairing or spin-triplet parity-odd pairing that is $\PT$-invariant.
We will first discuss the symmetries of the BdG Hamiltonian, which are essential to the study of nodes of the BdG Hamiltonian.
Then, we discuss the nodal superconductivity originating from a dominant Euler obstructed $\Delta_{\shpa}$.

\subsection{More Details on Symmetries of the BdG Hamiltonian}
\label{app:PandT_BdG_sym}

In this subsection, we will present more details on symmetries of the BdG Hamiltonian, which are essential to the discussion of nodal SC in the next subsection.

Based on the symmetry representations \eqnref{eq:reps_PT_T_P}, the most general normal-state Hamiltonian with spin $\SU(2)$ symmetry on $\M_\pm$ reads
\eq{
\label{eq:H_pm_normal_state}
H_\pm-\mu \hat{N}_{\pm}=\sum_{\bsl{q}\in\M_\pm} c^\dagger_{\pm,\bsl{q}} \left[\epsilon(\pm \bsl{q}) \tau_0 + m \bsl{g}_{\shpa}(\pm\bsl{q}) \cdot\bsl{\tau}_{\shpa} \right]\otimes s_0 c_{\pm,\bsl{q}}\ ,
}
where $\hat{N}_{\pm}=\sum_{\bsl{q}} c^\dagger_{\pm,\bsl{q}}c_{\pm,\bsl{q}} $.
Combined with the pairing form \eqnref{eq:H_pairing_TR}, the BdG Hamiltonian on $\M_\pm$ reads
\eq{
H_{BdG}= H_{BdG,+} + H_{BdG,-}  \ ,
}
where
\eq{
H_{BdG,+}=\frac{1}{2}\sum_{\bsl{q}\in\M_+} \mat{ c^{\dagger}_{+,\bsl{q}} &  c^{T}_{-,-\bsl{q}} } h_{BdG,+}(\bsl{q})\mat{ c^{\dagger}_{+,\bsl{q}}\\ (c^{\dagger}_{-,-\bsl{q}})^T }\ ,
}
\eq{
H_{BdG,-}=\frac{1}{2}\sum_{\bsl{q}\in\M_-} \mat{ c^{\dagger}_{-,\bsl{q}} &  c^{T}_{+,-\bsl{q}} } h_{BdG,-}(\bsl{q}) \mat{ c^{\dagger}_{-,\bsl{q}}\\ (c^{\dagger}_{+,-\bsl{q}})^T }\ ,
}
\eql{
\label{eq:h_BdG_+}
h_{BdG,+}(\bsl{q})=\mat{ [\epsilon(\bsl{q})\tau_0 +  m \bsl{g}_{\shpa}(\bsl{q})\cdot\bsl{\tau}_{\shpa}]\otimes s_0 & \Delta(\bsl{q})\otimes\Pi \\
\Delta^\dagger(\bsl{q})\otimes\Pi^\dagger & -[\epsilon(\bsl{q})\tau_0+ m \bsl{g}_{\shpa}(\bsl{q})\cdot\bsl{\tau}_{\shpa}] \otimes s_0
} \ .
}
$h_{BdG,-}(\bsl{q})$ can be obtained with the particle-hole symmetry (or particle-hole redundancy) as 
\eq{
\label{eq:h_BdG_PH_TR}
h_{BdG,-}(\bsl{q})=-\rho_x h_{BdG,+}^T(-\bsl{q})\rho_x
}
with $\rho$'s are Pauli matrices for the particle-hole index.
Owing to \eqnref{eq:h_BdG_PH_TR}, we only need to study the zero-energy nodes of $h_{BdG,+}(\bsl{q})$ since those of $h_{BdG,-}(\bsl{q})$ can be obtained with the particle-hole symmetry, serving as one simplification for the study of the nodes.

Another simplification for the study of the nodes exploits the momentum-independent spin part $\Pi$.
As mentioned in the last subsection, we can consider spin-singlet and spin-triple channels separately.
For spin-singlet pairing \eqnref{eq:Pi_form_spin_singlet}, $H_{BdG,+}$ has $\SU(2)$ spin-rotation symmetry.
For the spin-triplet pairing \eqnref{eq:Pi_form_spin_triplet}, we can first rotate the spin to make $\Pi=- s_z s_y$, and then $H_{BdG,+}$ is invariant under spin-$\U(1)$ along $z$.
Thus, in both cases, we can block diagonalize $H_{BdG,+}$ according to the conserved spin $z$.
Explicitly, we have
\eq{
H_{BdG,+}=H_{BdG,+,\uparrow}+H_{BdG,+,\downarrow}\ ,
}
where
\eq{
H_{BdG,+,s}= \frac{1}{2}\sum_{\bsl{q}\in\M_+} c_{BdG,+,s}^\dagger(\bsl{q})  \H_s(\bsl{q}) c_{BdG,+,s}(\bsl{q})\ ,
}
and
\eqa{
 & c_{BdG,+,s}^\dagger(\bsl{q}) = ( c^{\dagger}_{+,\bsl{q},s},  c_{-,-\bsl{q},-s}^T ) \text{ for spin singlet} \\ 
 & c_{BdG,+,s}^\dagger(\bsl{q}) = ( c^{\dagger}_{+,\bsl{q},s},  -\ii c_{-,-\bsl{q},-s}^T ) \text{ for spin-triplet} \ . 
}
In particular, $\H_\uparrow(\bsl{q})=\H(\bsl{q})$ is given in \eqnref{eq:H_cal}, and 
$\H_\downarrow(\bsl{k})$ is related to $\H_\uparrow(\bsl{k})$ as shown below.
For spin-singlet, we have the two-fold spin-rotation symmetry along $y$ direction $C^{spin}_{2,y}$, resulting in 
\eqa{
& C^{spin}_{2,y} H_{BdG,+} (C^{spin}_{2,y})^{-1} = H_{BdG,+} \\
& \Rightarrow \H_\downarrow(\bsl{k})=\rho_z \H_\uparrow(\bsl{q})\rho_z  \text{ for spin-singlet}\ .
}
For spin-triplet, we have the combined symmetry $ e^{\ii \hat{N} \pi/2}C^{spin}_{2,y}$ with $\hat{N}$ the particle number operator, leading to 
\eqa{
& e^{\ii \hat{N} \pi/2}C^{spin}_{2,y} H_{BdG,+} (e^{\ii \hat{N} \pi/2}C^{spin}_{2,y})^{-1} = H_{BdG} \\
& \Rightarrow \H_{BdG,+,\downarrow}(\bsl{k})=\H_{BdG,+,\uparrow}(\bsl{k})  \text{ for spin-triplet}\ .
}

According to the above discussion, we know the spin-down block has the same BdG bands as the spin-up block, and thus we only need to study the spin-up block $\H$ for the zero-energy gapless nodes, and the spin down block is given by  $C^{spin}_{2,y}$ or $e^{\ii \hat{N} \pi/2}C^{spin}_{2,y}$.
The spin-up block $\H$ has effective $\PT$ symmetry as \eqnref{eq:H_cal_PT},
and combined with the form of the pairing matrix \eqnref{eq:ssPeven_stPodd}, it also has the effective chiral symmetry as \eqnref{eq:H_cal_chiral}.
The effective $\PT$ and effective chiral symmetries anticommutes with each other.
Therefore, for both spin-singlet parity-even and spin-triplet parity-odd pairings, $\H$ belongs to the $CI$ class of the nodal classification in \refcite{Bzdusek2017AZInversionNodal}, which can support stable zero-energy line nodes in the 3D momentum space.
Owing to the effective $\PT$ symmetry, the line nodes must have $\pi$ Berry phase, and are classified by the $\dsZ_2$ monopole charge; the zero energy of the line nodes is protected by the effective chiral symmetry.

As discussed in \appref{app:Euler_Obstructed_Pairing_General}, the pairing operator \eqnref{eq:H_pairing_TR} is given by projecting the pairing of the full BdG Hamiltonian to the FSs of interest like \eqnref{eq:pairing_mat_project}.
Ultimately, what we want to derive is the zero-energy gapless nodes of the full BdG Hamiltonian.
Then, an essential question is whether the effective $\PT$ and chiral symmetries still exist in the full BdG Hamiltonian or are just artifacts of the form of $\H$ (\eqnref{eq:H_cal}).
It is important because if those symmetries were artifacts of \eqnref{eq:H_cal}, the zero-energy gapless nodes derived from $\H$ are not trustworthy, since the full BdG Hamiltonian does not have the corresponding symmetries and thus those nodes would be gapped out if we go from $\H$ to the full BdG Hamiltonian.
In the following, we will show that the full Hamiltonian does have the corresponding effective $\PT$ and chiral symmetries, and thus the zero-energy gapless nodes derived from \eqnref{eq:H_cal} are trustworthy.

In general, the translationally invariant full BdG Hamiltonian reads
\eq{
\label{eq:H_BdG_full}
\widetilde{H}_{BdG}=\frac{1}{2} \sum_{\bsl{k}}\mat{ \Psi^{\dagger}_{\bsl{k}} & \Psi^{T}_{-\bsl{k}} } 
\widetilde{h}_{BdG}(\bsl{k})
\mat{ \Psi_{\bsl{k}} \\ (\Psi^{\dagger}_{-\bsl{k}})^{T} }\ ,
}
where $\Psi^{\dagger}_{\bsl{k}}=(...\Psi^{\dagger}_{\bsl{k},a',s}...)$ is the basis of the normal-state Hamiltonian derived from Fourier transformation of atomic orbitals, $a'$ stands for all degrees of freedom (like the sublattices and orbitals) other than the momentum $\bsl{k}$ and spin $s$.
In particular, $\widetilde{h}_{BdG}(\bsl{k})$ has the form
\eq{
\label{eq:H_BdG_full_mat}
\widetilde{h}_{BdG}(\bsl{k})=\mat{ 
[h(\bsl{k})-\mu] \otimes s_0 & \widetilde{\Delta}(\bsl{k})\otimes\Pi \\
\widetilde{\Delta}^\dagger(\bsl{k})\otimes\Pi^\dagger &  -[h(-\bsl{k})-\mu]^T \otimes s_0 
}\ ,
}
where $h(\bsl{k})$ is the spinless part of the normal-state matrix Hamiltonian.
$\widetilde{h}_{BdG}(\bsl{k})$ has the particle-hole symmetry as 
\eq{
\label{eq:H_BdG_full_mat_PH}
\rho_x\widetilde{h}_{BdG}^T(-\bsl{k})\rho_x= -\widetilde{h}_{BdG}(\bsl{k})\ ,
}
where $\rho$'s are the Pauli matrices for the particle-hole index.
The zero-energy gapless nodes of the BdG Hamitlonian that we care about are those near the TR-related pairing-relevant manifolds $\M_\pm$, since their origin might be traced to the obstructed pairing.
\eqnref{eq:H_BdG_full_mat_PH} suggests that we only need to study the nodes near $\M_+$ since the nodes near $\M_-$ can be obtained with the particle-hole symmetry, coinciding with the simlification based on \eqnref{eq:h_BdG_PH_TR}.

Now we consider the spin blocks.
For spin-singlet pairing \eqnref{eq:Pi_form_spin_singlet}, $\widetilde{H}_{BdG}$ has $\SU(2)$ spin-rotation symmetry.
For the spin-triplet pairing \eqnref{eq:Pi_form_spin_triplet}, we can first rotate the spin to make $\Pi=- s_z s_y$, and then $\widetilde{H}_{BdG}$ is invariant under spin-$\U(1)$ along $z$.
Thus, in both cases, we can block diagonalize $\widetilde{H}_{BdG}$ according to the conserved spin $z$.
Explicitly, we have
\eq{
\widetilde{H}_{BdG}=\widetilde{H}_{BdG,\uparrow}+\widetilde{H}_{BdG,\downarrow}\ ,
}
where
\eq{
\widetilde{H}_{BdG,s}= \frac{1}{2}\sum_{\bsl{k}} \Psi_{BdG,s}^\dagger(\bsl{k})  \widetilde{\H}_s(\bsl{k}) \Psi_{BdG,s}(\bsl{k})\ ,
}
$\Psi_{BdG,s}^\dagger(\bsl{k})= ( \Psi^{\dagger}_{\bsl{k},s}, (-\ii)^{\Pi}  \Psi_{-\bsl{k},-s} )$ with $(-\ii)^{\Pi}=1$ for spin-singlet and $(-\ii)^{\Pi}=-\ii$ for spin-triplet.
In particular, we have
\eqa{
\label{eq:H_tilde_cal}
& \widetilde{\H}_\uparrow(\bsl{k})=\widetilde{\H}(\bsl{k}) =\mat{ 
h(\bsl{k})-\mu & \widetilde{\Delta}(\bsl{k}) \\
\widetilde{\Delta}^\dagger(\bsl{k}) &  -[h(-\bsl{k})-\mu]^T
} \ ,
}
and $\widetilde{\H}_\downarrow(\bsl{k})$ is related to $\widetilde{\H}_\uparrow(\bsl{k})$ as shown below.
For spin-singlet, we have the two-fold spin-rotation symmetry along $y$ direction $C^{spin}_{2,y}$, resulting in 
\eqa{
& C^{spin}_{2,y} \widetilde{H}_{BdG} (C^{spin}_{2,y})^{-1} = \widetilde{H}_{BdG} \\
& \Rightarrow \widetilde{\H}_\downarrow(\bsl{k})=\rho_z \widetilde{\H}_\uparrow(\bsl{q})\rho_z  \text{ for spin-singlet}\ .
}
For spin-triplet, we have the combined symmetry $ e^{\ii \hat{N} \pi/2}C^{spin}_{2,y}$ with $\hat{N}$ the particle number operator, leading to 
\eqa{
& e^{\ii \hat{N} \pi/2}C^{spin}_{2,y} \widetilde{H}_{BdG} (e^{\ii \hat{N} \pi/2}C^{spin}_{2,y})^{-1} = \widetilde{H}_{BdG} \\
& \Rightarrow \widetilde{\H}_\downarrow(\bsl{k})=\widetilde{\H}_\uparrow(\bsl{k})  \text{ for spin-triplet}\ .
}
According to the above discussion, we know the spin-down block has the same BdG bands as the spin-up block, and thus we only need to study the spin-up block $\widetilde{\H}$ for the zero-energy gapless nodes, and the spin down block is given by  $C^{spin}_{2,y}$ or $e^{\ii \hat{N} \pi/2}C^{spin}_{2,y}$.
These analysis also agrees with the block-diagonalization for the projected BdG Hamiltonian.

Now we show that $\widetilde{\H}$ also has effective $\PT$ and chiral symmetries.
%
First of all, as the $\Psi$ basis is derived from the local atomic orbitals, \appref{app:review_NLSM_Euler} suggests that we can always choose the following momentum-independent symmetry reps 
\eqa{
\label{eq:rep_T_P_PT_Psi}
& \TR \Psi^{\dagger}_{\bsl{k}} \TR^{-1} = \Psi^{\dagger}_{-\bsl{k}} U_{\TR}\otimes \ii s_y \\
& \P \Psi^{\dagger}_{\bsl{k}} \P^{-1} = \Psi^{\dagger}_{-\bsl{k}} U_{\TR}\otimes s_0 \\
& \PT \Psi^{\dagger}_{\bsl{k}} (\PT)^{-1} = \Psi^{\dagger}_{\bsl{k}} \mathds{1}\otimes \ii s_y \ ,
}
where $U_{\TR}$ is a real orthogonal matrix.
Since we only care about the $\PT$-invariant pairing, meaning that
\eq{
 \widetilde{\Delta}^*(\bsl{k})= \widetilde{\Delta}(\bsl{k}) \ .
}
As a result, $\widetilde{H}_{BdG}$ has $\PT$ symmetry.
But $\PT$ is not a symmetry of one spin block since $\PT$ flips spin as shown in \eqnref{eq:rep_T_P_PT_Psi}.
Nevertheless, $\widetilde{H}_{BdG,\uparrow}$ has an effective $\PT$ symmetry, which is the $C^{spin}_{2,y} \PT $ symmetry for spin-singlet or the $ e^{\ii \hat{N} \pi/2} C^{spin}_{2,y} \PT $ symmetry for spin-triplet.
In both cases, the effective $\PT$ symmetry results in
\eq{
\label{eq:H_tilde_cal_PT}
\widetilde{\H}^*(\bsl{k})=\widetilde{\H}(\bsl{k})\ .
}
Furthermore, as discussed in the last subsection, we only consider the spin-singlet parity-even pairing and the spin-triplet parity-odd pairing.
Then, in both cases, the chiral symmetry for $\widetilde{\H}(\bsl{k})$ reads
\eq{
\label{eq:H_tilde_cal_chiral}
\rho_y U_{\TR}\widetilde{\H}(\bsl{k})\rho_y U_{\TR}=-\widetilde{\H}(\bsl{k})\ .
}

Next, we project $\widetilde{\H}(\bsl{k})$ to $\M_+$.
As discussed in \appref{app:Euler_Obstructed_Pairing_General}, on the two TR-related pairing relevant manifolds $\M_\pm$, the creation operator $c^\dagger_{\pm,\bsl{q},a,s}$ in \eqnref{eq:H_pairing_TR} can be expressed as
\eq{
\label{eq:c_Psi_proj}
c^{\dagger}_{\pm,\bsl{q},s} = \Psi^{\dagger}_{\bsl{K}_{\pm}+\bsl{q},s}\xi(\bsl{K}_{\pm}+\bsl{q})\ ,
}
where $\xi(\bsl{K}_{\pm}+\bsl{q})=(\xi_1(\bsl{K}_{\pm}+\bsl{q}),\xi_2(\bsl{K}_{\pm}+\bsl{q}))$ as the othornormal linear combinations of the eigenvectors of $\widetilde{h}(\bsl{k})$ for the top two occupied bands on $\M_\pm$.
The symmetry representations \eqnref{eq:reps_PT_T_P} furnished by $c^{\dagger}_{\pm,\bsl{q}}$ can be satisfied by choosing  
\eqa{
\label{eq:xi_convention}
& \xi(\bsl{K}_{-}-\bsl{q})=U_{\TR}\xi^*(\bsl{K}_{+}+\bsl{q}) \\
& \xi^*(\bsl{K}_{\pm}+\bsl{q})=\xi(\bsl{K}_{\pm}+\bsl{q}) \ .
}
Then, based on the projection relation of the creation operator \eqnref{eq:c_Psi_proj}, we should have
\eq{
\label{eq:H_cal_H_tilde_cal_proj}
\H(\bsl{q})=  \xi_{BdG,+}^\dagger(\bsl{q}) \widetilde{\H} (K_++\bsl{q}) \xi_{BdG,+}(\bsl{q})\ ,
}
where
\eq{
\label{eq:xi_BdG}
\xi_{BdG,+}(\bsl{q})=\mat{ \xi(K_++\bsl{q}) & \\ & \xi^*(K_- -\bsl{q})}\ .
}
In other words, one way to get the form of $\H(\bsl{q})$ is through the projection of $\widetilde{H}$ onto $\M_+$. 
Now we show that the projection indeed reproduces form in
\eqnref{eq:H_cal} with \eqnref{eq:ssPeven_stPodd} and the effective $\PT$, and chiral symmetries of \eqnref{eq:H_cal} are the projection of those of \eqnref{eq:H_tilde_cal}.

First note $U_{\TR} h^*(\bsl{K}_{-}-\bsl{q}) U_{\TR}^\dagger = h(\bsl{K}_{+}+\bsl{q})$ suggests that 
\eqa{
& \xi^\dagger(\bsl{K}_{+}+\bsl{q}) [h(\bsl{K}_{+}+\bsl{q})-\mu] \xi(\bsl{K}_{+}+\bsl{q}) \\
& = \xi^\dagger(\bsl{K}_{-}-\bsl{q}) [h(\bsl{K}_{-}-\bsl{q})-\mu] \xi(\bsl{K}_{-}-\bsl{q})\ ,
}
and owing to the real $h$ and $\xi$, it should be a symmetric $2\times 2$ matrix of the form
\eq{
\epsilon(\bsl{q})\tau_0 + m \bsl{g}_{\shpa}(\bsl{q})\cdot\bsl{\tau}_{\shpa} \ ,
}
meaning that the projection reproduces the form of the normal-state part of $\H$.
On the other hand, for both spin-singlet parity-even pairing and spin-triplet parity-odd pairing, the projected pairing matrix, which is defined as 
\eq{
\Delta(\bsl{q})=\xi^\dagger(\bsl{K}_{+}+\bsl{q}) \widetilde{\Delta}(\bsl{q})  \xi^*(\bsl{K}_{-}-\bsl{q})\ ,
}
satisfies 
\eqa{
& [\Delta(\bsl{q})]^*=\Delta(\bsl{q}) \\
& [\Delta(\bsl{q})]^T=\Delta(\bsl{q})\ .
}
The above equation suggests that $\Delta(\bsl{q})$ must have the form of \eqnref{eq:ssPeven_stPodd}. 
Therefore, \eqnref{eq:H_cal_H_tilde_cal_proj} precisely reproduces the form of \eqnref{eq:H_cal} with \eqnref{eq:ssPeven_stPodd}.
Combined with
\eqa{
& \xi_{BdG,+}^*(\bsl{q})=\xi_{BdG,+}(\bsl{q}) \\
& \rho_y U_{\TR}\xi_{BdG,+}(\bsl{q})= \xi_{BdG,+}(\bsl{q})\rho_y \tau_0\ ,
}
we can see the effective $\PT$ and chiral symmetries of $\H$ (\eqnref{eq:H_cal_PT} and \eqnref{eq:H_cal_chiral}) are projection of those of $\widetilde{\H}$ (\eqnref{eq:H_tilde_cal_PT} and \eqnref{eq:H_tilde_cal_chiral}) .

In sum, for both spin-singlet parity-even and spin-triplet parity-odd pairings, both $\widetilde{\H}$ and $\H$ have (i) an effective $\PT$ symmetry that squares to 1 (\eqnref{eq:H_tilde_cal_PT} and \eqnref{eq:H_cal_PT}) and (ii) an effective chiral symmetries that anticommutes with the effective $\PT$ symmetry (\eqnref{eq:H_tilde_cal_chiral} and \eqnref{eq:H_cal_chiral}).
It means that both $\widetilde{\H}$ and $\H$ belongs to the CI nodal class in \refcite{Bzdusek2017AZInversionNodal}, which can support stable zero-energy line nodes in the 3D momentum space.
Owing to the effective $\PT$ symmetry, the line nodes are classified by the $\dsZ_2$ monopole charge; the zero energy of the line nodes is protected by the chiral symmetry.
Although the chiral symmetry enhances the topological invariant along a circle around each zero-energy line node to $\dsZ$~\cite{Bzdusek2017AZInversionNodal}, we still focus on the $\pi$ Berry phase protected by the $\PT$ symmetry, since it can exist even without the chiral symmetry. 
Furthermore, since the effective $\PT$ and chiral symmetries of $\H$ are the projection of those of $\widetilde{\H}$, all zero-energy gapless nodes of $\H$ protected by the effective $\PT$ and chiral symmetries is allowed to exist in the full $\widetilde{\H}$.
In the next subsection, we will study the zero-energy gapless nodes of $\H$ and $\widetilde{\H}$ that originate from the Euler obstructed $\Delta_{\shpa}$.

\subsection{More Details on Nodal SC}
\label{app:PandT_NodalSC}

In this subsection, we discuss the zero-energy gapless nodes of the projected $\H$ in \eqnref{eq:H_cal}, as well as the full $\widetilde{\H}$.
We emphasize that $\bsl{q}$ in \eqnref{eq:H_cal} takes values on a sphere-like FS $\M_+$ around $\bsl{K}_+$ and is also allowed to take value slightly away from $\M_+$.
We will focus on the weak pairing---the pairing amplitude maximum $|\Delta|$ is much smaller than the superconductivity cutoff $\epsilon_c$---and we choose $\epsilon_c$ to be much smaller than the chemical potenital $|\mu|$ and the minimum $E_g$ of the gaps above and below the normal-state top two occupied bands on FSs:
\eq{
|\Delta|\ll \epsilon_c \ll |\mu| \sim E_g\ .
}

We will first demonstrate that when the FS splitting $m$ is small, a sufficiently-dominant Euler obstructed $\bsl{d}_\shpa$ always leads to nodal SC with $\dsZ_2$ monopole nodal rings/points per spin.
To do so, let us first consider a special case where $m=0$ and $d_0=0$ on all pairing-relevant manifolds.
In this case, the two FSs that enclose $K_+$ have no splitting, and correspond to the $\epsilon(\bsl{q})=0$ manifold.
Furthermore, $\H$ is simplified to 
\eq{
\label{eq:H_cal_m0_d00}
\mat{ \epsilon(\bsl{q})\tau_0  &  \bsl{d}_\shpa(\bsl{q})\cdot\bsl{\tau}_\shpa  \\
 \bsl{d}_\shpa(\bsl{q})\cdot\bsl{\tau}_\shpa & -\epsilon(\bsl{q})\tau_0
} \ ,
}
which has two doubly-degenerate bands 
\eq{
\pm\sqrt{\epsilon(\bsl{q})^2+\bsl{d}_\shpa(\bsl{q})^2}\ .
}
The bands suggest that the gapless nodes only occur for $\bsl{q}_i$ that satisfies $\epsilon(\bsl{q}_i)=0$ and $\bsl{d}_\shpa(\bsl{q}_i)=0$, and must be four-fold degenerate.
The existence of the four-fold degenerate gapless nodes are guaranteed by the nonzero Euler index of $\Delta_\shpa$.
To see this, note that for $\bsl{q}\in\M_+$, we have $\epsilon(\bsl{q})=0$.
%
Then, the nonzero Euler index requires $|\Delta_\shpa(\bsl{q})|=|\bsl{d}_\shpa(\bsl{q})|$ to have nodes on $\M_+$, which correspond to the four-fold degenerate BdG nodes mentioned above.

In particular, each four-fold degenerate zero-energy BdG node on the FS in this case has nonzero $\dsZ_2$ monopole charge protected by the effective $\PT$ symmetry of $\H$.
To see this, first note that each BdG node on $\M_+$ should be isolated unless further finely tuning certain parameters.
Then, for each BdG node $\bsl{q}_i$, we can always choose $\xi(K_++\bsl{q})$ and $\xi(K_--\bsl{q})$ to be smooth around $\bsl{q}_i$ while keeping \eqnref{eq:xi_convention}, since the normal-state nodal points are far away from $\bsl{q}_i$.
In this case, $\H(\bsl{q})$ is smooth around $\bsl{q}_i$, and we can expand $\H(\bsl{q})$ near $\bsl{q}_i$ to the leading order of $\bsl{p}=\bsl{q}-\bsl{q}_i$ as 
\eq{
\H_i(\bsl{p})=\bsl{C}_1\cdot\bsl{p} \rho_z\tau_0+ \bsl{C}_2\cdot\bsl{p} \rho_x\tau_x+ \bsl{C}_3\cdot\bsl{p}\rho_x\tau_z\ .
}
$\bsl{C}_1$, $\bsl{C}_2$ and $\bsl{C}_3$ are generally linearly independent, and thus $ (\bsl{C}_1\cdot\bsl{p})^2 + (\bsl{C}_2\cdot\bsl{p})^2 +(\bsl{C}_3\cdot\bsl{p})^2 =\lambda^2 $ with $\lambda>0$ gives a sphere-like manifold (more precisely, ellipsoid) $\M_i$ that encloses $\bsl{q}_i$.
Then, $\H_i(\bsl{p})$ has only two occupied bands on $\M_i$, and we can calculate the Euler number of them to determine the $\dsZ_2$ monopole charge of the BdG node $\bsl{q}_i$ within $\H_i(\bsl{p})$, which turns out to be 1.
Therefore, the zero-energy BdG nodal point at $\bsl{q}_i$ should have $\dsZ_2$ monopole charge being 1.
Owing to the weak pairing and the existence of the corresponding effective $\PT$ and chiral symmetries in the full $\widetilde{\H}$, the zero-energy monopole-charged BdG nodal point at $\bsl{q}_i$ in $\H$ must manifest itself as a zero-energy BdG MNLs near $\bsl{q}_i$ in the full $\widetilde{\H}$.
The zero-energy BdG MNLs cannot be destroyed even if we add an infinitesimal FS splitting $m$ and an infinitesimal trivial channel $d_0$, since they preserve the chiral and $\PT$ symmetries.
The BdG MNLs may persist even if we increase the FS splitting $m$, as long as the change of $m$ does not bring any two BdG MNLs together.
Thus, we know that if the FS splitting is small, a sufficiently-dominant Euler obstructed $\Delta_\shpa$ channel always lead to nodal SC with BdG MNLs.

Next we show that even if the FS splitting is large, a dominant Euler obstructed $\Delta_\shpa$ on one FS still leads to nodal SC, though the $\dsZ_2$ monopole charge of the nodal rings might be trivial.
Here we have assumed that even for the large FS splitting, there is one FS on which the top two occupied bands have odd Euler number.
The Euler obstructed $\Delta_\shpa$ must have nodes on the FS.
Let us start with the special case where $d_0=0$ near the FS.
Then, the nodes of $\Delta_\shpa$ on the FS must lead to zero-energy BdG nodal points of $\H$ on the FS.

If the BdG nodes of $\Delta_\shpa$ coincide with the FS touching points, they lead to four-fold degenerate zero-energy BdG gapless points of $\H$ on the FS.
The four-fold degenerate BdG nodal points can either be isolated or connected to the zero-energy BdG nodal lines in the 3D momentum space.
If they are isolated, they should have nonzero $\dsZ_2$ monopole charge, since continuously decreasing $m$ to zero (while fixing $\bsl{d}_\shpa$ and $\epsilon$) would keep the node isolated and gapless.
As we know the BdG point node should have nonzero $\dsZ_2$ monopole charge for $\H$ with $m=0$, the isolated BdG nodal point should also have nonzero $\dsZ_2$ monopole charge for $\H$ for large FS splitting.
Thus, the isolated BdG nodal points should manifest themselves as BdG MNLs in the full $\widetilde{\H}$.
If they are connected to nodal lines, the nodal line should still exist in the full $\widetilde{\H}$ since they are protected by the effective $\PT$ symmetry or chiral symmetries.
Therefore, in both cases, there are still zero-energy BdG nodal rings in the full $\widetilde{\H}$, with zero or nonzero $\dsZ_2$ monopole charge.
As a result, adding a small nonzero $d_0$ cannot destroy those nodal rings, owing to the stable topological invariant of the BdG nodal line.

A more common situation is that the nodes of $\Delta_\shpa$ do not coincide with FS touching points, which lead to two-fold degenerate zero-energy BdG nodal points of $\H$ on the FS.
Without further fine-tuning, those BdG nodes should be isolated on the FS, and we can project $\H$ to the two zero-energy eigenvectors at $\bsl{q}_i$ and expand the projected model to the linear order of $\bsl{p}=\bsl{q}-\bsl{q}_i$, leading to 
\eq{
\H_i^{2B}= (C_{x1} p_1+C_{x2} p_2)\tau_x+(C_{z1} p_1+C_{z2} p_2)\tau_z\ ,
}
where $(p_1,p_2)$ parametrizes $\bsl{p}$ on the FS, and choose the two zero-energy eigenvectors at $\bsl{q}_i$ to be real. 
Again, $(C_{x1} , C_{x2} )$ and $(C_{z1}, C_{z2} )$ are generally linearly independent, and then $(C_{x1} p_1+C_{x2} p_2)^2+ (C_{z1} p_1+C_{z2} p_2)^2 = \lambda^2$ gives a circle-like loop $\gamma_i$ that surrounds $\bsl{q}_i$ on the FS.
Direct calculation shows that $\det[W_{\gamma_i}]=-1$ in $\H_i^{2B}$, which should still hold for the two occupied bands of $\H$ along a small $\gamma_i$.
Again owing to the weak pairing and the corresponding symmetries in the full $\widetilde{\H}$, we have $\det[W_{\gamma_i}]=-1$ for all occupied bands of $\widetilde{\H}$ along $\gamma_i$.
%
It means that $\gamma_i$ must enclose a zero energy BdG nodal line in the 3D momentum space, and $\bsl{q}_i$ is given by the intersection between the nodal line and the FS.
The nodal lines cannot disappear even if we add a small $d_0$ since it is protected by $\det[W_{\gamma_i}]=-1$.
Thus, regardless of the FS splitting, a sufficiently-dominant Euler obstructed $\Delta_\shpa$ on one FS still leads to nodal SC.

In conclusion, a sufficiently-dominant Euler obstructed $\Delta_\shpa$ on FSs like \figref{fig:FS}(b-c) leads to nodal SC with zero nodal lines.
The nodal lines have two-fold degeneracy per spin, and are near the pairing nodes of $\Delta_\shpa$ on the FS.
When the FS splitting is small (compared to the momentum distances among BdG nodal points for zero splitting), the nodal SC contains BdG MNLs that have nonzero $\dsZ_2$ monopole charge, and the BdG MNLs originate from the pairing nodes of $\Delta_\shpa$ on the FS.

\subsection{More Details on Linearized Gap Equation}
\label{app:PandT_LGE}

The above discussion on the nodal SC requires the following three assumptions for the pairing operator.
First, different spin channels (spin-singlet and spin-triplet) and different parities cannot mix into each other, allowing us to study them separately.
Second, we are physically allowed to have the pairing operator \eqnref{eq:H_pairing_TR} that is $\PT$-invariant and have a momentum-independent spin part.
Third, we are physically allowed to have \eqnref{eq:ssPeven_stPodd} for the pairing operator, corresponding to either spin-singlet parity-even or spin-triplet parity-odd.
Fourth, the Euler obstructed $\Delta_\shpa$ dominates.
In this subsection, we will use the linearized gap equation to study the SC transition, and justify that near the SC transition, the first and second hold for all values of FS splitting, and the third holds for small FS splitting.
For the fourth assumption, we will show a dominant $\Delta_\shpa$ depends on the interaction strength and leaves further justification to the next section.

The normal-state Hamiltonian that we consider is \eqnref{eq:H_pm_normal_state}, where $\M_\pm$ are replaced by $\Omega_{\pm}$, labelling the 3D region that consists of all pairing-relevant momenta.
The most general two-body interaction that accounts for the zero-total-momentum Cooper pairings reads
\begin{widetext}
\eq{
\label{eq:H_int}
H_{int}=\sum_{\bsl{q},\bsl{q}'\in\Omega_+}\sum_{s_1,s_2,s_3,s_4} \sum_{a_1, a_2, a_3, a_4} \widetilde{V}_{a_1 a_2 a_3 a_4}^{\bsl{q}\bsl{q}',s_1 s_2 s_3 s_4}c^\dagger_{+,\bsl{q},a_1,s_1} c^\dagger_{-,-\bsl{q},a_2,s_2} c_{-,-\bsl{q}',a_3,s_3} c_{+,\bsl{q}',a_4,s_4}\ ,
}
where at least certain components of $\widetilde{V}_{a_1 a_2 a_3 a_4}^{\bsl{q}\bsl{q}',s_1 s_2 s_3 s_4}$ are negative (or attractive) such that superconductivity is allowed.
In general, the interaction can always be expended as 
\eq{
\widetilde{V}_{a_1 a_2 a_3 a_4}^{\bsl{q}\bsl{q}',s_1 s_2 s_3 s_4}=\frac{1}{2}\sum_{j,j',j_s,j_s'=0,x,y,z}V_{j j',j_s j_s'}^{\bsl{q}\bsl{q}'} [\widetilde{\tau}_{j}]_{a_1a_2}(\widetilde{s}_{j_s}\ii s_y)_{s_1 s_2} [\widetilde{\tau}_{j'}]^\dagger_{a_3a_4} (\widetilde{s}_{j_s'}\ii s_y)^\dagger_{s_3 s_4}\ ,
}
\end{widetext}
where 
\eq{
\widetilde{\tau}_{j}=\left\{ 
\begin{array}{ll}
\tau_j  &,\  j=0,x,z\\
\ii \tau_y & ,\  j= y
\end{array}
\right.
}
satisfies $\widetilde{\tau}_{j}^*=\widetilde{\tau}_{j}$, and
\eq{
\widetilde{s}_{j_s}=\left\{ 
\begin{array}{ll}
s_0 & \ , j_s=0\\
\ii s_{j_s} & \ , j_s= x,y,z
\end{array}
\right.
}
satisfies $\ii s_y \widetilde{s}_{j_s}^* (\ii s_y)^T =\widetilde{s}_{j_s}$.
As the result, the interaction can be written as
\eq{
H_{int}=\sum_{\bsl{q},\bsl{q}'\in\Omega_+}\sum_{j,j',j_s,j_s'} \frac{V_{j j',j_sj_s'}^{\bsl{q}\bsl{q}'}}{2} O_{jj_s}(\bsl{q}) O_{j'j_s'}^\dagger(\bsl{q}')\ ,
}
where 
\eqa{
O_{jj_s}(\bsl{q})=  c^\dagger_{+,\bsl{q}} \widetilde{\tau}_{j}\otimes( \widetilde{s}_{j_s}\ii s_y) (c^\dagger_{-,-\bsl{q}})^T\ .
}
Then, the total Hamiltonian reads
\eq{
H-\mu \hat{N}=H_+ -\mu \hat{N}_+ + H_- -\mu \hat{N}_- + H_{int}\ .
}

Symmetries impose constrains on $V_{j j'}^{\bsl{q}\bsl{q}',j_sj_s'}$.
The interaction that accounts for superconductivity is usually derived from Coulumb interaction and electron-phonon interaction, which both preserve the spin $\SU(2)$ symmetry.
The spin $\SU(2)$ symmetry requires 
\eq{
V_{j j',j_sj_s'}^{\bsl{q}\bsl{q}'} =\delta_{j_s j_s'} V_{j j',j_s}^{\bsl{q}\bsl{q}'}\ ,
}
where
\eq{
\label{eq:V_js}
V_{j j',j_s}^{\bsl{q}\bsl{q}'}=\left\{
\begin{array}{ll}
 V_{j j',\text{SS}}^{\bsl{q}\bsl{q}'}    &,\ j_s = 0 \\
 V_{j j',\text{ST}}^{\bsl{q}\bsl{q}'}    &,\ j_s = x,y,z
\end{array}
\right.
}
and $\text{SS}$ and $\text{ST}$ are short for spin-singlet and spin-triplet, respectively.
As a result, the interaction can be further simplified to 
\eq{
\label{eq:H_int_sim}
H_{int}=\sum_{\bsl{q},\bsl{q}'\in\Omega_+}\sum_{j,j',j_s,} \frac{V_{j j',j_s}^{\bsl{q}\bsl{q}'}}{2} O_{jj_s}(\bsl{q}) O_{j'j_s}^\dagger(\bsl{q}')\ .
}
Hermiticity requires
\eq{
[V_{j' j,j_s}^{\bsl{q}'\bsl{q}}]^* =V_{j j',j_s}^{\bsl{q}\bsl{q}'}\ .
}
Combined with the symmetry representations \eqnref{eq:reps_PT_T_P}, the $\PT$ symmetry requires
\eqa{
\label{eq:int_PT}
 [V_{j j',j_s}^{\bsl{q}\bsl{q}'}]^*= V_{j j',j_s}^{\bsl{q}\bsl{q}'}\ ,
}
and the inversion symmetry requires
\eq{
\label{eq:int_P}
V_{y j,j_s}^{\bsl{q}\bsl{q}'}=V_{j y,j_s}^{\bsl{q}\bsl{q}'}=0 \text{ for $j\neq y$}\ .
}

Now we discuss the mean-field linearized gap equation. 
The mean-field Cooper pairing order parameter is defined as
\eq{
\label{eq:Delta_MF}
 \Delta_{jj_s}(\bsl{q})=\left\langle \sum_{\bsl{q}'\in\Omega_+} \sum_{j'} \frac{V_{j j',j_s}^{\bsl{q}\bsl{q}'}}{4} O_{j'j_s}^\dagger(\bsl{q}') \right\rangle\ ,
 }
where $\left\langle X \right\rangle= \Tr[e^{-\beta (H-\mu \hat{N})} X ]/\Tr[e^{-\beta (H-\mu \hat{N})} ] $, $\beta=1/T$, $T$ is the temperature, and recall that $k_B=1$ in our unit system.
The mean-field pairing operator reads
\eqa{
& H_{pairing}= \sum_{\bsl{q}\in \Omega_+} \sum_{j,j_s}  [ \Delta_{jj_s}(\bsl{q})O_{jj_s}(\bsl{q}) + h.c. ] \\
& = \sum_{\bsl{q}\in \Omega_+}  c^\dagger_{+,\bsl{q}} D(\bsl{q})  c^\dagger_{-,-\bsl{q}} + h.c.
}
with pairing matrix 
\eq{
\label{eq:D_pairing_mat}
D(\bsl{q})=\sum_{j,j_s}\Delta_{jj_s}(\bsl{q})\widetilde{\tau}_{j}\otimes( \widetilde{s}_{j_s}\ii s_y)\ ,
}
and the mean-field Hamiltonian then reads
\eq{
H_{MF}= H_+ -\mu \hat{N}_+  + H_- -\mu \hat{N}_- + H_{pairing}\ .
}
Replacing $H-\mu \hat{N}$ in the \eqnref{eq:Delta_MF} by the mean-field Hamiltonian $H_{MF}$, we arrive the self-consistent equation for the superconducting order parameter, which reads
\eq{
\label{eq:Delta_MF_SelfConsis}
\Delta_{jj_s}(\bsl{q}) = \frac{\Tr\left[ \sum_{\bsl{q}'\in\Omega_+} \sum_{j'} \frac{\widetilde{V}_{j j',j_s}^{\bsl{q}\bsl{q}'}}{4} O_{j'j_s}^\dagger(\bsl{q}') e^{-\beta H_{MF}}\right] }{ \Tr\left[ e^{-\beta H_{MF}}\right] }\ .
}
Near the superconductivity phase transition, the superconducting order parameter is very small, and therefore we can keep only the first order of the \eqnref{eq:Delta_MF_SelfConsis}, resulting in the linearized gap equation
\eql{
\label{eq:Delta_LGE}
\Delta_{jj_s}(\bsl{q}) = -\sum_{j',\omega'}\sum_{\bsl{q}'\in \Omega_+}\frac{V_{jj',j_s}^{\bsl{q}\bsl{q}'}}{2\beta}\sum_{j''}\Tr\left[\widetilde{\tau}_j'^\dagger G(\omega',\bsl{q}') \widetilde{\tau}_{j''} G^T(-\omega',\bsl{q}') \right] \Delta_{j'' j_s} (\bsl{q}') \ ,
}
where $\omega=(2n+1)\pi/\beta$ is the fermionic Matusbara frequency, and
\eq{
G(\omega,\bsl{q})= [ (\ii \omega + \epsilon(\bsl{q}))\tau_0 + m \bsl{g}_{\shpa}(\bsl{q})\cdot \bsl{\tau}_{\shpa} ]^{-1}
}
is the normal-state Green function.
Solving \eqnref{eq:Delta_LGE} for $\Delta_{jj_s}(\bsl{q})$ and $T$ gives us the SC critical temperature $T=T_c$.
In the following, we will show that the highest-$T_c$ can always be achieved by $\PT$-invariant pairing form \eqnref{eq:H_pairing_TR} with a momentum-independent spin part.

According to the $\PT$ symmetry \eqnref{eq:int_PT}, \eqnref{eq:Delta_LGE} holds after performing $\Delta_{jj_s}(\bsl{q})\rightarrow \Delta_{jj_s}^*(\bsl{q})$, meaing that if $\Delta_{jj_s}(\bsl{q})$ yields $T=T_c$,  $\Delta_{jj_s}^*(\bsl{q})$ yields the same $T=T_c$.
Then, the highest $T_c$ can be achieved with a real $\Delta_{jj_s}(\bsl{q})$, which makes the pairing matrix \eqnref{eq:D_pairing_mat} $\PT$-invariant.
On the other hand, the linearized gap equation \eqnref{eq:Delta_LGE} is decoupled for different spin components, meaning that the spin-singlet and spin-triplet pairings are decoupled.
The spin-singlet and spin-triplet channels generally have different SC critical temperatures, meaning that the highest $T_c$ should be achieved by either spin-singlet or spin-triplet but not the mixing of them.
For spin-singlet pairing, we have $\Delta_{jx}=\Delta_{jy}=\Delta_{jz}=0$, and by defining $\Delta_{j0}=d_j$, the pairing matrix 
\eqnref{eq:D_pairing_mat} reads
\eq{
\label{eq:D_pairing_mat_sim}
D(\bsl{q})=(\sum_{j } d_j(\bsl{q}) \widetilde{\tau}_j)\otimes \Pi
}
with $\Pi=\ii s_y$, which has the same form as \eqnref{eq:H_pairing_TR}.
For spin-triplet pairing ($\Delta_{j0}=0$), \eqnref{eq:Delta_LGE} together with \eqnref{eq:V_js} suggests that different spin directions of the spin-triplet SC have the same linearized gap equation, which makes sense since different spin-directions are related by the spin $\SU(2)$ symmetry.
As a result, for spin-triplet pairing, the highest $T_c$ can always be achieved by a pairing order parameter with a momentum-independent spin direction.
Therefore, we can always choose $\Delta_{jj_s}=d_j(\bsl{q}) \hat{n}_{j_s}$ for the spin-triplet pairing with $\hat{\bsl{n}}=(\hat{n}_x,\hat{n}_y,\hat{n}_z)$ a real $\bsl{q}$-independent unit vector to achieve the highest $T_c$.
Such choice would make the pairing matrix have the form \eqnref{eq:D_pairing_mat_sim} with $\Pi=\ii (\hat{\bsl{n}}\cdot\bsl{s})\ii s_y$, also coinciding with \eqnref{eq:H_pairing_TR}.
We conclude that at the SC transition, different spin channels cannot mix into each other, and the highest superconducting critical temperature $T_c$ can always be achieved by the $\PT$-invariant pairing with momentum-independent spin part, justifying \eqnref{eq:H_pairing_TR}.

\begin{widetext}
With the simplified pairing \eqnref{eq:H_pairing_TR}, we now derive what form of $d_j$ is required to have to achieve the highest $T_c$.
First, owing to the simplification \eqnref{eq:D_pairing_mat_sim}, the linearized gap equation for $d_j$ reads
\eq{
\label{eq:d_LGE}
d_{j}(\bsl{q}) = -\sum_{\omega'}\sum_{\bsl{q}'\in \Omega_+}\sum_{j'}\frac{V_{jj'}^{\bsl{q}\bsl{q}'}}{2\beta}\sum_{j''}\Tr\left[\widetilde{\tau}_j'^\dagger G(\omega',\bsl{q}') \widetilde{\tau}_{j''} G^T(-\omega',\bsl{q}') \right] d_{j''} (\bsl{q}') \ ,
}
where $V_{jj'}^{\bsl{q}\bsl{q}'}=V_{j j',\text{SS}}^{\bsl{q}\bsl{q}'}$ for spin-sinplet pairing and $V_{jj'}^{\bsl{q}\bsl{q}'}=V_{j j',\text{ST}}^{\bsl{q}\bsl{q}'}$ for spin-triplet pairing.
To proceed, we need to reexpress the normal-state Green function as
\eq{
G(\omega,\bsl{q}) = G_1(\omega,\bsl{q})\tau_0 + \hat{\bsl{g}}_\shpa(\bsl{q}) \cdot \bsl{\tau}_{\shpa} G_2(\omega,\bsl{q})\ ,
}
where $\hat{\bsl{g}}_\shpa=\bsl{g}_\shpa(\bsl{q})/|\bsl{g}_\shpa(\bsl{q})|$,
\eqa{
\label{eq:G1_G2}
& G_1(\omega,\bsl{q}) = \frac{1}{2} \left( \frac{1}{\ii \omega + \epsilon(\bsl{q}) + m |\bsl{g}_\shpa(\bsl{q})|} + \frac{1}{\ii \omega + \epsilon(\bsl{q}) - m |\bsl{g}_\shpa(\bsl{q})|} \right) \\ 
& G_2(\omega,\bsl{q}) = \frac{1}{2} \left( \frac{1}{\ii \omega + \epsilon(\bsl{q}) + m |\bsl{g}_\shpa(\bsl{q})|} - \frac{1}{\ii \omega + \epsilon(\bsl{q}) - m |\bsl{g}_\shpa(\bsl{q})|} \right) \ ,
}
Then, substituting them into \eqnref{eq:d_LGE} results in 
\eqa{
& d_{j}(\bsl{q}) = -\frac{1}{\beta}\sum_{\omega'}\sum_{\bsl{q}'\in \Omega_+}\{ \sum_{j'} V_{jj'}^{\bsl{q}\bsl{q}'} d_{j'} ( G_1 G_1 + G_2 G_2) + (G_2 G_1 + G_1 G_2) ( V_{j0}^{\bsl{q}\bsl{q}'} \hat{\bsl{g}}_\shpa \cdot \bsl{d}_\shpa + \bsl{V}_{j\shpa}^{\bsl{q}\bsl{q}'} \cdot \hat{\bsl{g}}_\shpa d_0)  \\
& + G_2 G_2 (-V^{\bsl{q} \bsl{q}'}_{j y } d_y+ 2 \bsl{V}_{j\shpa}^{\bsl{q}\bsl{q}'}\cdot \hat{\bsl{g}}_\shpa\ \hat{\bsl{g}}_\shpa \cdot \bsl{d}_\shpa-2 \bsl{V}_{j\shpa}^{\bsl{q}\bsl{q}'}\cdot \bsl{d}_\shpa) \} \ ,
}
where $\bsl{V}_{j\shpa}^{\bsl{q}\bsl{q}'}=(V_{jz}^{\bsl{q}\bsl{q}'},V_{jx}^{\bsl{q}\bsl{q}'})$, 
\eq{
G_a G_{a'}= G_a(\omega',\bsl{q}') G_{a'}(-\omega',\bsl{q}') \ ,
}
$\bsl{q}'$ dependence of $\hat{\bsl{g}}_\shpa$ and $d_{j}$ is implied, and we have used the fact that $G_1 G_{2}-G_2 G_{1}$ is odd in $\omega'$,r resulting in
\eq{
\sum_{\omega'}(G_1 G_{2}-G_2 G_{1})=0\ .
}
Combined with the inversion symmetry \eqnref{eq:int_P}, the above equation can be decoupled into
\eqa{
\label{eq:d_LGE_sim}
& d_y (\bsl{q}) = -\frac{1}{\beta}\sum_{\omega'}\sum_{\bsl{q}'\in \Omega_+} V_{yy}^{\bsl{q}\bsl{q}'} d_{y} ( G_1 G_1 - G_2 G_2) \\
& d_{j}(\bsl{q}) = -\frac{1}{\beta}\sum_{\omega'}\sum_{\bsl{q}'\in \Omega_+}\{ \sum_{j'\neq y} V_{jj'}^{\bsl{q}\bsl{q}'} d_{j'} ( G_1 G_1 + G_2 G_2) + (G_2 G_1 + G_1 G_2) ( V_{j0}^{\bsl{q}\bsl{q}'} \hat{\bsl{g}}_\shpa \cdot \bsl{d}_\shpa + \bsl{V}_{j\shpa}^{\bsl{q}\bsl{q}'} \cdot \hat{\bsl{g}}_\shpa d_0)  \\
& + G_2 G_2 ( 2 \bsl{V}_{j\shpa}^{\bsl{q}\bsl{q}'}\cdot \hat{\bsl{g}}_\shpa\ \hat{\bsl{g}}_\shpa \cdot \bsl{d}_\shpa-2 \bsl{V}_{j\shpa}^{\bsl{q}\bsl{q}'}\cdot \bsl{d}_\shpa) \} \text{for }j\neq y\ .
}
\end{widetext}
Therefore, owing to the inversion symmetry, the linearized gap equation is decoupled for $d_y$ and $d_{0,z,x}$, coinciding with the fact that they have opposite parities as shown in \tabref{tab:parity_pairing}.
In other words, different parities cannot mix into each other, and the highest $T_c$ is generally given by either (i) $d_y$ with $d_{0,z,x}=0$ or (ii) $d_{0,z,x}$ with $d_y=0$, justifying that we can consider them separately.

In the following, we will show that when the FS splitting is small $|m|\ll |\mu|$, it is very likely that the highest $T_c$ is given by $d_{0,z,x}$.
Besides the above approximation, it is also common to have $T_c\ll \epsilon_c$ with $\epsilon_c$ the SC cutoff.
Then, omitting all orders of $O(|m/\mu|)$ and $O(1/(\beta\epsilon_c))$, we have the following three useful formula arrive at
\eql{
&  \sum_{\omega',\bsl{q}'\in\Omega_+} \widetilde{f}(\bsl{q}')(G_1 G_2 + G_2 G_1 ) = 0 \\
& \sum_{\omega',\bsl{q}'\in\Omega_+} \widetilde{f}(\bsl{q}')(G_1 G_1 + G_2 G_2 ) = N(0)\beta\ln(\frac{2 e^{\gamma} \epsilon_c}{\pi T})\left\langle \widetilde{f}(\bsl{q}') \right\rangle_{\bsl{q}'\in FS_0} \\
& \sum_{\omega',\bsl{q}'\in\Omega_+} \widetilde{f}(\bsl{q}') G_2 G_2  = -\frac{N(0)\beta}{2}\left\langle \widetilde{f}(\bsl{q}') f(\rho_{\bsl{q}'})\right\rangle_{\bsl{q}'\in FS_0} \ ,
}
where $FS_0=\{ \bsl{q}\in\Omega_+ | \epsilon(\bsl{q})=0 \}$, $N(0)=\sum_{\bsl{q}}\delta(\epsilon(\bsl{q}))$ is the zero-energy density of state for $FS_0$, $\widetilde{f}(\bsl{q}')$ is strongly localized near $FS_0$,
\eq{
\left\langle \widetilde{f}(\bsl{q}') \right\rangle_{\bsl{q}'\in FS_0}=\frac{1}{N(0)}\sum_{\bsl{q}'}\delta(\epsilon(\bsl{q}'))\widetilde{f}(\bsl{q}')
}
is the average over $FS_0$, and $\gamma=0.5772...$ is the Euler's constant.
Moreover, 
\eq{
f(\rho_{\bsl{q}'})=\Re[\psi(\frac{1}{2})-\psi(\frac{1}{2}+\ii \rho_{\bsl{q}'})]\leq 0
}
with $ \rho_{\bsl{q}'}=\beta m |\bsl{g}_\shpa(\bsl{q}')|/(2\pi)$, $\psi$ is the di-gamma function.

With the above expressions, we can derive the form of $T_c$.
Let us start with $m=0$.
In this case, the critical temperature $T_{c,0}$ for $d_y^{(0)}$ reads 
\eq{
\ln (\frac{2 e^{\gamma} \epsilon_c}{\pi T_{c,0} } ) = \frac{1}{\lambda_y}
}
with  
\eq{
-N(0)\left\langle V_{yy}^{\bsl{q}\bsl{q}'} d_y^{(0)}(\bsl{q}') \right\rangle_{\bsl{q}'\in FS_0} = \lambda_y d_y^{(0)}(\bsl{q})\ ,
}
and the critical temperature $T_{c,0}$ for $d_{0,z,x}$ reads 
\eq{
\ln (\frac{2 e^{\gamma} \epsilon_c}{\pi T_{c,0} } ) = \frac{1}{\lambda_{0\shpa}}
}
with  
\eq{
-N(0)\sum_{j'\neq y}\left\langle V_{jj'}^{\bsl{q}\bsl{q}'} d_{j'}^{(0)}(\bsl{q}') \right\rangle_{\bsl{q}'\in FS_0} = \lambda_{0\shpa} d_{j}^{(0)}(\bsl{q}) \text{ for }j\neq y\ .
}
Here, the superscript $(0)$ means this is for $m=0$.

Then, for small nonzero $m$, the leading-order correction to the critical temperature can be included by substituting the above four expressions into \eqnref{eq:d_LGE_sim}.
We eventually arrive at the corrected $T_c$ for $d_y$ as \eqnref{eq:Tc_dy} with $f_{\bsl{q}}= f(\rho_{\bsl{q}})$,
while the corrected $T_c$ for $d_{0,z,x}$ is \eqnref{eq:Tc_d0zx} with $f_{\bsl{q}}= f(\rho_{\bsl{q}})$.
\eqnref{eq:Tc_dy} and \eqnref{eq:Tc_d0zx} suggests that while the nonzero $m$ would typically suppress the critical temperature for $d_y$, the critical temperature for $d_{0,z,x}$ can be un-suppressed by (i) aligning the nonzero $\bsl{d}_\shpa^{(0)}$ to $\hat{g}_{\shpa}^{(0)}$ or (ii) having a zero $\bsl{d}_\shpa^{(0)}$.
Therefore, given that the FS splitting should generally exist, $d_{0,z,x}$ channel tends to have higher $T_c$ than the $d_y$ channel.
In other words, the parity-even spin-singlet and the pairing-odd spin-triplet channels tend to have higher $T_c$ than the parity-even spin-triplet and the pairing-odd spin-singlet channels.

In conclusion, the linearized gap equation suggests that  (i) different spin channels (spin-singlet and spin-triplet) and different parities cannot mix into each other at the SC transition, allowing us to study them separately; (ii) the $\PT$-invariant pairing matrix \eqnref{eq:H_pairing_TR} can achieve the highest critical temperature, justifying \eqnref{eq:H_pairing_TR}; (iii) for small FS splitting $|m|\ll |\mu|$, the parity-even spin-singlet and the pairing-odd spin-triplet channels tend to have the highest critical temperature, justifying \eqnref{eq:ssPeven_stPodd} for small $m$.
The only remaining assumption that was made in the above discussion on the nodal SC is the dominant $\Delta_\shpa$.
As shown above, whether $\Delta_\shpa$ depends on the interaction strength, and we cannot show any tendency without knowing the interaction.
In the next section, we will use an effective model to show that $\bsl{d}_\shpa$ can dominate in a large regime of parameter values.
We emphasize that we only focus on the SC transition here, and have not considered the zero-temperature ground state of the SC.
In other words, although the highest $T_c$ can be achieved by the pairing \eqnref{eq:H_pairing_TR}, whether it is the zero-temperature ground state of the SC requires further discussion, which we leave as a future work.

\subsection{Extra $C_6$ Symmetry}

In this subsection, we include an extra $C_6$ symmetry to study the Euler obstructed Cooper pairing and the resultant nodal SC.
We focus on the zero-total-momentum parity-even spin-singlet $C_6$-odd weak pairing between two small sphere-like FS $\M_\pm$ that have nonzero Euler numbers, implying that each FS contains one MSDP.
As discussed above, we only focus on the nodes of the spin-up block \eqnref{eq:H_cal} near $\M_+$.
We will show that (i) the pairing matrix \eqnref{eq:H_pairing_TR} must vanish at two $C_6$ invariant points on $\M_+$, resulting in two zero-energy BdG MSDP in the full BdG Hamiltonian, and (ii) $\Delta_\shpa\sim O(q_r)$ and $\Delta_\perp\sim O(q_r^3)$ generally hold, meaning that the Euler obstructed pairing channel dominates over the trivial pairing channel, where $q_r$ roughly measures the radius of the FS.

Let start with a more general scenario where we consider a generic point group operation $g$ instead of just $C_6$.
We focus on the case where $g$ does not relate $\M_+$ to $\M_-$, since if it does, then we can consider $g \P$ instead of $g$.
Then, $g$ is represented as 
\eq{
g c^\dagger_{\pm,\bsl{q}}g^{-1} = c^\dagger_{\pm,g \bsl{q}} U_{\pm,g}(\bsl{q})\otimes e^{\ii \hat{n}_g\cdot\bsl{s} \theta_g}\ ,
}
where $\hat{n}_g$ is a real unit vector that depends on $g$, and $\theta_g\in \dsR$.
As $g$ commutes with TR and inversion symmetries, we have 
\eqa{
& U_{\pm,g}^*(\bsl{q}) = U_{\mp,g}(-\bsl{q})\\
& U_{\pm,g}^*(\bsl{q}) = U_{\pm,g}(\bsl{q})\ ,
}
where the representations of $\TR$ and $\P$ \eqnref{eq:reps_PT_T_P} have been used.
If we consider $g$-odd pairing, which means 
\eq{
g H_{pairing} g^\dagger = - H_{pairing}\ ,
}
we would have the following constrain for the pseudo-spin part 
\eq{
\label{eq:Delta_g_constraint}
-\Delta(\bsl{q}) = U_{+,g}(g^{-1} \bsl{q})\Delta(g^{-1} \bsl{q}) U_{+,g}^\dagger (g^{-1} \bsl{q})\ .
} 
Combined with \eqnref{eq:ssPeven_stPodd}, we have
\eqa{
\label{eq:d0_dpara_g}
& d_0(g^{-1} \bsl{q}) = -d_0(\bsl{q})\\
&  U_{+,g}(g^{-1} \bsl{q})(\bsl{d}_{\shpa}(g^{-1} \bsl{q})\cdot\bsl{\tau}_\shpa)  U_{+,g}^\dagger (g^{-1} \bsl{q})= -\bsl{d}_{\shpa}(\bsl{q})\cdot\bsl{\tau}_\shpa\ .
}

Now we restrict $g=C_6$.
Without loss of generality, we choose the $C_6$-invariant axis to align along $z$ direction.
According to \figref{fig:SpinlessDirac_NLSM}(c), there are two $C_6$ invariant points on $\M_+$, labelled as the north pole ($\bsl{q}=\bsl{q}_N$) and south pole ($\bsl{q}=\bsl{q}_S$). 
At each $C_6$ invariant point $\bsl{q}_0$, $U_{+,C_6}(\bsl{q}_0)$ should correspond to either $\Lambda_1$ or $\Lambda_2$, meaning that $U_{+,C_6}^3(\bsl{q}_0)=\pm \tau_0$.
As a result, \eqnref{eq:Delta_g_constraint} suggests that 
\eq{
-\Delta(\bsl{q}_0) = \Delta(\bsl{q}_0) \Rightarrow  \Delta(\bsl{q}_0)=0\ ,
}
meaning that the pairing matrix \eqnref{eq:H_pairing_TR} must vanish at two $C_6$ invariant points on $\M_+$.
At each $C_6$ invariant point, $\H$ in \eqnref{eq:H_cal} is zero, and thus the basis becomes the eigenstates.
Owing to the $C_6$-odd pairing, the BdG Hamiltonian does not have $C_6$ symmetry; instead, it have the modified $\widetilde{C}_6= C_{2z}^{spin}e^{\ii \hat{N} \pi/2} C_6 (C_{6z}^{spin})^{-1}$ symmetry, where we have used the fact that the BdG Hamiltonian has spin $SU(2)$ symmetry.
Then, we have 
\eq{
\widetilde{C}_6(c^\dagger_{+,\bsl{q},\uparrow}, c^T_{-,-\bsl{q},\downarrow}) \widetilde{C}_6^{-1} = (c^\dagger_{+,C_6\bsl{q},\uparrow}, c^T_{-,-C_6\bsl{q},\downarrow}) \rho_z U_{+,C_6}(\bsl{q})\ ,
}
resulting in 
\eq{
\rho_z U_{+,C_6}(C_6^{-1}\bsl{q})  \H(C_6^{-1}\bsl{q}) \rho_z U_{+,C_6}(C_6^{-1}\bsl{q}) = \H(\bsl{q})\ .
}
The modified $\widetilde{C}_6$ symmetry commutes with the effective $\PT$ symmetry, and $\widetilde{C}_6^6=1$ .
Therefore, the ICRs of the symmetry group spanned by the $\widetilde{C}_6$ and the effective $\PT$ symmetries are the same as \eqnref{eq:A_B_Lambda12}.
At $\bsl{q}_0$,  since $U_{+,C_6}(\bsl{q}_0)$ corresponds to either $\Lambda_1$ or $\Lambda_2$, we know $\rho_z U_{+,C_6}(C_6^{-1}\bsl{q})$ contains both $\Lambda_1$ and $\Lambda_2$, meaning that the four-fold degenerate point of $\H(\bsl{q})$ at $\bsl{q}_0$ is given by the crossing between $\Lambda_1$ and $\Lambda_2$.
Therefore, the four-fold degenerate point of $\H(\bsl{q})$ at $\bsl{q}_0$ is a MSDP, and it must have zero energy owing to the chiral symmetry.
As the $\widetilde{C}_6$, effective $\PT$ and chiral symmetries still hold in the full BdG Hamiltonian, then we have two zero-energy BdG MSDPs around near $\M_+$ per spin even in the full BdG Hamiltonian.

Now we show that the Euler obstructed pairing channel must dominate over the trivial pairing channel, unless fine-tuning is invoked.
To show this, we first split $\M_+$ into two patches---including north and south hemispheres---and we choose the gauges such that (i) $\bsl{d}_\shpa$ and $d_0$ are smooth in each patch and (ii) $U_{+,C_6}(\bsl{q})=U_{+,C_6}(\bsl{q}_N)$ in the north hemisphere and $U_{+,C_6}(\bsl{q})=U_{+,C_6}(\bsl{q}_S)$ in the south hemisphere.
According to \eqnref{eq:d0_dpara_g}, we have $d_0(\bsl{q})=-d_0(C_6^{-1}\bsl{q})$ in each patch, and thus $d_0(\bsl{q})= C_1 q_x (q_x^2-3 q_y^2)+C_2 q_y (q_y^2-3 q_x^2) + O(q^4)$, resulting that $d_0\sim O(q_r^3)$, unless $C_1$ and $C_2$ are finely tuned to zero.
Furthermore, \eqnref{eq:d0_dpara_g} also suggests $U_{+,C_6}(\bsl{q}_\alpha)(\bsl{d}_{\shpa}(C_6^{-1} \bsl{q})\cdot\bsl{\tau}_\shpa)  U_{+,C_6}^\dagger (\bsl{q}_\alpha)= -\bsl{d}_{\shpa}(\bsl{q})\cdot\bsl{\tau}_\shpa$ for the $\alpha$ path with $\alpha=N,S$.
Since $U_{+,C_6}(\bsl{q}_\alpha)$ takes the expression for either $\Lambda_1$ or $\Lambda_2$ in \eqnref{eq:A_B_Lambda12}, $\bsl{d}_{\shpa}(\bsl{q})=C (q_y,q_x)+O(q^2)$ for $\Lambda_1$ and $\bsl{d}_{\shpa}(\bsl{q})=C (q_x,q_y)+O(q^2)$ for $\Lambda_2$. resulting in $\bsl{d}_\shpa\sim O(q_r)$, unless $C$ are finely tuned to zero.
Then, $|\Delta_\perp|=|d_0|\sim O(q_r^3)$ and $|\Delta_\shpa|=|\bsl{d}_\shpa|\sim O(q_r)$ for this special gauge, and since $|\Delta_\perp|$ and $|\Delta_\shpa|$ are gauge-invariant, the statement holds for all gauges, verifying the dominant $\Delta_\shpa$.

\section{More Details on Effective and Tight-Binding Models}
\label{app:model}

In this section, we provide more details on the effective and tight-binding models.

\subsection{Effective Model}

As discussed in the main text, we consider the effective model similar to that of the 3D graphdiyne~\cite{Nomura2018GraphdiyneNLSM, Ahn2018MonopoleNLSM} as \eqnref{eq:H_pm_eff}.
The four spin-doubly-degenerate bands of $H_+^{eff}$ read
\eq{
\pm \sqrt{q_z^2+(\sqrt{q_x^2+q_y^2}\pm |m|)^2}\ .
}
Thus, the MNL exists at $q_z=0$ and $\sqrt{q_x^2+q_y^2}=|m|$, and its $\dsZ_2$ monopole charge is shown in \refcite{Nomura2018GraphdiyneNLSM, Ahn2018MonopoleNLSM}.
MNL of $H_-^{eff}$ can be obtained through TR symmetry.
For $m=0$, every band is doubly degenerate (in addition to the spin degeneracy), meaning that all FSs have no splitting, and the MNL shrinks to a nodal point. 
Therefore, $m$ measures the value of the FS splitting and the size of the nodal ring.
To verify the existence of the Euler obstructed Cooper pairing and the resultant nodal SC, we consider the pairing \eqnref{eq:pairing_eff_Podd_st}.

\subsubsection{Dominant Euler Obstructed Cooper Pairing and Nodal SC}

In this part, we provide more details about the nodal SC resulted from the dominant Euler obstructed Cooper pairing for the effective model. 

According to \appref{app:PandT_BdG_sym}, the BdG Hamiltonian can be splitting into two parts according to $\bsl{K}_\pm$, which are related by the particle-hole symmetry, and we should only need to consider the BdG Hamiltonian around $\bsl{K}_+$.
For the BdG Hamiltonian around $\bsl{K}_+$, it can also be splitted into two spin blocks as
\eq{
\label{eq:H_BdG_eff}
H_{BdG,+}^{eff}=H_{BdG,+,\uparrow}^{eff}+ H_{BdG,+,\downarrow}^{eff}\ ,
}
where 
\eq{
H_{BdG,s}^{eff}=\frac{1}{2}\sum_{\bsl{q}}\Psi_{BdG,+,s}^\dagger(\bsl{q}) h_{BdG,+,s}^{eff}(\bsl{q})\Psi_{BdG,+,s}(\bsl{q})\ ,
}
and $\Psi_{BdG,+,s}^\dagger(\bsl{q})= ( \Psi^{\dagger}_{\bsl{q},+,s}, -\ii \Psi_{-\bsl{q},-,-s} )$.
In particular, we have $h_{BdG,+,\uparrow}^{eff}(\bsl{q})=h_{BdG,-,\downarrow}^{eff}(\bsl{q})=\H_{eff}(\bsl{q})$, where
\eqa{
\label{eq:H_cal_eff}
& \H_{eff}(\bsl{q}) =\mat{ 
h_+^{eff}(\bsl{q})-\mu & \Delta_{eff}  \tau_{y}\sigma_y U_{\TR}^T\\
(\Delta_{eff}  \tau_{y}\sigma_y U_{\TR}^T)^\dagger &  -[h_-^{eff}(-\bsl{q})-\mu]^T
} 
}
where \eqnref{eq:H_pm_eff} and \eqnref{eq:pairing_eff_Podd_st} are used.
Then, we only need to study the spin-up Block $\H_{eff}$, and the plot the nodal structure of $\H_{eff}$ in \figref{fig:EOP_Nodal}(d,h).
We can see that the SC is nodal for both small and large $m$.
Furthermore, the odd linking number between the nodal rings and the degenerate lines between the top two occupied bands indicating the nodal rings have nonzero $\dsZ_2$ monopole charge.

\begin{figure*}[t]
    \centering
    \includegraphics[width=1.5\columnwidth]{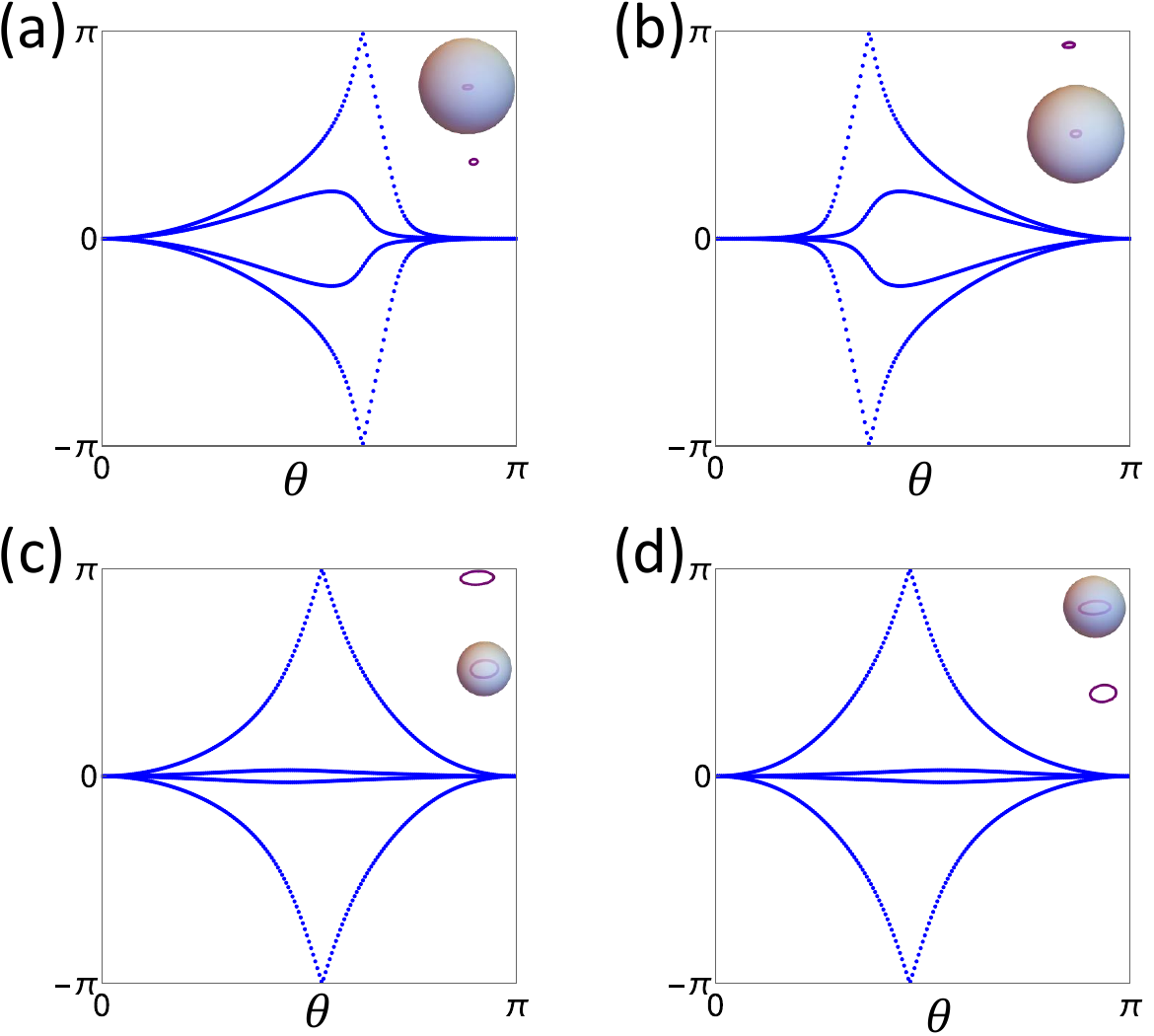}
    \caption{The phases of the eigenvalues of the Wilson loop $W_{\gamma_\theta}$ with the path $\gamma_\theta$ defined in \eqnref{eq:WL_path_theta}.
    In each plot, the blue lines are the phases as functions of the $\theta$, the red lines in the inset are the nodal rings, and the blue sphere in the inset marks the sphere formed by $\gamma_\theta$ with $\theta\in[0,\pi]$. 
    (a-b) are for the two nodal rings in \figref{fig:EOP_Nodal}(d), and (c-d) are for the two nodal rings in \figref{fig:EOP_Nodal}(h).
    }
    \label{fig:WL_spectrum}
\end{figure*}

We verify the nonzero $\dsZ_2$ monopole charge by directly calculating of the Wilson loop spectrum.
Explicitly, we choose the path for the Wilson loop
\eqa{
\label{eq:WL_path_theta}
& \gamma_{\theta}= \\
&\{\bsl{q}_0 + q (\sin(\theta)\cos(\phi), \sin(\theta)\sin(\phi), \cos(\theta)) | \phi\in [0,2\pi] \}\ ,
}
which is nothing but a circle with fixed polar angle $\theta$ as shown in \figref{fig:SpinlessDirac_NLSM}(d). 
We set $\bsl{q}_0 = (0,0,0.3), (0,0,-0.3)$ and $q=0.3$ for \figref{fig:EOP_Nodal}(d), and $\bsl{q}_0 = (0,0,0.36), (0,0,-0.36)$ and $q=0.2$ for \figref{fig:EOP_Nodal}(h).
With this choice, by varying $\theta$ from $0$ to $\pi$ in $\gamma_{\theta}$, the path form a sphere that encloses each nodal ring, as shown in \figref{fig:WL_spectrum}.
Then, we plot the phases of the eigenvalues of the Wilson loop $W_{\gamma_{\theta}}$ as functions of $\theta$ in \figref{fig:WL_spectrum}(a-d), which shows the Wilson loop winding number being $1$ for all the four nodal rings, verifying their nonzero $\dsZ_2$ monopole charges.

\subsubsection{Linearized Gap Equation}

In the above discussion, we assume the pairing form \eqnref{eq:pairing_eff_Podd_st}, which provides dominant Euler obstructed channels on FSs.
In this part, we discuss whether having \eqnref{eq:pairing_eff_Podd_st} is possible by studying the linearized gap equation for the effective model with small FS splitting.
We will demonstrate that \eqnref{eq:pairing_eff_Podd_st} is promising in a large region of parameter values by compairing \eqnref{eq:pairing_eff_Podd_st} to a trivial spin-singlet channel.

The interaction that we use reads
\eql{
\label{eq:H_eff_int}
& H^{eff}_{int} =V_1 \sum_{\bsl{q}}\Psi_{+,\bsl{q}}^\dagger  \tau_0 \sigma_0 U_T^T\otimes \ii s_y (\Psi_{-,-\bsl{q}}^\dagger)^T \sum_{\bsl{q}'}\Psi_{-,-\bsl{q}'}^T  (\tau_0 \sigma_0 U_T^T)^\dagger\otimes \ii s_y \Psi_{+,+\bsl{q}} \\
& + V_2 \sum_{j_s=x,y,z} \sum_{\bsl{q}}\Psi_{+,\bsl{q}}^\dagger  \tau_y \sigma_y U_T^T\otimes \ii \widetilde{s}_{j_s}\ii s_y (\Psi_{-,-\bsl{q}}^\dagger)^T \sum_{\bsl{q}'}\Psi_{-,-\bsl{q}'}^T  (\tau_y \sigma_y U_T^T)^\dagger\otimes (\ii \widetilde{s}_{j_s}\ii s_y)^\dagger\Psi_{+,+\bsl{q}}\ ,
}
where $\tau_0 \sigma_0 U_T^T$ is the trivial channel and $V_2$ is for \eqnref{eq:pairing_eff_Podd_st}.
By projecting both channels to the occupied bands of $H_{\pm}^{eff}$ at $m=0$, we find that the $\tau_0 \sigma_0 U_T^T$ channel has zero $\Delta_\shpa$ and a dominant $\Delta_\perp$, while the $\tau_y \sigma_y U_T^T$ channel  has zero $\Delta_\perp$ and a dominant $\Delta_\shpa$.
We also find that both channels cannot be suppressed by the small $m$, and thus we can focus on the critical temperatures at $m=0$.
Using \eqnref{eq:Tc_d0zx}, we arrive at
\eq{
\log(\frac{T_{c,1}^{(0)} \pi}{2e^\gamma \epsilon_c}) = \lambda_1
}
for the $\tau_0 \sigma_0 U_T^T$ channel, and
\eq{
\log(\frac{T_{c,2}^{(0)} \pi}{2e^\gamma \epsilon_c}) = \frac{2}{3}\lambda_2 
}
for the $\tau_y \sigma_y U_T^T$ channel, where $\lambda_{1,2}=- N(0) V_{1,2}$.
Then, when $\lambda_1> 2 \lambda_2/3$, the $\tau_0 \sigma_0 U_T^T$ channel gives higher $T_c$, resulting in dominant $\Delta_{\perp}$ for small $m$.
When $\lambda_1< 2 \lambda_2/3$, \eqnref{eq:pairing_eff_Podd_st} gives higher $T_c$, resulting in dominant $\Delta_{\shpa}$ for small $m$.
Therefore, for the interaction chosen as \eqnref{eq:H_eff_int}, \eqnref{eq:pairing_eff_Podd_st} can dominate in a large parameter region, as shown in \figref{fig:LGE}.

\subsubsection{Hinge Majorana Modes}

In this part, we show more details on the hinge Majorana zero modes in each spin block of the effective BdG Hamiltonian.
Similar as above, two spin blocks of the BdG Hamiltonian have the same dispersion, and thus we only need to study the spin-up block $\H_{eff}$ (\eqnref{eq:H_cal_eff}).
We solve for the hinge Majorana zero modes by performing $q_x\rightarrow -\ii \partial_x$ and $q_y\rightarrow -\ii \partial_y$ in \eqnref{eq:H_cal_eff}.
Without loss of generality, we focus on the BdG MNL near $q_z=\sqrt{\mu^2+\Delta_2^2}$.
The zero mode equation reads
\eq{
\H_{eff}(-\ii\partial_x,-\ii\partial_y,\sqrt{\mu^2+\Delta_2^2}+m_z) \chi(x,y,m_z)=0\ .
}
\begin{widetext}
However, exactly solving this equation is difficult, and we resort to an approximate equation as 
\eq{
\label{eq:zero_modes_approx}
\chi(x,y,m_z)^\dagger \H_{eff}(-\ii\partial_x,-\ii\partial_y,\sqrt{\mu^2+\Delta_2^2}+m_0) \chi(x,y,m_z) =0\ .
}
To solve \eqnref{eq:zero_modes_approx}, first note that $\H_{eff}(0,0,\sqrt{\mu^2+\Delta_2^2})$ with $m=0$ has four zero-energy eigenvectors forming a matrix $U_{proj,1}$.
Then, the projection of $\H_{eff}(-\ii\partial_x,-\ii\partial_y,\sqrt{\mu^2+\Delta_2^2}+m_z)$ onto this zero-energy subspace reads
\eqa{
\label{eq:H_eff_proj}
\H_{eff}^{proj}(-\ii\partial_x,-\ii\partial_y,m_z)
=U_{proj,1}^\dagger \H_{eff}(-\ii\partial_x,-\ii\partial_y,\sqrt{\mu^2+\Delta_2^2}+m_z) U_{proj,1} 
= m_z \tau_z \sigma_0 -\ii \partial_x \tau_x \sigma_x -\ii \partial_y \tau_x\sigma_z + m \tau_y \sigma_y\ ,
}
where we have perform $(x,y)\rightarrow (x,y) \frac{\Delta}{\sqrt{\Delta_2 ^2+\mu ^2}}$ and $m\rightarrow m \sqrt{\Delta^2 +\mu^2}/\Delta$.
For solving boundary modes, we need to introduce the boundary, which is usually done by making the mass term spacial dependent.
The mass term here is $m_z$ term, since it anticommutes with the $\partial_{x}$ and $\partial_{y}$ terms, and we choose 
\eq{
m_z(r) = \left\{ 
\begin{array}{ll}
m_0      & r<R \\
\infty     & r\geq R 
\end{array}
\right.\ ,
}
where $m_0$ is the bulk value of $m_z$, $r=\sqrt{x^2+y^2}$, and $R\gg 1/m_0$.
It stands for a circular open boundary condition with boundary at $r=R$.
Then, by converting to the polar coordinate $(x,y)=r (\cos\theta, \sin\theta)$ and choosing $\chi(x,y,m_0)=U_{proj,1}\chi_1(x,y,m_0)$, the approximate zero mode equation becomes
\eq{
\label{eq:chi_1}
\chi_1^\dagger(x,y,m_0) [m_z(r) \Gamma_3 -\ii \partial_r \Gamma_1 -\ii\Gamma_2  \frac{1}{r}\partial_\theta+ m \tau_y \sigma_y] \chi_1(x,y,m_0) =0\ ,
}
where $\Gamma_3=\tau_z \sigma_0$, $\Gamma_1=\cos\theta \tau_x\sigma_x + \sin\theta \tau_x\sigma_z$, and $\Gamma_2=-\sin\theta \tau_x\sigma_x + \cos\theta \tau_x\sigma_z$.
\end{widetext}

To solve the above equation, we first solve 
\eq{
[m_z(r) \Gamma_3 -\ii \partial_r \Gamma_1] f(r) \chi'_{\theta} = 0\ .
}
In general, we can make $\chi'_{\theta}$ the eigenvector of  $\ii \Gamma_1 \Gamma_3$ as  $\ii \Gamma_1 \Gamma_3\chi'_{\theta} = \pm \chi'_{\theta}$, resulting in
\eq{
f(r)=\frac{1}{A}\exp[\mp\int_0^r m_z(r') dr']
}
with $A$ the normalization factor.
Since we require $f(r\rightarrow \infty) =0 $, we should choose $\ii \Gamma_1 \Gamma_3\chi'_{\theta} =  \chi'_{\theta}$, and thus we have 
\eq{
f(r)=\frac{1}{A}\exp[-\int_0^r m_z(r)]
}
which is localized at $r=R$ for $m_0<0$ and at $r=0$ for $m_0>0$.
Furthermore, $\chi'_{\theta}$ can takes two values, forming a matrix $U'_{\theta}$.

Now we consider $m_0<0$, for which is $f(r)$ localized at $r=R$.
We can express the solution of $\chi_1(x,y,m_0)$ as $ f(r) U'_{\theta} \chi_2(r,\theta)$, and the equation for $\chi_2(r,\theta)$ is 
\eq{
 \chi_2^\dagger (r,\theta) (-\ii \tau_z \partial_\theta -\frac{1}{2}) \chi_2(r,\theta) =0\ .
}
The equation has two solutions as $(e^{\ii \theta/2},0)$ and $(0,e^{-\ii \theta/2})$, standing for two counter-propagating chiral ``edge" modes.
As a result, we arrive at two solutions to \eqnref{eq:chi_1} as two columns of the following matrix
\eq{
U_{proj}(r,\theta) =  f(r) U_{proj,1} U'_{\theta} \mat{ e^{\ii \theta/2} & 0 \\ 0 & e^{-\ii \theta/2} }\ .
}

However, the two counter-propagating chiral surfaces modes for $m_0<0$ are artifacts of the special form of the $\H_{eff}$ in \eqnref{eq:H_cal_eff}.
If we add a symmetry-preserving mass term to the normal-state Hamiltonian as \eqnref{eq:H_eff_extra}, then the solution to the modified \eqnref{eq:zero_modes_approx} has the form $U_{proj}(r,\theta)V_{r,\theta}$, and the equation for $V_{r,\theta}$ is given in \eqnref{eq:zero_mode_eq}.
The above equation suggests that $V_{r,\theta}$ can be made independent of $r$, and $V_{r,\theta}$ has two zero-mode solutions localized at $\theta=\pi/4$ and $5\pi/4$, where $(-\cos\theta+ \sin\theta)$ changes sign.
In other words, the extra mass term turns the counter-propagating chiral surface modes into localized zero-energy hinge modes for $m_0<0$.

Although those hinge zero modes have nonzero momentum along $z$, they still have the Majorana nature.
It is because if a zero mode exist at $k_z$ with creation operator $b^\dagger_{k_z}$, we have $[b^\dagger_{k_z}, H_{BdG}]=0$, meaning that $[b_{k_z}, H_{BdG}]=0$.
Owing to the particle hole redundancy, we can combine $b_{k_z}$ and $b^\dagger_{k_z}$ to get two Majorana operators and assign $k_z$ and $-k_z$ to them, resulting in two Majorana modes at two momenta.
Thus, those hinge zero modes can be called hinge MZMs.

\subsection{Tight-Binding Model}

In this subsection, we will verify the existence of the hinge MZMs with a tight-binding model.
We will further demonstrate that the hinge MZMs are stable against weak chemical potential and magnetic disorder that preserves the symmetries on average.

We start with a normal-state tight-binding model that precisely reproduces the normal-state effective model \eqnref{eq:H_pm_eff} together with the extra term \eqnref{eq:H_eff_extra} at low energy.
We consider a 3D lattice with primitive lattice vectors as $\bsl{a}_1=(\frac{\sqrt{3}}{2},\frac{1}{2},0)$, $\bsl{a}_2=(-\frac{\sqrt{3}}{2},\frac{1}{2},0)$, and $\bsl{a}_3=(0,0,1)$, where we have adopted the unit system in which the lattice constant is 1.
On each lattice site, we include four atomic orbitals.
Together with spin, the real-space basis of the tight-binding model can be labelled as $\ket{W_{\bsl{R},a,s}}$, where $\bsl{R}\in\dsZ \bsl{a}_1 + \dsZ \bsl{a}_2 + \dsZ \bsl{a}_3$ the lattice vector, $a=1,2,3,4$ labels the four orbitals, and $s=\uparrow,\downarrow$ labels the spin.
Specifically, we choose the $a=1,2$ atomic orbitals to be imaginary parity-even, and choose the $a=3,4$ atomic orbitals to be real parity-odd.
Then, the TR and inversion symmetries are represented as 
\eqa{
& \TR \ket{W_{\bsl{R}}} = \ket{W_{\bsl{R}}}U_{\TR}\otimes \ii s_y \\
& \P \ket{W_{\bsl{R}}} = \ket{W_{-\bsl{R}}}U_{\TR}\otimes s_0\ ,
}
where $U_{\TR}=\tau_z\sigma_0$,  $\ket{W_{\bsl{R}}}=(...,\ket{W_{\bsl{R},a,s}},...)$, and $\tau$ and $\sigma$ are Pauli matrices for the index $a$.
We can define $\Psi^\dagger_{\bsl{R}}=(...,c^\dagger_{\bsl{R},a,s},...)$ as the creation operators $\ket{W_{\bsl{R}}}$, and then derive the creation operators for the Bloch basis with the Fourier transformation 
\eq{
\Psi^\dagger_{\bsl{k}}=\frac{1}{\sqrt{N_R}}\sum_{\bsl{R}} e^{\ii \bsl{k}\cdot \bsl{R}} \Psi^\dagger_{\bsl{R}}\ ,
}
where $N_R$ is the number of lattice sites.
With $\Psi^\dagger_{\bsl{k}}$, the TR and inversion symmetries are represented as 
\eqa{
& \TR \Psi^\dagger_{\bsl{k}} (\TR)^{-1} = \Psi^\dagger_{-\bsl{k}}U_{\TR}\otimes  \ii s_y\\
& \P \Psi^\dagger_{\bsl{k}} (\P)^{-1}= \Psi^\dagger_{-\bsl{k}}U_{\TR}\otimes s_0\ .
}

With the basis, we construct the normal-state tight-binding model using onsite terms and nearest-neighboring hopping as
\eq{
\label{eq:H_TB}
H_{TB}=\sum_{\bsl{k}} \Psi^\dagger_{\bsl{k}} h_{TB}(\bsl{k})\otimes s_0 \Psi_{\bsl{k}}\ ,
}
where
\eq{
h_{TB}(\bsl{k}) = h_1(\bsl{k}) \tau_z\sigma_0 + h_2(\bsl{k}) \tau_y \sigma_y + h_3(\bsl{k}) \tau_x \sigma_x + h_4(\bsl{k}) \tau_x \sigma_z + M\ ,
}
\eqa{
& h_1 = m_0 + t_0 \cos k_z  + t_3 [2  \cos (\frac{\sqrt{3} k_{x}}{2}) \cos (\frac{k_{y}}{2})+ \cos k_{y} ] \\
& h_2 = t_1 \sin k_z \\
& h_3 = -\sqrt{3} t_2  \cos(\frac{k_y}{2})\sin(\frac{\sqrt{3}k_x}{2}) \\
& h_4=t_2 [\cos(\frac{\sqrt{3} k_x}{2})\sin(\frac{k_y}{2})+\sin(k_y)]\ ,
}
and 
$M=m_1 (\tau_0\sigma_x+\tau_0\sigma_z)$ is the extra mass term introduced in \eqnref{eq:H_eff_extra} for the study of the hinge modes.
$H_{TB}$ preserves both inversion and TR symmetries for real $m_{0,1}$ and $t_{0,1,2,3}$.
In the following, we choose 
\eq{
t_0 = -1,\ m_0 =3,\ t_1 = m,\ t_2 = -\frac{2}{3},\ t_3 = -1\ .
}
With these parameter values, we can choose $\bsl{K}_\pm=(0,0,\pm \frac{\pi}{2})$, and expanding $h_{TB}(\bsl{q}+\bsl{K}_\pm)$ in \eqnref{eq:H_TB} around $\bsl{K}_\pm$ to the first order of $\bsl{q}$ exactly reproduces \eqnref{eq:H_pm_eff} together with \eqnref{eq:H_eff_extra}.

We further introduce an on-site pairing term as 
\eq{
\label{eq:H_TB_pairing}
H_{TB,pairing}=\frac{1}{2}\sum_{\bsl{k}} \Psi^\dagger_{\bsl{k}}\widetilde{\Delta} (\tau_y\sigma_y U_{\TR}) \otimes \ii s_x \Psi_{-\bsl{k}}+h.c.\ ,
}
which reproduces the pairing in the effective model \eqnref{eq:pairing_eff_Podd_st} and preserves $\PT$ symmetry for real $\widetilde{\Delta}$.
Then, the tight-binding BdG Hamiltonian reads
\eql{
\label{eq:H_TB_BdG}
H_{TB,BdG}=\frac{1}{2}\sum_{\bsl{k}}\Psi^\dagger_{BdG,\bsl{k}}
\mat{
(h_{TB}(\bsl{k})-\mu)\otimes s_0 & \widetilde{\Delta} (\tau_y\sigma_y U_{\TR}) \otimes \ii s_x\\
[\widetilde{\Delta} (\tau_y\sigma_y U_{\TR}) \otimes \ii s_x]^\dagger& -(h_{TB}(-\bsl{k})-\mu)^T\otimes s_0
}
\Psi_{BdG,\bsl{k}}\ .
}
\eqnref{eq:H_TB_BdG} serves as a legitimate UV completion of the effective BdG Hamiltonian \eqnref{eq:H_BdG_eff} with the extra mass term \eqnref{eq:H_eff_extra}.

According to \appref{app:PandT_BdG_sym}, the tight-binding BdG model can be separated into two equivalent spin blocks as $h_{TB,BdG,\uparrow}(\bsl{k})=h_{TB,BdG,\downarrow}(\bsl{k})=\H_{TB}(\bsl{k})$ with
\eq{
\label{eq:H_cal_TB}
\H_{TB}(\bsl{k})= \mat{
h_{TB}(\bsl{k})-\mu & \widetilde{\Delta} (\tau_y\sigma_y U_{\TR}) \\
[\widetilde{\Delta} (\tau_y\sigma_y U_{\TR})]^\dagger& -(h_{TB}(-\bsl{k})-\mu)^T
}\ .
}
$\H_{TB}(\bsl{k})$ has particle-hole symmetry
\eq{
\rho_x \H_{TB}^T(-\bsl{k}) \rho_x = - \H_{TB}(\bsl{k})\ ,
}
chiral symmetry
\eq{
\rho_y U_{\TR} \H_{TB}(\bsl{k}) (\rho_y U_{\TR})^\dagger = -\H_{TB}(\bsl{k})\ ,
}
effective $\PT$ symmetry
\eq{
\H_{TB}^*(\bsl{k}) = \H_{TB}(\bsl{k})\ .
}
\eqnref{eq:H_cal_TB} is a UV completion of \eqnref{eq:H_cal_eff} with the extra mass term \eqnref{eq:H_eff_extra}.

\subsubsection{Hinge MZMs}

In this part, we will use \eqnref{eq:H_TB_BdG} to demonstrate the existence of the hinge Majorana zero modes for
\eq{
m=0.1, \mu=-0.3, \widetilde{\Delta}=0.2,\ m_1 =0.05\ .
}
According to the above discussion on the effective model, if we choose open boundary condition along $x$ and $y$ but keep periodic boundary condition along $z$, we should find the BdG Hamiltonian should have hinge MZMs with $k_z$ near $K_{\pm,z}$ as it lies in between two BdG MNLs with nonzero $\dsZ_2$ monopole charge.
Furthermore, the hinge modes should appear at $\theta=\pi/4$ and $\theta=5\pi/4$ on the boundary, where $\theta$ measures the angle from the $x$ axis in the $x-y$ plane.
To verify this, we choose a square geometry with length 50 in the $x,y$ plane, and plot the energy dispersion of one spin block \eqnref{eq:H_cal_TB} of the tight-binding BdG Hamiltonian in \figref{fig:Hinge}(a), which shows a Majorana flat band.
Here the length does not represent the number of lattice sites on one edge since the lattice is triangular in the $x,y$ plane, and we only show the dispersion for one block because the energy dispersion is the same for two spin blocks.
We further plot the zero-energy LDOS of one spin block \eqnref{eq:H_cal_TB} on the $x-y$ plane with $k_z=\pi/2$ in \figref{fig:Hinge}(b), which shows two zero-energy hinge modes localized at $\theta=\pi/4$ and $\theta=5\pi/4$.
%
%
In the above discussion, we only care about one spin blocks.
In total, we should have two hinge MZMs with opposite spins for one value of $k_z$ on one hinge, and they are related by the combined charge and spin rotation.

\subsubsection{Stability Against Disorder}

In this part, we study the stability of the hinge MZMs against disorder.
Unlike the above discussion, we will focus on the spinful tight-binding BdG Model \eqnref{eq:H_TB_BdG} since we will need to introduce the magnetic disorder. 
To introduce disorder, we need to consider a 3D finite system with open boundary condition along $x$, $y$ and $z$.
Specifically, we choose the lengths of the finite systems along $x$, $y$ and $z$ to be $L_x=20$, $L_y=20$ and $L_z=10$, respectively.
Again, the lengths along $x$ and $y$ do not stand for the number of lattice sites owing to the triangular lattice in the $x-y$ plane, while the length along $z$ does represent of the number of layers along $z$.

Let us first consider the 3D finite system in the clean limit.
We diagonalize the 3D clean system, and only keep 40 eigenvectors with the smallest energies, where being small means having small absolute value.
Then, from the 40 eigenvectors $V_i$ and their energies $E_i$ with $i=1,...,40$, we can approximately derive the zero energy Green function as 
\eq{
G=\sum_{i=1,...,40}\frac{1}{-E_i+\ii 0^+} V_i V_i^\dagger\ .
}
Here $V_i$ is a column vector; it carries the index $(x,y,z)$ that takes values from the 3D finite lattice, the orbital index $a$, and the spin index $s$.
The Green function $G$ is actually a matrix with entry as
\eq{
G(x,y,z,a,s;x',y',z',a',s')
}
We derive the zero-energy LDOS on $(x,y)$ plane with fixed $k_z$ from the Green function as~\cite{Wilson2018WSMDisorder}
\eql{
A_{x,y,k_z} = - \frac{1}{\pi}\Im[ \sum_{a,s}\frac{1}{L_z}\sum_{z,z'=1,...,L_z} e^{\ii k_z (z-z')} G(x,y,z',a,s;x,y,z,a,s)]\ ,
}
and we plot the LDOS in \figref{fig:Hinge_Disorder}(a).
Clearly, we can see the zero-energy hinge MZMs similar as \figref{fig:Hinge}(b), meaning that the procedure above can reproduce the above result for periodic boundary condition $z$.

We now add the chemical potential disorder as
\eq{
H_{cd}=\sum_{\bsl{R}} V(\bsl{R}) \Psi^\dagger_{\bsl{R}} \Psi_{\bsl{R}}\ ,
}
and consider $H_{TB,BdG}+H_{cd}$.
The values of $V(\bsl{R})$ at each $\bsl{R}$ obeys a Gaussian distribution with zero mean and standard deviation $W=0.1$, and the distributions at different $\bsl{R}$'s are independent of each other. 
We have $V(\bsl{R})=0$ on average, and thus the disorder preserves all symmetries of $H_{TB,BdG}$ on average.
For each disorder configuration $V(\bsl{R})$, we derive the zero-energy LDOS on the $x-y$ plane with $k_z=\pi/2$.
Then, we average the LDOS over 10 configurations and plot the result in \figref{fig:Hinge_Disorder}(b), which shows that the hinge MZMs are stable against the chemical disorder.

We now consider the magnetic disorder as
\eq{
H_{md}=\sum_{\bsl{R}}  \Psi^\dagger_{\bsl{R}} \tau_0\sigma_0\otimes(V_x(\bsl{R})s_x+V_y(\bsl{R})s_y+V_z(\bsl{R})s_z) \Psi_{\bsl{R}}\ .
}
and consider $H_{TB,BdG}+H_{md}$.
The values of $V_{j}(\bsl{R})$ ($j=x,y,z$) at each $\bsl{R}$ obeys a Gaussian distribution with zero mean and standard deviation $W=0.1$, and the distributions at different $\bsl{R}$'s and for different $j$'s are independent of each other. 
In average, we have $V_{j}(\bsl{R})=0$, and thus the disorder preserves all symmetries of $H_{TB,BdG}$.
We plot the zero-energy LDOS on the $x-y$ plane with $k_z=\pi/2$ averaged over 10 configurations in \figref{fig:Hinge_Disorder}(c).
We can see that the hinge modes in \figref{fig:Hinge_Disorder}(c) have smaller LDOS than those in \figref{fig:Hinge_Disorder}(b), meaning that the magnetic disorder has a stronger effect on the hinge MZMs than the chemical potential disorder.

Yet, compared to the actual symmetry breaking, the hinge MZMs can still be treated to be stable against the magnetic disorder.
To show this, we, in \figref{fig:Hinge_Disorder}(d), also plot the zero-energy LDOS on the $x-y$ plane with $k_z=\pi/2$ for the clean system but in the presence of a Zeeman field term as
\eq{
B \sum_{\bsl{R}} \Psi^\dagger_{\bsl{R}} \tau_0\sigma_0
\otimes \s_z\Psi_{\bsl{R}}\ ,
}
with $B=0.02$.
The Zeeman field breaks the chiral symmetry in each spin subspace, and thus gap out the two hinge modes with opposite spins, resulting in very small zero-energy LDOS at the hinges.
Although the Zeeman field strength $B=0.02$ is much smaller than the disorder strength $\sim 0.15$, the zero-energy LDOS at the corner in \figref{fig:Hinge_Disorder}(d) is much smaller than that in \figref{fig:Hinge_Disorder}(c), showing the stability of the hinge MZMs against the magnetic disorder. 

\end{document}